\documentstyle[11pt,aaspp4]{article}

\def\HI{\ion{H}{1}}
\def\sb{\ifmmode{\;{\rm mag}\;{\rm arcsec}^{-2}}\else{~mag~arcsec$^{-2}$}\fi}
\def\linsb{\ifmmode{\;L_{\sun}{\rm pc}^{-2}}\else{~$L_{\sun}{\rm pc}^{-2}$}\fi}
\def\csb{\ifmmode{\mu_0}\else{$\mu_0$}\fi}
\def\lincsb{\ifmmode{\Sigma_0}\else{$\Sigma_0$}\fi}
\def\solar{\ifmmode{_{\sun}}\else{$_{\sun}$}\fi}
\def\eg{e.g.,\ }
\def\ie{i.e.,\ }

\def\vs{vs.\ }
\def\etal{{et al}.\ }
\def\MLo{\ifmmode{\Upsilon_o}\else{$\Upsilon_o$}\fi}
\def\ML*{\ifmmode{\Upsilon_*}\else{$\Upsilon_*$}\fi}
\def\MLT{\ifmmode{\Upsilon_T}\else{$\Upsilon_T$}\fi}
\def\mass{\ifmmode{\cal M}\else{${\cal M}$}\fi}
\def\kms{\ifmmode{\;{\rm km}{\rm s}^{-1}}\else{~${\rm km}{\rm s}^{-1}$}\fi}

\begin{document}
\title{Testing the Dark Matter Hypothesis
       with Low Surface Brightness Galaxies and Other Evidence}
\author{Stacy S. McGaugh\altaffilmark{1}}
\affil{Department of Terrestrial Magnetism \\ Carnegie Institution of
Washington \\ 5241 Broad Branch Road, NW \\ Washington, DC 20015}
\and
\author{W. J. G. de Blok\altaffilmark{2}}
\affil{Kapteyn Astronomical Institute \\
Postbus 800 \\ 9700 AV Groningen \\ The Netherlands}

\altaffiltext{1}{Present Address: Physics and Astronmy, Rutgers University,
136 Frelinghuysen Road, Piscataway, NJ 08854-8019}
\altaffiltext{2}{Present Address: Astrophysics Groups, School of Physics,
University of Melbourne, Parkville, Victoria 3052, Australia}


\begin{abstract}

The severity of the mass discrepancy in spiral galaxies is
strongly correlated with the central surface brightness of the disk.
Progressively lower surface brightness galaxies have ever larger mass
discrepancies.  No other parameter (luminosity, size, velocity, morphology)
is so well correlated with the magnitude of the mass deficit.

The rotation curves of low surface brightness disks thus provide
a unique data set with which to probe the dark matter distribution in
galaxies.  The mass discrepancy is apparent from $R = 0$ giving a
nearly direct map of the halo mass distribution.
The luminous mass is insignificant.

Interpreting the data in terms of dark matter leads to troublesome fine-tuning
problems.  Different observations require contradictory amounts of dark
matter.  Structure formation theories are as yet far from able to explain the
observations.

\end{abstract}


\begin{quote}
\raggedleft
What gets us into trouble is not what we don't know. \\
It's what we know for sure that just aint so. \\
--- Yogi Berra
\end{quote}


\section{The Problem}

The evidence for the existence of dark matter is clear
in a great variety of data (Trimble\markcite{VT} 1987).  These include
the Oort discrepancy in the disk of the Milky Way (Kuijken \&
Gilmore\markcite{KG} 1989; Bahcall\markcite{BFG} \etal 1992), the
velocity dispersions of dwarf Spheroidal galaxies (\eg
Vogt\markcite{VMOK} \etal 1995), the flat rotation curves of spiral
galaxies (Rubin\markcite{RFT} \etal 1980; Bosma\markcite{Bosma} 1981),
the statistics of satellite galaxy orbits (Zaritsky \&
White\markcite{ZW} 1994), the timing argument in the Local Group (Kahn
\& Woltjer\markcite{KW} 1959), the velocity dispersions of clusters of
galaxies (Zwicky \& Humason\markcite{ZH} 1964), bulk flows on large
scales (Lynden-Bell\markcite{7sam} \etal 1988; Mould\markcite{MABHHRS}
\etal 1993), the excess of mass density in the universe over that in
visible in galaxies ($\Omega_{dyn} \gg \Omega_{gal}$; Ostriker \&
Steinhardt\markcite{OS} 1995), and gravitational lensing
(Tyson\markcite{TWV} \etal 1990).

What these data really demonstrate is that the
observed distribution of luminous matter together with the usual
dynamical equations can not reproduce the observations.  This can be
interpreted {\it either\/} to require dark matter, {\it or\/} as a
need to modify the equations we use (\eg the law of gravity).  In
this series of papers we examine and compare the pros and cons of
both alternatives.  Here we shall examine the
dark matter hypothesis.  In companion papers [McGaugh \& de
Blok\markcite{pII} 1998 (paper II); de Blok \& McGaugh\markcite{pIII}
1998 (paper III)], we examine alternative dynamical theories.

The \HI\ rotation curves of disk galaxies provide powerful tests of
the various hypotheses (\eg Kent\markcite{K87} 1987; Begeman\markcite{BBS}
\etal 1991).  These share the general characteristic of becoming
asymptotically {\it flat\/} at large radii.  One would expect
$V(R) \propto R^{-1/2}$ without dark matter.

The major advantage of using \HI\ rotation curves is that the geometry of
the orbits is obvious.  Dissipation in the gas enforces circular orbits.
Thus it is possible to directly equate the centripetal acceleration
\begin{equation}
a_c = \frac{V^2}{R}
\end{equation}
with the gravitational acceleration
\begin{equation}
g_N = - \frac{\partial \varphi}{\partial R}
\end{equation}
determined from the Poisson equation
\begin{equation}
\nabla^2 \varphi = 4 \pi G \rho
\end{equation}
in order to predict the expected form of the rotation curve $V(R)$ from
any given mass distribution $\rho(R)$.
In no other type of system are tests so direct and free of assumptions.

An important aspect of any test is probing a large dynamic range in
the relevant parameters.  It turns out that the luminous surface
density of disks is a critical parameter (see \S 4).  So the
important thing is to have a large dynamic range in surface
brightness.  This is
exemplified by the fact that the few good rotation curves of low
surface brightness dwarf galaxies which exist are constantly being
reanalyzed (\eg Flores \& Primack\markcite{FP} 1994;
Moore\markcite{Moore} 1994) precisely because they provide
leverage for testing ideas
principally motivated by data for high surface brightness (HSB) spirals.
Here, we augment existing data with our own data for low surface
brightness (LSB) disk galaxies (van der Hulst\markcite{vdH} \etal
1993; de Blok\markcite{BMH} etal 1996) to
probe the lower extremes of surface brightness with many more galaxies
than previously available thus extending the dynamic range in the critical
parameter of luminous surface density.

One important result is the universality of the Tully-Fisher relation
across this increased dynamic range in surface brightness
(Zwaan\markcite{ZHBM} \etal 1995; Sprayberry\markcite{lsbtf} \etal 1995).
This requires a surprising fine-tuning between the optical properties
(central surface brightness) of a galaxy and its halo (mass-to-light
ratio).  Zwaan\markcite{ZHBM} \etal (1995) suggested that galaxies
must become progressively more dark
matter dominated towards lower surface brightnesses.  This conclusion
was confirmed by \HI\ rotation curves (de Blok\markcite{BMH} \etal 1996)
which require a greater ratio of dark to luminous mass at any given radius
in LSB relative to HSB galaxies
(de Blok \& McGaugh\markcite{dBM1} 1997).  Another essential result
is the way in which the shape of rotation curves changes systematically
with surface brightness from steeply rising curves in HSB galaxies to
slowly rising curves in LSB galaxies.

The challenge is to fit these and other observations into a
consistent and coherent picture of the formation and evolution of
dark matter halos and their associated spiral disks. In this
paper we examine the difficulties encountered in undertaking this task.
As yet, it is impossible to develop such a picture
without resorting to a large number of fine-tuned
relations between supposedly independent galaxy properties.

In section 2, we describe the data.  In section 3, we define symbols
and clarify terms.  Section 4 gives a summary of the most relevant
empirical facts.  The physical interpretation of the Tully-Fisher relation
is discussed in \S 5.  In \S 6 we introduce and test a variety of
galaxy formation models.  Section 7 tests various dark matter candidates
and \S 8 discusses the implications of baryon fractions determined from a
variety of data.

We adopt a Hubble constant of $H_0 = 75\kms{\rm Mpc}^{-1}$ throughout.

\section{Data}

We employ data we have obtained for low surface brightness disk galaxies
in the 21 cm line as described in van der Hulst\markcite{vdH} \etal
(1993) and de Blok\markcite{BMH} \etal (1996), combined with optical
surface photometry presented
in those papers and in McGaugh \& Bothun\markcite{MB} (1994) and
de Blok\markcite{BHB} \etal (1995).  We augment our own data with
published data for dwarfs and high surface brightness spirals for which
both 21 cm rotation curves and adequate surface photometry are available,
as compiled by Broeils\markcite{broeils} (1992)
and de Blok\markcite{BMH} \etal (1996).

The data are listed in Table~1.  The line demarcates
our own data from that drawn from the literature.  The columns give
(1) the name of the galaxy, (2) its $B$-band absolute magnitude $M_B$,
(3) the inclination corrected central surface brightness \csb\ 
the disk in $B\sb$, (4) the scale length of the disk $h$ in kpc,
(5) the circular velocity $V_c$ in the flat part of the rotation
curve, (6) the inclination $i$ in degrees, and (7) the dynamical mass to light
ratio \MLo\ measured at $R = 4h$ in $\mass\solar/L\solar$ (see \S 3).

\placetable{data}

A requirement for inclusion in Table 1 is that both an \HI\ rotation curve
and surface photometry exist.  Most data of the latter sort are in $B$ so
we use that as standard.  Though $R$ or even $K$ might seem preferable,
this severely reduces the amount of available data.
The choice of bandpass makes no difference to the interpretation.
In collecting data from the literature, we have attempted to be as inclusive as
possible, in some cases transforming photographic surface photometry from
different bands to $B$ when a color is available.

We correct optical surface brightnesses to face-on values assuming
that the disk is optically thin.  This is supported by statistical
studies which show that spiral galaxies are semi-transparent
throughout their disks, except for the innermost regions (\eg
Huizinga \& van Albada\markcite{HvA} 1992;
Giovanelli\markcite{GHSWCF} \etal 1994).  In general the
outer parts of galaxies are optically thin.  Measurements of
extinction in foreground spirals obscuring a background galaxy suggest
extinction values of $\sim$ 0.3 mag in $B$ in the interarm regions of the
outer parts of a spiral (Andredakis \& van der Kruit\markcite{AvdK} 1992;
White \& Keel\markcite{WK} 1992). As \csb\ is not a measure
of the actual surface brightness in the center of a galaxy, but the
intercept of a fit to the disk profile which depends more
on the outer than the inner points, extinction is not a serious concern.
This is especially true in LSB galaxies where the metallicity and dust
content are low (McGaugh\markcite{myOH} 1994).  The assumption of optically
thin disks is only likely to be invalid in edge-on galaxies.

Bulges have little effect on the derived \csb\ of LSB galaxies (McGaugh \&
Bothun\markcite{MB} 1994; de Blok\markcite{BHB} 1995; see also de
Jong\markcite{dJ2} 1996a; Courteau\markcite{Cour} 1996).
A bigger problem is the great inhomogeneity
of sources of the surface photometry amongst the literature data.
However, the range is surface brightness in our sample ($\sim 5\sb$)
is much larger than the most pessimistic error estimates on
the surface brightnesses.

Sometimes, the rotation curves themselves are lacking.
For completeness, we published all
our synthesis data in de Blok\markcite{BMH} \etal (1996), but not all galaxies
are useful for dynamical analyses (Table~2).  We observed galaxies of a
variety of inclinations, and some are too face-on to be of use here.
We nonetheless prefer to accept large errors than to arbitrarily limit the
data, and so impose a very liberal inclination limit:  $i > 25\arcdeg$.

As well as very face-on galaxies, there are LSB galaxies with such
slowly rising rotation curves that even the synthesis observations do
not reach the flat portion of the rotation curve.  In others,
substantial asymmetries are present.  This is not an uncommon feature
of galaxies generally (Richter \& Sancisi\markcite{RS} 1994), and
calls into question the assumption of circular orbits on which
dynamical analyses are based.  Hence these galaxies are also excluded.

\placetable{notused}

The list of excluded galaxies in Table~2 is not precisely identical to
that for our rotation curve fits (de Blok \& McGaugh\markcite{dBM2}
1997; paper III) because that is a different sort of analysis.
The fits depend to a large extent on the shape of the rotation curve
which can be sensitive to the resolution of the observations.  In
this analysis we are only concerned with the amplitude $V_c$ of the
rotation curve so we can proceed if $V_c$ is reasonably well defined.

Some tests are less sensitive to these limitations than others, so it
will sometimes be possible to make use of the galaxies in Table~2.
However, unless so noted, the sample is restricted to those galaxies
listed in Table~1.  This sample is not complete in any volume-limited
sense, but that is not the goal of this work.  Our aim here is to
represent as broad as possible a range of the relevant physical
parameters.  The collected data span a factor of 4000 in luminosity,
30 in size, and nearly 100 in surface brightness.
This large dynamic range in surface brightness with good sampling
of the LSB regime is the essence of the new contribution of this work.

\section{Definitions}

Here we explicitly define the symbols and notation we will use.
The optical luminosity of a
galaxy will be denoted in the usual way by $L_O$ and the corresponding
absolute magnitude by $M_O$, where $O$ denotes the relevant band pass.
Unless otherwise specified, all optical quantities are $B$-band
measurements.  To symbolically distinguish between absolute magnitude
and mass, the latter will be denoted by \mass.

The light distribution of an exponential disk is characterized by the
central surface brightness \csb\ and the scale length $h$ (de
Vaucouleurs 1959).  As the intercept and slope of a fit to the light
profile, these describe the global luminous surface density and size
of disks. 
It is sometimes convenient to discuss surface brightness
in\sb, and sometimes in linear units\linsb.  In general, we will use $\mu$
to specify a surface brightness, $\Sigma$ to specify a luminous
surface density, and $\sigma$ to specify a mass surface density.
The total luminosity of the disk component is of course
simply related to the disk parameters by $L = 2\pi\lincsb h^2$.

An important quantity is the mass to light ratio,
which we will denote by $\Upsilon$.  We need
to distinguish several different mass to light ratios:
\begin{enumerate}
\item \MLT, the total mass to light ratio,
\item \MLo, the observable dynamical mass to light ratio, and
\item \ML*, the mass to light ratio of the stars.
\end{enumerate}
The latter relates stellar mass to observable luminosity; it includes
any remnants that occur in the natural course of stellar evolution
(\eg white dwarfs) but not baryons in the gas phase.  The total mass
to light ratio \MLT\ includes all mass, and encompasses the entirety
of any halo.  Since $\mass \propto R$ for a flat rotation curve,
and there exists no clear evidence of an edge to any halo mass
distribution, \MLT\ is a purely theoretical construct.  It is
nonetheless useful to keep in mind when discussing the observable mass
to light ratio \MLo.  This includes all mass (including dark
matter) within some finite observable radius $R_o$.
For the usual assumption of sphericity, the mass within $R_o$ is
\begin{equation}
\mass(R_o) = R_o V^2(R_o)/G.
\end{equation}
Deviations from sphericity can complicate this equation, but only by
geometrical factors.  They do not alter the basic functional dependence
which is the issue of relevance here.

The value of $G$ is known and $L$, $R$ and $V$ are directly measured.
The quantities $\mass(R_o)$ and \MLo\ are thus uniquely determined by
equation~(4).  If \MLo\ exceeds what is reasonable for \ML*,
we infer the presence of dark matter.

The absolute mass inferred depends on the distance scale which enters through
$R$ ($V$ is distance independent for the redshifts that concern us).
The choice of $H_0$ is not important to the shape derived for trends
of \MLo\ with optical properties.  The absolute value of \MLo\ will
of course shift with changes to the distance scale.

The choice of $R_o$ is important.  It is tempting simply to take this
as the largest radius measured.  However, this depends more on the
details of the observations than anything intrinsic to a galaxy.
It is therefore necessary to make a sensible choice for $R_o$ which
can systematically be applied to all disk galaxies.
Since $V(R) \rightarrow V_c$ as $R$ becomes large,
one obvious stipulation is that $R_o$ be sufficiently
large that the rotation curves have become essentially flat.  The rotation
curves of LSB galaxies are observed to rise slowly (de Blok\markcite{BMH}
\etal 1996), and in some cases the flat portion has not yet been
reached at the last measured point.  These galaxies must be excluded
from analysis.  In most cases, the rotation curves are still rising
slightly but becoming asymptotically flat.
We proceed if the gradient in the
outer slope is small (following Broeils\markcite{broeils}
1992; see Fig.~11 of de Blok\markcite{BMH} \etal 1996).
The actual observed value is used, not the apparent asymptotic value.
Another stipulation in defining $R_o$
is that it contain essentially all of the luminosity of the galaxy, so that
\MLo\ has the obvious meaning.  A further requirement is that $R_o$ not be
so large as to exceed the last measured point of the observations.

Ideally, we would like to relate $R_o$ to the extent of the mass
distribution.  Presumably dark halos have some sort of edge, which
we will denote by $R_H$.  Since we can not see the dark matter, it is
impossible to relate $R_o$ to $R_H$ without somehow assuming the answer.
Observations are incapable of uniquely deconvolving the
contribution of luminous and dark mass into any useful scale radius
(\eg van Albada \& Sancisi\markcite{vAS} 1986; Lake \& Feinswog\markcite{lake}
1989; de Blok \& McGaugh\markcite{dBM2} 1997).

The only fair measure of a disk's size observationally available is
the scale length $h$.  Isophotal radii like $R_{25}$ are meaningless
for LSB galaxies: the dimmest objects studied here have $\csb \approx
25$, so $R_{25} \approx 0$ by definition.  The effective (half light)
radius is proportional to the scale length for pure disk systems, but
depends on the bulge component in general.  We are concerned here with
defining a radius for the rotating disk, so the scale length is
preferable to the effective radius.

It might seem preferable to define a half mass radius of the
observable matter by combining stars and gas.  However, this requires
that molecular gas be mapped as well as the \HI\ and optical light.
Even if such data were available, some assumption about \ML*\ and
the $CO/H_2$ conversion factor would still have to be made.

We therefore choose to define $R_o$ in units of $h$, in particular
$R_o = 4h$.  Most rotation curves have reached their asymptotic
velocities by 3 scale lengths, while the last measured point for the
bulk of the data is around $5 h$.  Four scale lengths contains 91\% of
the luminosity of a pure exponential disk extrapolated to infinity,
and disks are generally thought to truncate around 4 or 5 scale
lengths (van der Kruit\markcite{vdK} 1987).  A choice of
$5 h$, or any other definition based on the disk light distribution,
simply inserts a multiplicative factor.

As a matter of nomenclature, we will refer to the observable gas and
stars as ``baryonic'' matter, and treat the dark matter as a distinct
component of unknown composition.

\section{Observational Facts}

Empirical results about the rotation curves and what they imply
for the dependence of the mass discrepancy on the optical properties
of galaxies can be summarized independently of
any theoretical framework by the following four facts:
\begin{enumerate}
\item All disk galaxies obey the same Tully-Fisher relation, irrespective
of surface brightness or band pass (Fig.~1; Sprayberry\markcite{lsbtf} \etal
1995; Zwaan\markcite{ZHBM} \etal 1995; Hoffman\markcite{HSFRWH} \etal 1996;
Tully \& Verheijen\markcite{TV} 1997).
\item At a given luminosity, the shape of rotation curves $V(R)$ varies
systematically with surface brightness (Fig.~2).  When $R$ is measured in
physical units (kpc), lower surface brightness galaxies have rotation
curves which rise more gradually.  When $R$ is measured in terms of disk
scale lengths, the shape of $V(R/h)$ is more similar though not necessarily
identical (see Fig.~8 of de Blok\markcite{BMH} \etal 1996).
\item The mass discrepancy manifests itself at progressively smaller radii
in dimmer galaxies (de Blok \& McGaugh\markcite{dBM1} 1996,
1997\markcite{dBM2}; Fig.~3; see also
Casertano \& van Gorkom\markcite{CvG} 1991;
Broeils\markcite{broeils} 1992; van Zee\markcite{vZHSB} \etal 1997).
\item The severity of the mass discrepancy is strongly correlated
with the central surface brightness of the disk (Fig.~4;
de Blok\markcite{BMH} \etal 1996).  This is the
$\Upsilon$-$\Sigma$ relation (Zwaan\markcite{ZHBM} \etal 1995).
\end{enumerate}

\subsection{Fact 1}

The relationship between luminosity and linewidth has long been known
(Tully \& Fisher\markcite{TF} 1977).  That the same relation holds for
all surface brightnesses has been obtained independently
by Sprayberry\markcite{lsbtf} \etal (1995), Zwaan\markcite{ZHBM} \etal (1995),
Hoffman\markcite{HSFRWH} \etal 1996, and Tully \& Verheijen\markcite{TV} 1997.
These different works all obtain the same observational result but offer
varying interpretations of it.

The interpretation depends entirely on the basic assumptions one must make.
One obvious assumption is that light traces mass, in which case
LSB galaxies should rotate slowly for their luminosity and {\it not\/}
obey the Tully-Fisher relation.  This occurs because $R$ is larger for the
same $L$, reducing $V$ in equation~(4) (Zwaan\markcite{ZHBM} \etal 1995).
That LSB galaxies are not expected to fall on the Tully-Fisher relation
was first discussed by Aaronson\markcite{oldtf} \etal (1979) and
Romanishin\markcite{roman} \etal (1982) using this assumption.
Other assumptions can also be made (discussed in detail \S 5) but generally
result in some sort of shift of LSB galaxies off the Tully-Fisher relation
(\eg Dalcanton\markcite{D95} \etal 1995).  The only scenario in which no
shift is expected is one in which halos of a given mass are identical.
This assumption is itself motivated by the Tully-Fisher relation, so it
offers no independent expectation value.  In cases where there is
a clear expectation of a shift, that expectation is not realized.

\placefigure{figTF}

Indeed, it is already difficult to
understand a luminosity-linewidth relation with little scatter.
One expects a great deal more scatter for
plausible initial conditions (\eg Eisenstein \& Loeb\markcite{EL} 1995).
The lack of surface brightness segregation in the luminosity-linewidth
plane complicates things much further (as discussed in \S 5).
For now, note that in Fig.~1 a Tully-Fisher relation is obtained
for both $W_{20}^c$ and $V_c$:  the velocity width is as expected an indicator
of the rotation velocity.  The scatter is greater when the linewidth is
used (0.9 \vs 0.6 mag.).  We can not say much about the intrinsic scatter of
the Tully-Fisher relation since we include low inclination galaxies.
This naturally increases the scatter, so the intrinsic scatter is presumably
smaller than what we measure.  The slopes obtained from Fig~1(a) and (b)
are marginally different, being $-7 \pm 2$ in (a)
and $-9.4 \pm 1$ in (b).  This may indicate that $W_{20}^c$ is not
a perfect indicator of the real quantity of interest, $V_c$.

The most important fact is that there is
a strong luminosity-rotation velocity relation with little intrinsic scatter
and a slope indistinguishable from the theoretical value.
The details of the band-pass dependence of the slope are not critical
(see Appendix) as
we are not trying to establish a calibration of the Tully-Fisher relation,
but rather trying to understand why LSB galaxies follow it at all.  In this
context, we are concerned with the relation of stellar {\it mass\/} to
the flat rotation velocity $V_c$.  The near infrared is thought to be the best
indicator of stellar mass, and both $H$ (Aaronson\markcite{oldtf}
\etal 1979) and $K$ (Tully \& Verheijen\markcite{TV} 1997) band Tully-Fisher
relations have slopes indistinguishable from the theoretical expectation $-10$.

\subsection{Fact 2}

It is well known that the shape of galaxy rotation curves is luminosity
dependent (Burstein\markcite{BRTF} \etal 1982; Persic\markcite{URC} \etal
1996).  At a given luminosity, the shape varies systematically with
surface brightness in that the rotation curves of
LSB galaxies rise more gradually than do those of high surface brightness
galaxies.  This behavior {\it is\/} expected if light traces mass and
surface mass density follows from surface brightness in the obvious way.
This is indeed the case at small radii, both for the specific example
illustrated in Fig.~2 (de Blok \& McGaugh\markcite{dBM1} 1996a) and in general.
Moreover, the relative similarity of $V(R/h)$ suggests that the dark
halo is strongly coupled to the properties of the disk (see also paper II).

If surface brightness is a good indicator of mass density,
LSB galaxies should not lie on the Tully-Fisher relation.  Yet they do.
Reconciling facts (1) and (2) proves to be very difficult, and leads to
fact (3).

\placefigure{figcomp}

\subsection{Fact 3}

In high surface brightness spiral galaxies, it is generally possible to
attribute essentially all of the observed rotation to the luminous
disk out to 3 scale lengths or so by scaling the disk contribution up
to the maximum value allowed by the amplitude of the inner part of
the rotation curve (Bosma\markcite{Bosma} 1981; Kent\markcite{K87} 1987;
Sancisi \& van Albada\markcite{SvA} 1987; Sanders\markcite{S90} 1990).
This ``maximum disk'' solution has led to a general picture in which
the luminous material dominates the inner part of spiral galaxies,
and dark matter only becomes dominant in the outer fringes near the
edge of the luminous disk.  An important aspect of this picture is
that the peak in the rotation curve of the exponential disk at $R = 2.2h$
is comparable in amplitude to the circular velocity due to the halo
[$V^{peak}_{disk}(2.2h) \approx V_H$], giving a smooth transition
between the two components (the disk-halo ``conspiracy'').

\placefigure{R2}

The disks of LSB spirals can not be maximal in this sense.
Because they are very diffuse, the Newtonian rotation curves of
LSB disks peak at very low amplitudes ($\ll V_c$) for plausible values of
the mass to light ratio of the stellar population.  From a purely
dynamical standpoint, there is some room to consider
values of \ML*\ which are unreasonably large for the blue stellar
populations of LSB galaxies.  However, even for $\ML* \sim 10$
(as opposed to reasonable values of $\sim 1$ or 2), it is still
not possible to attribute a velocity comparable to $V_c$ to the disk.
Even larger \ML*\ are not allowed by the slow rate of rise of the
rotation curves.  The disk simply can not dominate
in these galaxies except at very small radii: the mass discrepancy sets in at
essentially $R = 0$ (Fig.~3).  Only the highest surface
brightness disks can be maximal in the sense that luminous disks dominates
the inner dynamics, with dark matter only becoming important at large radii.
There is no reason to believe that
a segregation between inner dominance by a disk and outer dominance by
a dark matter halo is a general property of galaxies.

\subsection{Fact 4}

The severity of the global mass discrepancy, as measured by
the observed mass to light ratio \MLo,
is correlated with both luminosity and surface brightness
(Fig.~4).  It is not correlated with the optical size of the galaxy.
The correlation with surface brightness is much stronger (${\cal R} = 0.88$)
than with absolute magnitude (${\cal R} = 0.46$).  The scatter in the former
case can plausibly be attributed entirely to observational error,
especially considering the very inhomogeneous sources of the surface
photometry.  In the case of $M_B$ there is substantial real scatter.
Absolute magnitude and surface brightness are themselves weakly correlated
(${\cal R} = 0.50$ in this sample), so it seems likely that the
relation between luminosity and \MLo\ is
due to the convolution of the strong \MLo-\lincsb\ relation
with the lack of a size relation through $L \propto \lincsb h^2$.
Surface brightness, not luminosity or morphological type, is the
variable of greatest interest.

\placefigure{figMLM}

\section{The Tully-Fisher and \MLo-$\Sigma$ Relations}

The relation between \MLo\ and \csb\ can be derived from the Tully-Fisher
relation (Zwaan\markcite{ZHBM} \etal 1995) assuming that the observed
velocity width $W$ is proportional to the asymptotic rotation
velocity $V_c$, and taking $R \propto \sqrt{L/\Sigma}$.  The latter is true
by definition; the former is empirically confirmed by the full rotation
curve data.  Various authors (\ie Sprayberry\markcite{lsbtf} \etal 1995;
Salpeter \& Hoffman\markcite{SH} 1996) give apparently conflicting results,
but these stem entirely from differing definitions of $R_o$.  The {\it data\/}
are all consistent.

To discuss possible interpretations of the Tully-Fisher
relation, we make the usual assumptions that the \HI\ gas is in circular
orbit ($a_c = V^2(R)/R$) and that halos are spherical and dominant
where $V(R) \rightarrow V_c$ so that $g_N = G\mass(R)/R^2$.
Setting $a_c = g_N$ gives
\begin{equation}
V^2(R) = \frac{G \mass(R)}{R}.
\end{equation}
These are very basic assumptions which are widely employed; plausible
deviations (\eg triaxial halos) will not alter the gist of the results.
Observationally, the Tully-Fisher relation is
\begin{equation}
L_j \propto V_c^{x_j}
\end{equation}
where $x_j$ is the slope in band pass $j$.
To relate equation (5) to (6), let us define a mass surface density
\begin{equation}
\sigma(R) \equiv \frac{\mass(R)}{R^2}
\end{equation}
and make a substitution of variables so that
\begin{equation}
V^4 = \frac{G^2 \mass^2(R)}{R^2} = G^2 \sigma \mass(\sigma).
\end{equation}
By definition, $\mass(R) = \Upsilon_j(R) L_j$ and $\sigma(R) =
\Upsilon_j(R) \Sigma_j(R)$.  Here, $\Sigma_j(R)$ is the average
surface brightness within $R$ (which will decrease as $R$ grows) and
$\Upsilon_j(R)$ is the dynamical mass to light ratio
within radius $R$.  Both $\Sigma_j$ and $\Upsilon_j$ are
strong functions of $R$.  However, we are concerned only with the
relative functional dependencies, so the actual value of $R$ drops out
where $V(R) = V_c$.

In theory,
\begin{equation}
V_c^4 \propto \Upsilon_j^2 \Sigma_j L_j,
\end{equation}
while in actuality
\begin{equation}
V_c^{x_j} \propto L_j.
\end{equation}
As discussed by
Zwaan\markcite{ZHBM} \etal (1995), one needs to arrange to keep the
term $\Upsilon_j^2 \Sigma_j$ constant in order to recover the observed
Tully-Fisher relation from basic physics.
This is often attributed to a universal constancy of the mass to
light ratio and Freeman's law (Freeman\markcite{F70} 1970) which
states that all spirals have essentially the same surface brightness,
$\csb \approx 21.65\sb$.  By definition LSB galaxies deviate from this
value, and {\it should not\/} fall on the same Tully-Fisher relation
as Freeman disks if the mass to light ratio is constant as generally
assumed.  Fig.~5 shows the relation expected from equation~(9) if
$\MLo$ is constant; the range of surface
brightnesses examined is large enough that this effect would be very
readily apparent.  Instead of following this relation, LSB galaxies
rigorously adhere to {\it the\/} Tully-Fisher relation.  The mass to
light ratio must somehow be fine tuned to compensate.

\placefigure{TFresid}

Consider the error budget required.  In magnitude units, the Tully-Fisher
relation is $M_j = -10\log V_c +C$ assuming $x_j = 4$.  Hence
\begin{equation}
\delta M = 4.34 \frac{\delta V_c}{V_c},
\end{equation}
so an intrinsic scatter of $< 0.6$ mag.\ stems from a modest
$< 14\%$ intrinsic variation in $V_c$.  Disk and halo must combine to
specify $V_c$ very precisely.  More generally, the variation is
\begin{equation}
\left|\delta M\right|^2 = \left|\delta \csb\right|^2
+ 18.86 \left|\frac{\delta V_c}{V_c}\right|^2
+ 1.18 \left|\frac{\delta \Upsilon}{\Upsilon}\right|^2.
\end{equation}
A one magnitude shift in \csb\ should translate directly into one
magnitude in $M$.  Even for disks which obey the Freeman (1970) Law,
the scatter in \csb\ (0.35 mag.) should propagate directly into the
Tully-Fisher relation.  Here we have disks spanning nearly 5
magnitudes in \csb.  The mass to light ratio must be fine
tuned to compensate, with essentially zero scatter.

\subsection{Implications}

The observed mass to light ratio \MLo\ as we have defined it includes
all mass within some arbitrary radius $R_o$.  The exact definition of $R_o$
is not essential, as long as we choose a fixed number
of scale lengths which encompasses the
luminous matter and reaches the flat part of the rotation curve.
Then, using equation~(9),
\begin{equation}
\MLo = \frac{\lambda \ML*}{f_b f_*}\propto \Sigma^{-1/2}
\end{equation}
where \ML*\ is the mass to light ratio of the stellar population,
$f_b = \mass_b/\mass_T$ is the baryonic mass fraction,
$f_* = \mass_*/\mass_b$ is
the fraction of baryonic disk mass in the form of stars, and
$\lambda = \mass(R_o)/\mass_T$ is the fraction of the
total mass encompassed by the edge of the disk.  Note that since
$\mass \propto R$, $\lambda \approx R_o/R_H$ where $R_H$ is the
halo radius (edge).  We thus decompose
\MLo\ into four fundamental parameters; two involve
galaxy evolution (\ML*, $f_*$) and two involve galaxy formation
($f_b$, $\lambda$).
The requirement $\MLo \propto \Sigma^{-1/2}$
now implies that a combination of four parameters must
be fine tuned with surface brightness. Let us examine each of them in turn.

\subsection{Evolution}

Zwaan\markcite{ZHBM} \etal (1995) attributed the observed adherence to
the Tully-Fisher relation to evolutionary regulation.  From a
theoretical perspective, not taking the surface
brightness effects into account, Eisenstein \&
Loeb\markcite{EL} (1995) came to the same conclusion in more general
terms.  They showed that it was very difficult to obtain a
Tully-Fisher relation with the small observed scatter from plausible
initial conditions.  There should always be some scatter in the
density structure of halos so that velocity is not so precisely
specified by total mass and luminosity. Some evolutionary regularizing
effects must thus reduce the scatter.  The physical mechanism by which this
occurs is unknown.  Feedback from star formation
activity is often invoked, but it is hard
to see how the stochastic energy input of massive stars will have a
strong {\it regularizing\/} effect.  Since the nature of the evolutionary
regulation is unclear, let us simply examine the consequences
for \ML*\ and $f_*$.

\subsubsection{Stellar Mass to Light Ratios}

The mass to light ratio of a stellar population is measured directly
in our own Milky Way, where
$\ML* \approx 1.7$ in $B$ (Kuijken \& Gilmore\markcite{KG} 1989).  This
appears to be typical of high surface brightness disks, with substantial
variation and uncertainty.  Bottema\markcite{Bott2} (1997) finds
$\ML* = 1.8 \pm 0.5$ from velocity dispersions.  For LSB galaxies,
one infers slightly smaller \ML*\ from their colors,
$\sim 1$ with great uncertainty
(McGaugh \& Bothun\markcite{MB} 1994; de Blok\markcite{BHB} \etal 1995).  
There is a lot of real scatter in color at any given surface
brightness, implying substantial scatter in \ML*.  

The inferred trend of \ML*\ with \lincsb\ 
is much smaller in magnitude than the \MLo-\lincsb\ relation and has
the opposite sign.  In addition to what the colors imply for \ML*,
the slow rate of rise of the rotation curves (de Blok
\& McGaugh 1996\markcite{dBM} 1997\markcite{dBM2}) dynamically rules out the
large \ML*\ values that are needed in LSB galaxies to explain
the \MLo-\lincsb\ relation.  The peak amplitudes of the rotation curves
of the stellar disk are much smaller than $V_c$ so the stellar disks
alone cannot cause the TF relation, let alone the \MLo-\lincsb\ fine-tuning.
The only way to alter this conclusion is to drop the usual assumption that
light traces {\it stellar\/} mass, and make
\ML*\ a strong function of radius [the unlikely form
$\ML* \propto R^{-1} exp(R/h)$ would be required].
It is therefore unlikely that the \MLo-\lincsb\ relation
can be caused by a \ML*-\lincsb\ relation.

\subsubsection{Stellar Mass Fractions}

The fraction of the baryonic disk mass which is in the form of stars $f_*$
{\it is\/} well correlated with surface brightness (Fig.~6; McGaugh \&
de Blok\markcite{MdB} 1997).  The correlation goes in the correct sense,
with LSB galaxies having relatively fewer stars.  The slope is
approximately correct, though the correlation is not as tight as that
with \MLo.

\placefigure{figFstar}

It is therefore tempting to attribute the fact that $\MLo \propto
\lincsb^{-1/2}$ to $f_* \propto \lincsb^{1/2}$.  This would mean a
regulatory mechanism which somehow maintains a mass fraction of stars
appropriate to both the Tully-Fisher and \MLo-$\lincsb$ relations.
It is possible to
check this since the gas mass is directly measurable from the 21 cm
flux.  That is, $\mass_{gas} = \eta \mass_{HI}$,
where $\eta = 1.4$ is appropriate for solar metallicity atomic gas.
Molecular gas does not seem to be common in LSB galaxies
(Schombert\markcite{CO} \etal 1990) and
is presumably distributed like the stars.  If this is the case, it
is subject to the same dynamical constraints as \ML*.  Moreover, the
approach to evolutionary regulation taken here is motivated by the
similarity of the \MLo-$\lincsb$ and $\mass(HI)/L$-$\lincsb$ relations.
This motivation would vanish if molecular gas were more important than
the \HI.

\placefigure{figMLMcorr}

We can correct \MLo\ for the atomic gas simply by subtracting its mass
from the total.  This removes the term $f_*$ from equation~(13):
\begin{equation}
\MLo^c = \frac{\mass(R_o) - \mass_{gas}}{L} = \frac{\lambda \ML*}{f_b}.
\end{equation}
Though this accounts for the $\mass(HI)/L$-$\lincsb$ relation, it
makes no real difference to the \MLo-$\lincsb$ problem (Fig.~7).
The baryonic mass is too small a part of the total mass to matter much
to the problem, even within the optical edge of LSB disks.
Evolutionary regulation simply can {\it not\/} solve the problem.

\subsection{Formation}

The evolutionary parameters just discussed fail to have an impact on
the problem because LSB galaxies are dark matter dominated throughout.
The evolution of the baryonic disk is almost irrelevant.
This leaves the conditions of formation.

\subsubsection{Variations in the Baryon Fraction}

To explain the \MLo-\lincsb\ relation using the baryon fraction
$f_b$, one can simply assume that halos contain different amounts of baryons.
This is of course just an ad-hoc assumption.  Nominally, one would expect
a universal baryon fraction with some modest amount of cosmic scatter.

This approach requires that the baryon fraction be uniquely related
to the disk central surface brightness:
\begin{equation}
f_b \propto \lincsb^{1/2}.
\end{equation}
Since the Tully-Fisher relation
must be maintained with $\mass \propto L \propto \lincsb h^2$, equation~(15)
implies $\mass \propto h^2 f_b^2$.  Halos of a given $\mass$ may have different
baryon fractions, but somehow form larger or smaller disks to
counterbalance this.  Halos with low $f_b$ must make up the deficiency
in luminosity they suffer from low surface brightness by forming larger disks:
$f_b \propto h^{-1}$.  Why this should be the case is unclear.
Beyond stating these requirements, little can be said because the
hypothesized solution is so arbitrary.

\subsubsection{Enclosed Mass and Spin}

The situation is only a little better with $\lambda$.  In
this case, the requirement is
\begin{equation}
\lambda \propto \lincsb^{-1/2}.
\end{equation}
Since $\lambda = \mass(R_o)/\mass_T \approx R_o/R_H$,
the difference in this case
is that the disks of lower surface brightness galaxies stretch further out
into their halos.  This has the built-in virtue of explaining the increasing
dark matter domination of lower surface brightness disks while keeping the
global baryon fraction fixed.  In effect, this asserts that the mass density
$\sigma(R)$ is unrelated to the surface brightness $\Sigma(R)$.  Then,
if $\sigma$ is a constant in equation~(8), the lack of a shift in the
Tully-Fisher relation follows and the \MLo-$\Sigma$ relation just reflects
the difference in $\sigma(R)$ and $\Sigma(R)$.  We measure increasing mass
to light ratios with decreasing surface brightness because
$\MLo(R_o)$ just measures a systematically
greater portion of the global \MLT: $\MLo/\MLT \sim R_o/R_H \sim \lambda \sim
\lincsb^{-1/2}$.

A consequence of this approach is that
measuring one parameter of the optical light
distribution, the central surface brightness, immediately tells one the
extent of the dark matter relative to the light.
Specifying the distribution of either $\lambda$ or $\Sigma$
immediately determines the distribution of the other.
By the variational principle,
\begin{equation}
\frac{\delta \Sigma}{\Sigma} = -2 \frac{\delta \lambda}{\lambda}.
\end{equation}

Our parameter $\lambda$ can be related to the theoretical concept of the spin
parameter
\begin{equation}
\lambda_s = \frac{{\cal J} {\vert E \vert}^{1/2}}{G \mass^{5/2}}
\end{equation}
(Peebles\markcite{firstlambda} 1971).  This parameterizes the angular
momentum acquired by protogalaxies from tidal torques.  The details of
this process are uncertain, but all that
matters here is that {\it within this framework\/} dissipative
baryonic collapse is halted by angular momentum.  Collapse
from the initial $\lambda_s$ of the halo and protogalaxy stops when
$\lambda_s^{disk} \rightarrow 1$.  This gives the condition
$\lambda \approx R_o/R_H \approx \lambda_s$.
Objects with high primordial spin $\lambda_s$ will collapse less than
low spin objects.  

Presuming that spin is the underlying reason for $\lambda$ provides a test.
Equations~(18) and (19) give the mapping from spin to surface brightness:
\begin{equation}
\csb = 5 \log\left(\frac{\lambda_s}{\lambda_s^*}\right) + \csb^*.
\end{equation}
This can be used 
to compare the observed distribution $\phi(\csb)$ (McGaugh\markcite{me}
1996a; de Jong\markcite{dJ3} 1996b) with the spin distribution predicted
by numerical models.  We are free to fit the parameters $\lambda_s^*$ and
$\csb^*$ in order to obtain a fit, but the shapes of the distributions
are specified by equation~(19) and hence provide a test.

There have been
various theoretical attempts to predict the distribution of the spin parameter
(\eg Efstathiou \& Jones\markcite{EJ} 1979; Barnes \& Efstathiou\markcite{BE}
1987; Eisenstein \& Loeb\markcite{EL} 1995; Steinmetz \&
Bartelmann\markcite{SB} 1995).  These are all rather uncertain.
Nevertheless, there seems to be general consensus that the distribution of
spins is broad and approximately symmetric about some typical value,
$\lambda_s \approx 0.07 \pm 0.03$.  Two typical examples are plotted together
with the surface brightness data in Fig.~8 for $\lambda_s^* = 0.03$ and
$\csb^* = 21.5$.  These values are chosen to put the theoretical
distributions in the same vicinity as the data, but we do not attempt
a formal fit because of the great uncertainty in the predicted spin
distribution.  A tolerable match is found given the large error bars.

\placefigure{spin}

At the moment, there is no test in the shape of the surface brightness
distribution at the low surface brightnesses.  Because of the stretching
imposed by equation~(19), the theoretical prediction of the spin distribution
needs to be made at quite high resolution in $\lambda_s$
to provide a useful test.  Similarly,
the observed surface brightness distribution needs to be determined to
much greater accuracy.

A sharp feature in the surface brightness distribution occurs
at high surface brightness which is not reflected in the spin distributions.
The $\lambda_s \rightarrow \csb$ picture predicts a large number of high
surface brightness disks which are not found
observationally.  In order to match the high surface brightness cut off,
one must invoke some critical phenomenon at $\lambda_s^*
\approx 0.03$.  For example, one might suppose that protogalaxies with
$\lambda_s < 0.03$ fail to form disks, perhaps becoming ellipticals
instead.  This might plausibly come about if destructive
disk instabilities set in when a critical surface density is exceeded.

The environmental dependence of \csb\ and $\lambda_s$ provides
a constraint which is independent of $\phi(\csb)$.  The numerical
simulations generally agree that $\lambda_s$ is only
weakly correlated with environment (Steinmetz \& Bartelmann\markcite{SB} 1995)
if at all (Barnes \& Efstathiou\markcite{BE} 1987).
In contrast, putatively high spin LSB galaxies very clearly
reside in low density environments (Schombert\markcite{LSBcat} \etal 1992;
Bothun\markcite{small} \etal 1993; Mo\markcite{Mo} \etal 1994).
It is thus difficult to understand the observed environments of LSB
galaxies if spin is the principal parameter determining the surface brightness.

\section{Rotation Curve Shapes}

So far, we have only discussed possible interpretations
of observational fact (1), the universality of the Tully-Fisher relation.
That is, we have only attempted to understand why $V_c$ is
correlated with $L$.  This proves difficult since $L \propto V_c^4$ does
not follow from $V^2 = G\mass/R$.
The full shapes of the rotation curves complicate matters even further.

\subsection{Expectation Values}

In order to interpret the shapes of rotation curves, it is necessary
to establish some prior expectation.  This requires a deeper
investigation of the subject of galaxy formation than we have made so
far.  Galaxy formation is a difficult subject, and even very
sophisticated modeling efforts have yet to reach the point
where the formation and evolution of an individual galaxy
can be studied in detail.  Building sophisticated models
requires specifying a cosmology and deciding what the dark matter is.
Our interest here is in placing constraints free of such assumptions, so
we take a different tack.  We construct two models which illustrate
the most obvious possibilities.  We intentionally keep these models
simple to illustrate the important physical ideas.  More complex models
are usually special cases of one or the other.

Disk galaxies are thought to be composed of two basic components:
a dynamically cold disk and a dynamically hot dark matter halo.
To construct our models, we will assume that the disk is
purely exponential.  This is not precisely true of course, but
it is an adequate approximation for our purposes.  We will also
make the usual assumption that dark matter halos are isothermal spheres.

The density distribution of the dark matter is taken to be
\begin{equation}
\rho(R) = \frac{\rho_C}{1 + (R/R_C)^2}
\end{equation}
where $\rho_C$ is the core density and $R_C$ is the core radius.
This density distribution is divergent with an infinite total mass,
so there must be an additional parameter $R_H$ to describe the outer edge
of the halo.  The effects of $R_H$ are not apparent in the collected data.
Experimenting with the models indicates that effects due to $R_H$
would be apparent if observations were anywhere near reaching it,
so it can be ignored for now.
The rotation curve resulting from this mass distribution is
\begin{equation}
V_{halo}(R) = V_H \sqrt{1-\frac{R_C}{R}\tan^{-1}\left(\frac{R}{R_C}\right)}
\end{equation}
where the asymptotic circular velocity is $V_H = \sqrt{4 \pi G \rho_C R_C^2}$.
This gives rise to asymptotically flat rotation curves with
$V_c \rightarrow V_H$ as $R \rightarrow \infty$.  Note, however, that
the flatness of observed rotation curves at finite $R$
depends on a delicate balance
between disk and halo components (the disk-halo ``conspiracy'').

The rotation curve of a thin exponential disk (Freeman\markcite{F70} 1970) is
\begin{equation}
V_{disk}(R) = \sqrt{4 \pi G \sigma_0 h y^2 \left[I_0(y) K_0(y) -
I_1(y) K_1(y)\right]}
\end{equation}
where $y = R/2h$, $\sigma_0$ is the mass surface density corresponding to
the central surface brightness, and $I_n$ and $K_n$ are modified
Bessel functions of the first and second kind.

\subsubsection{Simple, General Model Galaxies}

We construct toy galaxy models based on these mass distributions and
two simple ideas.  One hypothesis is that LSB galaxies are diffuse
because their dark matter halos are diffuse.  We will refer to this
hypothesis as ``density begets density,'' or DD for short.  In DD,
differences in surface brightness are driven by the differences in the
amplitude of the original density fluctuations $\delta$ from which
galaxies arise.  The other hypothesis is that galaxies of the same
mass live in the same halos, but that the relative extent of the disk
varies.  This is essentially the same as the $\lambda$-$\Sigma$
relation discussed above, and we will refer to this as the ``same
halo'' hypothesis, or SH for short.  In SH, differences in surface
brightness are driven by differences in the spin parameter $\lambda$.

In both cases, there is a spectrum of masses which are the main factor
determining the total luminosity.  The surface brightness is principally
determined by $\delta$ and $\lambda$ in DD and SH respectively.  This is
not to say that mass maps directly to luminosity and $\delta$ or $\lambda$
directly to surface brightness, merely that these are the strongest
relations.  In principle, there are theoretical distribution functions
$\Phi(\mass,\delta)$ (\eg Mo\markcite{Mo} \etal 1994) and
$\Phi(\mass,\lambda$) (\eg Dalcanton\markcite{D97} \etal 1997)
which somehow transform to the observable bivariate distribution
$\Phi(M,\csb)$.  The transformation need not be free of axial twists;
in both cases one expects some weak correlation between luminosity and
surface brightness.  A complete theory should convolve the distributions
of $\mass$, $\delta$, and $\lambda$ and incorporate the physics of star
formation to obtain luminosity and surface brightness from the calculated
baryonic mass distribution.  This is beyond current abilities.
However, we are concerned here merely with establishing a basic expectation
for the dependence of $V(R)$ on surface brightness.  For this simple purpose,
it suffices to fix the principal luminosity determinant (mass) and look at
the effects of variation in density and spin.

\paragraph{Density begets Density:}
The DD hypothesis is motivated by a number of observations
(McGaugh 1992\markcite{mythesis}, 1996b\markcite{IAU}).  Principle
among these is that the observed surface density of gas and stars is
low in LSB galaxies.  This will occur naturally if LSB galaxies arise
from low density peaks in the initial fluctuation spectrum.  At a
given mass, small $\delta$ peaks will turn around and collapse at a
later time.  They should also have a larger virialization radius and
hence have a lower density of dark as well as luminous matter.
A collapse time which is somewhat later for LSB than HSB galaxies
is suggested by the ages inferred from their colors (McGaugh \&
Bothun\markcite{MB} 1994; R\"onnback \& Bergvall\markcite{RB} 1994;
de Blok\markcite{BHB} \etal 1995).
That LSB galaxies should be associated with low ($\sim 1 \sigma$)
density peaks is often asserted as obvious,
and is indeed well motivated by all the physical properties of
LSB galaxies (colors, metallicities, and
gas contents) as well as actual densities
(de Blok \& McGaugh\markcite{dBM2} 1997).
The DD hypothesis makes three important predictions which have
subsequently been tested.  One is that LSB galaxies should be less strongly
clustered than HSB galaxies.  There should be a shift in the amplitude of
the correlation function $\xi(r)$ by an amount corresponding to the
difference in the amplitudes of the perturbations $\delta$ from which the
galaxies arose.  This was confirmed by Mo\markcite{Mo} \etal
(1994).  The second prediction is that LSB galaxies
should have slowly rising rotation curves, in qualitative agreement with
the data.  The third prediction is that LSB galaxies
should {\it not\/} fall on the same Tully-Fisher relation as HSB galaxies,
in flat contradiction to the observations.

\paragraph{Same Halo:}
The SH hypothesis is motivated by the Tully-Fisher relation.  It
presumes that $V_c$ is determined by $V_H$; the halos are identical
regardless of surface brightness so a single Tully-Fisher relation is
assured.  LSB disks are more dark matter dominated than HSB disks
simply because they extend further out into their halos.  Aside from
the Tully-Fisher relation, SH enjoys none of the observational
successes which motivated the DD model.  There should be no age
difference: all halos of the same mass collapse at the same time.
This is not a strong objection, as it is difficult to distinguish
age from a slow evolutionary rate (\eg McGaugh \& de Blok{MdB} 1997).
A more serious problem with SH is that it does not predict the
observed shift in the correlation function: spin and environment
are not correlated (Steinmetz \& Bartelmann\markcite{SB} 1995;
Barnes \& Efstathiou\markcite{BE} 1987).  An additional environmental
effect must be invoked to produce this.  Whatever mechanism causes LSB
galaxies to be isolated must be quite strong, as their isolation is
clear out to $> 1$ Mpc (Bothun\markcite{small} \etal 1993).
This is much too large a
range for tidal forces to be effective, especially as the dark matter
domination of LSB galaxies makes them quite robust against even strong
tidal encounters (Mihos\markcite{MMB} \etal 1997).  Yet so compelling is the
constraint imposed by the Tully-Fisher relation that a number of
models for the formation of LSB galaxies have already appeared which
are primarily motivated by this observation (\eg
Dalcanton\markcite{D97} \etal 1997; Navarro\markcite{sesto} 1996a).
Though more complicated than our illustrative SH model, these models
reduce to the same basic scenario, and enjoy the same virtues and
suffer the same vices as SH.

The level $V_c$ of the flat part of the rotation curve is only one piece
of the information contained in $V(R)$.  The full shape of the rotation
curves provide the most complete test of both DD and SH hypotheses.
To see what these predict, let us construct a nominal HSB model with
$h = R_C = 1$ (Fig.~9).  In general this is approximately true in HSB galaxies
(\eg Athanassoula\markcite{ABP} \etal 1987; Broeils\markcite{broeils} 1992;
Rhee\markcite{Rhee} 1996; de Blok \& McGaugh\markcite{dBM2}
1997) which are frequently modeled in this way
(\eg Hernquist\markcite{H93} 1993).
We assume the disks are maximal with $V_{disk}^{peak} \approx V_H$,
as suggested by the disk-halo conspiracy.  This is a conservative assumption
for testing the shapes of rotation curves, since the differences between
HSB and LSB halos are minimized in this case (de Blok \& McGaugh\markcite{dBM2}
1997).  In practice, this means $V_H = \sqrt{\pi G \sigma_0 h} = 1$.
This results in $V_{disk}^{peak} \approx 0.8 \, V_{total}(2.2h)$ as found
empirically in HSB galaxies (de Blok \& McGaugh\markcite{dBM2} 1997).

\placefigure{Toygal}

The resulting null HSB model is plotted in the top panels of Fig~9.
The expected shape of the rotation curve of an LSB galaxy (bottom panels
of Fig~9) follows from the null HSB model and each hypothesis by simple
scaling.  We assume that the mass of the HSB and LSB galaxy are the same,
both disk and halo, so that the baryon fraction is the same.  The same
disk mass imposes the requirement that $\sigma_0 h^2$ be the same.  This
gives the scaling required to calculate the rotation curves under each
hypothesis.  For illustration, we choose $h({\rm LSB}) = 3$ (the difference
between NGC 2403 and UGC 128) which gives
$\sigma_0({\rm HSB})/\sigma_0({\rm LSB)} = 9$ to conserve mass.

Under the DD hypothesis, the halo should be diffuse like the disk.
We assume $R_C = h({\rm LSB}) = 3$.  This is not the only possibility,
but all that matters for this test is that there is some stretching of
the halo.  The resulting rotation curve is plotted
in the lower left panel of Fig.~9.  This is essentially just a stretched
version of the HSB model, sharing the same disk-halo conspiracy but not
the same $V_c$ since $R$ is larger in $V^2 = G\mass/R$.

Under the SH hypothesis, the halo is the same so $R_C = 1$ even though
$h({\rm LSB}) = 3$.  For the baryon fractions to be the same, the LSB
galaxy must reach to much nearer the edge of the halo than the HSB, though
this is presumably still a long way off.  The resulting $V(R)$ is dominated
by the halo at all radii, and by construction reaches the same $V_c$.

We can now test the models against the data.  The HSB model has been
constructed to typify a disk with the Freeman central surface
brightness $\csb = 21.65$.  The LSB model is a factor of 9 lower in
surface density, corresponding to $\csb = 24.04$ assuming \ML*\ is
similar.  There are four galaxies with reasonable data
within 0.2 mag of this surface brightness.  These are plotted in
Fig.~10 by scaling $R$ by the observed scale length $h$ and $V$ so that
$V_c = 1$.

\placefigure{cruelreality}

The data are not well predicted by either model.  The DD model predicts a
gradual rise of $V(R)$ which is indeed seen at small $R$.  However,
$V(R)$ continues to rise to a level $V_c$ well in excess of that
anticipated by DD, which does not predict a universal Tully-Fisher
relation.  The SH model has the correct $V_c$ (indeed, it was
constructed to do just this), but predicts that $V(R)$ should rise
much more rapidly than it actually does.  To be consistent with
observations, the halos of LSB galaxies must indeed by more diffuse
(the basis of the DD model).  Yet somehow they must also attain the $V_c$
dictated by the Tully-Fisher relation (the basis of SH).

\placetable{prosandcons}

Neither DD nor SH alone adequately explain the observations.
It seems that some hybrid is required.  Perhaps a convolution of
the $\delta$ and $\lambda$ distributions is needed, with both initial
density and spin contributing to determine the surface brightness.
Indeed, a proper theory should account for both effects; these simple
models assume one or the other dominates.

The hybrid approach has its own problems.  To match the environments
of LSB galaxies and the slow rate of rise of their rotation curves,
DD must make a substantial contribution to the hybrid.  However,
any mixing of DD with SH degrades the latter's ability to explain the
Tully-Fisher relation.  A perceptible shift in the normalization with
surface brightness would be introduced, and the scatter would increase.

Having argued that neither model nor a hybrid is adequate,
we note that the DD model was able to explain many of the observations
of LSB galaxy densities and ages, and did successfully predict the slow
rate of rise of LSB galaxy rotation curves and the shift
in the correlation function (Table~3).  Yet it is made very implausible
by LSB galaxies obeying the Tully-Fisher relation.
Why a model should be successful in many
respects and yet fail on one important point is a puzzle.  But fail it
does, so in the following section we examine more elaborate models to
see if they fare any better.

\subsubsection{Complex, Specific Model Galaxies}

Until now, we have intentionally kept our models simple
so that genuine predictions can be identified and tested.
More complex models necessarily have more parameters and generally assume a
specific cosmogony.  Many make no specific predictions which are actually
testable.  However, some do, and we examine these here.

A more elaborate version of the SH model is given by
Dalcanton\markcite{D97} \etal (1997).  They give a detailed model which
specifically incorporates LSB galaxies and makes testable predictions about
them.  In essence this is an elaboration of SH and shares the same virtues
and vices.  As such, it does have the correct Tully-Fisher relation, with
LSB disks reaching further out into the halo than HSB disks.
Consistency with the Tully-Fisher relation is not a prediction,
it is a construction motivated by the data.  The initial prediction of the
same physical scenario does predict a shift in the Tully-Fisher relation
for LSB galaxies (Dalcanton\markcite{D95} \etal 1995).

Nevertheless, the model of Dalcanton\markcite{D97} \etal (1997) does have
the virtue of making a number of specific, bona-fide predictions
about the bivariate distribution $\Phi(\csb,h)$ (their Fig.~2), the
differential luminosity density as a function of surface brightness
$J(\csb)$ (their Fig.~7), and the shapes of the rotation curves (their
Figs.~1 and 8).  At this time, the bivariate distribution is too
ill-determined observationally to provide a good test.  The best available
estimate of $\Phi(\csb,h)$ is provided by de Jong\markcite{dJ3} (1996).
It has sharp features which are not obviously consistent
with the smooth prediction of $\Phi(\csb,h)$, but the error bars are
sufficiently large that it is not obviously inconsistent except for very
high surface brightnesses.  Here, many more galaxies are predicted than
observed.  This is the same problem already discussed for $\lambda$-$\Sigma$
in \S 5.2 and Fig.~8.

The differential luminosity distribution $J(\csb)$, though still rather
uncertain observationally, currently provides a test which is less sensitive
to assumptions about the size distribution.  The
predictions of Dalcanton\markcite{D97} \etal (1997) are reproduced
together with the available data (as per McGaugh\markcite{me} 1996a)
in Fig.~11.  The model overpredicts the luminosity density in HSB
galaxies just as it overpredicts their number.
It also overpredicts the luminosity density due to LSB galaxies.
The predicted distribution is broad and continuous unlike the sharp
featured $J(\csb)$ which is observed.  However, the disagreement at
the faint end is not terribly severe, so the model might be salvaged by
imposing a cut off at the bright end.  Disk stability might provide
such a mechanism, since at some high surface density, disk self gravity
will become so dominant that the usual instabilities set in.

\placefigure{DalcJ}

A stronger test is provided by the shape of the rotation curves.  The
model makes specific choices for the baryon fraction and mass to light
ratio which need not be precisely correct.  To circumvent this and
obtain a test which is independent of these assumptions, we use the
information provided by Fig.~1 of Dalcanton\markcite{D97} \etal (1997)
to construct directly observable quantities.  For each model of a
given mass and spin ($\mass,\lambda_s$), we extract the disk scale
length $h$ and the radius $R_{34}$ where $V(R_{34}) = \threequarters
V_c$.  These can be combined to give a dimensionless ratio $R_{34}/h$
which measures the rate of rise of the rotation curve independent of
most model assumptions and directly is relatable to observations.

The choice of $R_{34}/h$ is motivated by Fig.~9, which shows that the
rate of rise of the inner portion of $V(R)$ provides the most
sensitive test of SH models.  We choose $\threequarters$ of $V_c$
rather than $\onehalf$ because resolution can affect the estimate of
$V$ at small $R$.  For this test we only use
galaxies resolved with at least 8 beams across the diameter
corresponding to $R_{34}$.  Note that this test is independent of
inclination, so we can use galaxies from Table~2 which are adequately
resolved and for which $V_c$ is well defined.

The data and model predictions are plotted in Fig.~12.
The rate of rise of $V(R/h)$ is clearly correlated
with both luminosity and surface brightness in the sense that brighter
galaxies have more rapidly rising rotation curves.  The correlation
with absolute magnitude is particularly clear, consistent with other
work (Persic\markcite{URC} \etal 1996).  This is not to say that we
confirm the precise functional form Persic\markcite{URC} \etal (1996)
suggest for the ``universal rotation curve,'' but it is true that to
first order luminosity is a good predictor of the shape of $V(R/h)$.
As with the Tully-Fisher relation, the same result applies to both HSB
and LSB galaxies, and is not a selection effect stemming from samples
limited only to one or the other.

Also shown in Fig.~12 are curves constructed from the models of
Dalcanton\markcite{D97} \etal (1997).  These can be shifted or
stretched in the abscissa by changing the assumed \ML*\ or inserting a
\mass-$\lambda_s$ correlation.  Curves are shown for \mass\ 
independent of $\lambda_s$ and for a relation which on average gives
$L \propto \Sigma^{1/3}$ as Dalcanton\markcite{D97} \etal (1997)
predict.  The model parameters have been well chosen to reproduce the
observed range of $M_B$ and \csb.  However, they are far too low in
$R_{34}/h$: $V(R)$ is predicted to rise much more rapidly than
observed, just as in the simple SH model.

\placefigure{ROR}

Though it is possible to shift the models along the abscissa, the same
is not true for the ordinate.  The nearly order of magnitude offset in
$R_{34}/h$ is a serious problem.  Worse, the predicted run of
$R_{34}/h$ with $M_B$ is orthogonal to the observations in Fig.~12(a).
This can not be cured by a simple offset.  The shape of the model
curve is also wrong in Fig.~12(b), with $R_{34}/h$ decreasing slightly
over most of the range of $\lambda_s$ where in fact it should continue
to increase.  Hence, the inner shape of the observed rotation curves
is a serious problem for the Dalcanton\markcite{D97} \etal (1997) model.

The severe normalization problem encountered by this $\lambda_s$ driven
model could be helped by using
isothermal halos rather than Hernquist or Navarro profiles as has been
done.  However, this provides an additional parameter, the core radius.
It is $R_C$ which controls the shape of rotation curves (equation~21),
so it will always be possible to obtain an adequate solution by making
$R_C$ an appropriate function of $M_B$.  The challenge then is to
understand why halos have finite core radii contrary to the expectations
of many numerical simulations (Dubinski \& Carlberg\markcite{DC} 1991;
Navarro\markcite{NFW} \etal 1996; Cole \& Lacey\markcite{CL} 1996),
and additionally why $R_C$ assumes the particular value required at each
$M_B$ (Casertano \& van Gorkom\markcite{CvG} 1991).

Flores\markcite{FPBF} \etal (1993) describe a $\lambda_s$ model which
employs isothermal halos.  They predict how the outer slope of
rotation curves should depend on $\lambda_s$ (their Figs.~5 and 6c)
which should lead to a segregation of the data by surface brightness
in these plots.  This is not apparent in our data, but the predicted
amplitude of the effect is too small relative to the uncertainties
for this to provide a useful test.  We can note that, by their own
stipulation, the model of Flores\markcite{FPBF} \etal (1993) is only
viable for a fairly narrow range of parameters:
$f_b \approx \langle \lambda_s \rangle \approx 0.05$ and $R_c/R_{tr} > 0.2$.
The truncation radius $R_{tr}$ is the radius beyond which no baryonic infall
occurs.  The low required baryon fraction is not consistent with that
determined from clusters
of galaxies (\eg Evrard\markcite{EMN} \etal 1996); they can not both be
correct.  In the context of LSB galaxies the required truncation radius $R_{tr}$
seems to be too small.
Flores\markcite{FPBF} \etal (1993) state that the cooling time
places an upper limit on the truncation radius so that
\begin{equation}
R_{tr} < 100\left(\frac{f_b}{0.1}\right)^{1/2}
\left(\frac{V_c}{300\kms}\right)^{1/2}
\left(\frac{H_0}{100\kms {\rm Mpc}^{-1}}\right)^{-1/2} {\rm kpc}.
\end{equation}
For a galaxy like Malin 1, this gives $R_{tr} \approx 58$ kpc which
is only one disk scale length (Bothun\markcite{malin1} \etal 1987).
The disk extends much further than
this, and the baryons should have originated at even larger radii in this sort
of collapse model.  Though Malin 1 is an extreme example and equation~(23)
is subject to modification, the problem is generic to LSB galaxies.
Upper limits on their baryon fractions (\S 7.2; de Blok \&
McGaugh\markcite{dBM2} 1997) are comparable to the value $f_b \approx
0.05$ preferred by Flores\markcite{FPBF} \etal (1993).  This means either
that the true $f_b$ is much lower than tolerable, or that LSB galaxies
have collapsed so little that they sample the initial baryon fraction:
$R_o \approx R_{tr}$.  This implies very high spin,
$\lambda_s \sim 1$.  The highest spin considered in any model
is around $\lambda_s \sim 0.2$.

Since sophisticated models are unable to explain any more than
the simple DD and SH models, one wonders if perhaps the answer lies
in the complex interplay between dark matter, gas hydrodynamics, and
star formation during galaxy formation.  A fundamental assumption of
$\lambda_s$ models is that angular momentum is conserved and not
transferred between dark and luminous components.  Numerical models
indicate that angular momentum transfer does indeed occur (Katz \&
Gunn\markcite{Katz} 1991).

There has been a great deal of progress on numerical models which
incorporate hydrodynamics.  One promising example which makes testable
predictions for galaxies is that of Cen \& Ostriker\markcite{CenOs} (1993).
Their Fig.~3 predicts the trend of total mass and cold gas mass with
stellar mass, and is reproduced here in Fig.~13 together with the data.

The data are plotted in Fig.~13(a) assuming $\ML* = 2$ and for a dynamical
mass within a radius of $R = 5h$.  The precise value assumed for \ML*\ makes no
difference on this logarithmic plot which spans many decades.  A different
choice for $R$ simply translates the dynamical mass vertically.  The trend
predicted by Cen \& Ostriker\markcite{CenOs} (1993) qualitatively mimics
that of Fig.~4(a).  However, they predict a much stronger change of
$\mass_T$ with $\mass_*$ than is actually observed.  The difference between
the slope of the data and that of the line of constant $f_b$ is barely
perceptible.

\placefigure{CenOsfig}

Presumably, halos extend further than $R = 5h$.  Some of the
data could be shifted into agreement with the models by an appropriate
choice of $R$, but not all of it at once.  This could only be done by
forcing $R_H$ to be an appropriate function of $\mass_*$, \ie by
inserting the answer.  Unfortunately, there is no test here.  The simulations
predict $\mass_T$ but not $\mass(R)$, while observations can only
probe the latter.

There is a test in the other panel of Fig.~3 of Cen \&
Ostriker\markcite{CenOs} (1993).  This is a prediction of the cold gas
mass as a function of stellar mass, and is reproduced in Fig.~13(b).
This is an important test of any hydrodynamical model which attempts
to model the gastrophysics of galaxy formation.  Data for spiral
galaxies are taken from McGaugh \& de Blok\markcite{MdB} (1997), and
that for elliptical galaxies from Wiklind\markcite{WCH} \etal (1995)
with \ML*\ computed following the relation of van der
Marel\markcite{vdM} (1991).  The model prediction is orthogonal to the
data, and off by five orders of magnitude at the bright end.

It would appear that numerical models are as yet a long way
from producing realistic galaxies
(see also Navarro \& Steinmetz\markcite{NSz} 1997).

\subsection{More Fine Tuning Conspiracies}

The fundamental problem is this: one needs pieces of both the DD and SH
models, yet grafting them together always results in a serious fine-tuning
conspiracy.  Recall that $V_H \propto \sqrt{\rho_C} R_C$.
To satisfy the Tully-Fisher relation (Fact 1), $V_c$, and hence the product
$\rho_C R_C^2$, must be the same for all galaxies of the same luminosity.
Lower surface brightness galaxies have rotation curves which rise more
gradually than those of higher surface brightness galaxies of
the same luminosity (Fact 2).
To decrease the rate of rise of the rotation curves as surface brightness
decreases, $R_C$ must increase (equation~21).
Together, these two require a fine-tuning
conspiracy to keep the product $\rho_C R_C^2$ constant at fixed luminosity
while $R_C$ must vary with the surface brightness.  That is,
$V_H \propto \sqrt{\rho_C} R_C$ stays fixed while $R_C$ and $\sqrt{\rho_C}$
oscillate up and down as dictated by \csb.  The \lincsb-\MLo\
conspiracy has grown into one tying the optical and
halo parameters intimately together.

Though both DD and SH and their variants do not work,
there is a very clear need to keep
$V_H$ fixed (as in SH) and an equally clear need to vary $R_C$ (as in DD).
The exact amount of variation depends on how the disk-halo decomposition
is treated (de Blok \& McGaugh\markcite{dBM2} 1997).  For example,
in the maximum disk case, much of the variation in $R_C$ is transferred
to \ML*, resulting in a systematic increase of \ML*\ as \lincsb\ decreases,
opposite the sense indicated by the colors.  The problem simply shifts from
one parameter to another.  In general, introducing more parameters just
increases the number of things which must be fine tuned.

To be successful, structure formation models must reproduce this
peculiar behavior.

\section{Flavors of Dark Matter}

In this section, we examine constraints that our data can place
on the various hypothesized forms of dark matter.  Ideally, this requires
testable predictions for each.  There are some, but not many.  The issue
of dark matter has always been data driven.  As a result, the process
is often more a matter of examining the sensibility of previous data-based
inferences with the expanded dynamic range of the present data.

\subsection{Cold Dark Matter}

Perhaps the leading candidate for the dark matter is CDM.  This DM candidate
is composed of dynamically cold massive particles.  Usually imagined
to be some hypothetical fundamental particle (\eg WIMPs or Axions), CDM
could also be massive black holes or some other entity which only interacts
gravitationally.  Here we are concerned
with only the dynamical effects, since this is all that defines CDM.
By stipulation, there is very little else about it that can be tested.
This is both a virtue and a vice, since DM candidates which are relatively
easy to detect (\eg faint stars) are more easily falsified.

Nevertheless, CDM has many genuine virtues.  A non-baryonic form of dark matter
is required to reconcile dynamical measures of the cosmic density ($\Omega
\gtrsim 0.2$; \eg Davis\markcite{omega} \etal 1996) with the low density of
baryons indicated by primordial nucleosynthesis ($\Omega_b \lesssim 0.03$;
Fields\markcite{omegab} \etal 1996; Copi\markcite{nucleo} \etal 1995).
An important aspect of CDM is that it does not respond to radiation
pressure.  It can therefore begin to clump and form structure early without
leaving too much of an imprint on the microwave background.  Indeed, this
was another motivation for inventing CDM, since it is not otherwise
possible to get from the very smooth universe that existed at the time
of recombination to the very clumpy one we see today.

These two facts, $\Omega \gg \Omega_b$ and the lack of structural imprint
on the microwave background demand CDM.  Indeed, the ability of CDM
models to hide anisotropies in the microwave background far below
the upper limits of various experiments was considered a great success
until fluctuations were actually measured by COBE (Bennett\markcite{COBE}
\etal 1994).
These are at a level much higher than expected in standard $\Omega = 1$ CDM.

The shape of the power spectrum $P(k)$ predicted by standard CDM has
been clearly falsified (\eg Fisher\markcite{powerspec} \etal 1993).
The expectation $P(k) \propto k^n$ with $n = 1$ is not so much a
bona-fide prediction as the physically most plausible case.  So it is
possible to repair CDM after the fact.  Ideas for doing this
include ``tilting'' the spectrum
(twiddling $n$), lowering $\Omega_{CDM}$, adding an admixture of hot dark
matter to boost $P(k)$ on large scales, or invoking the cosmological
constant.  Indeed, it appears necessary to do several of these things
(Ostriker \& Steinhardt\markcite{OS} 1995).

Even so, there are no dark matter candidates more viable than CDM, which
does at least make some testable predictions.  For example, the mass function
of galaxies is directly calculable, and can be used to predict the
observed luminosity function.  It is very difficult to reconcile the two
(Heyl\markcite{heyl} \etal 1995).  One expects a much higher ($> 300\kms$)
pairwise velocity dispersion than is observed ($< 100\kms$;
Cen \& Ostriker\markcite{CenOs} 1993; Governato\markcite{GMCSLQ} \etal 1997).
Though structure forms rapidly in CDM, CDM scenarios have considerable
difficulty in explaining the presence of large disks (Prochaska \&
Wolfe\markcite{PW} 1997) and hot 
clusters (Donahue\markcite{DGLHS} \etal 1997) at high redshift.
It also anticipates (Kaiser\markcite{nick} 1986)
rapid evolution in the cluster X-ray luminosity function
which is not observed (Scharf\markcite{warps} \etal 1997;
Burke\markcite{sharc} \etal 1997; Rosati\markcite{rosat} \etal 1997).

Another important test is whether LSB galaxies
are biased relative to HSB galaxies.  This clearly is not as strong an
effect as would be required if $\Omega = 1$ (Pildis\markcite{PSE} \etal 1997),
but does occur in the expected sense (Mo \etal 1994).  This is a definite
success, but not one specific to CDM.  The tendency for
low density galaxies to reside in low density regions is a generic
consequence of bottom-up structure formation.

A very important, testable, bona-fide prediction of CDM has recently been made
by Navarro\markcite{NFW} \etal (1996) and Cole \& Lacey\markcite{CL} (1996);
see also Dubinski \& Carlberg\markcite{DC} (1991).
They show that individual CDM halos have a universal structure profile.
These take the form
\begin{equation}
\rho_{CDM}(R) = \frac{\rho_i}{\left(R/R_s\right)
\left(1+ R/R_s\right)^2}
\end{equation}
where $R_s$ is the characteristic radius of the halo and
$\rho_i$ is related to the density of the universe at the time of collapse.
These parameters are not independent and are set
by the cosmology.  The concentration of the resultant halo is encapsulated
in the concentration parameter $c = R_{200}/R_s$.  $R_{200}$ is the radius
where the density contrast exceeds 200, roughly the virial radius.
This establishes a clear expectation value for the mass distribution
and resultant rotation curves of CDM halos, which are
\begin{equation}
V(R) = V_{200} \left[\frac{\ln(1+cx)-cx/(1+cx)}
{x[\ln(1+c)-c/(1+c)]}\right]^{1/2},
\end{equation}
where $x = R/R_{200}$
(Navarro\markcite{NFW} \etal 1996).  The velocity $V_{200}$ is characteristic
of the halo, and is defined in the same way as $R_{200}$.  The halo
rotation curve is thus specified by two parameters, $V_{200}$ and $c$,
which give the total halo mass and the degree of concentration of that
mass.

Low surface brightness galaxies are a good place to test this prediction
as their disks are dynamically insignificant: the rotation curves provide
a direct map of the dark mass distribution.  The strongest test is afforded
by the lowest surface brightness galaxy with the best resolved rotation
curve.  Surface brightness is important because it is related to the rate
of rise of the rotation curve which constrains the concentration parameter.
Resolution is important because we wish to test the shape of the
rotation curve specified by equation~(25).  The galaxy which best suits
these requirements is F583--1.  

In order to compare observations with theory, we need to do several things.
Observationally, we wish to remove the influence of the known baryonic
component and isolate the dark matter.  This is
done by assuming the maximum disk mass of de Blok \& McGaugh (1997).
Both full and corrected rotation curves are shown in Fig.~14, where it
can be seen that the assumption about the optical disk mass is irrelevant.
This is the virtue of using LSB galaxies for this sort of test.  

Theoretically, we need to specify an appropriate $c$ and $V_{200}$.
The concentration parameter depends on the cosmology, for which we consider
several possibilities using the halo characterization code provided by
Navarro (1997, private communication).  We will refer to three basic cases:
standard CDM with $\Omega = 1$ (SCDM), open CDM with $\Omega = 0.3$ (OCDM),
and a flat $\Lambda$-dominated cosmology with $\Omega_{\Lambda} = 0.7$
($\Lambda$CDM).
Lower $\Omega$ generally results in lower $c$, as does lower $H_0$.
The latter is not a strong effect, so we retain a fixed $H_0 = 75\kms
{\rm Mpc}^{-2}$.  Other parameters matter fairly little, except the
normalization of the power spectrum which we fix to the COBE observations.
Adopting another normalization, like that for rich clusters, has the fairly
minor effect of interchanging the relative concentrations of the OCDM and
$\Lambda$CDM cases:  OCDM is the least concentrated model with a COBE
normalization, but $\Lambda$CDM is the least concentrated
for a lower normalization.  This distinction is unimportant.

We shall see that the observations require extremely low $c \lesssim 5$.
Such low concentrations are readily obtained only by lowering
the power spectrum normalization to $\sigma_8 < 0.2$.  This is
completely inconsistent with either the COBE or the rich cluster
normalization.

The last item we need is an estimate $V_{200}$ or the mass of the halo, which
is also a minor factor in determining $c$.  This can
be done in a variety of ways.  Perhaps the most obvious is to use the
observed baryonic disk mass as an indicator of the halo mass.
The mass of the disk is reasonably well constrained by
the maximum disk solution ($\mass_* \le 4.5 \times 10^8 \;\mass\solar$)
and the fact that most of the baryons are in
directly observable atomic gas ($\mass_g = 2.4 \times 10^9 \;\mass\solar$).
It is unlikely that any significant additional baryonic mass is in undetected
forms (either molecular or ionized gas), and even factors of two here will
not much affect the arguments which follow.  The total baryonic mass of
F583--1 is thus $\mass_{bar}
\approx 2.9 \times 10^9 \;\mass\solar$.  This can be combined with a
baryon fraction to give a halo mass.  The universal baryon fraction is thought
to be well measured by rich clusters of galaxies, giving $f_b \approx 0.09$
(for $H_0 = 75\kms {\rm Mpc}^{-2}$; \eg White \& Fabian 1995).  This then
implies a halo mass of $\mass_H \approx 2.9 \times 10^{10} \;\mass\solar$.
Note that the maximum disk decomposition already implies $\mass_H \ge
2.2 \times 10^{10} \;\mass\solar$ (de Blok \& McGaugh\markcite{dBM2} 1997),
so either this disk is observed to very near the edge of its halo, or
the baryon fraction determined in clusters is not universal (\S 8).

\placetable{Navconc}

The concentration indices derived for these cases are listed in Table~4
and the results plotted in Fig.~14(a).  The results are disastrous.  The
SCDM model grossly overpredicts the rate of rise of the rotation curve.
The less concentrated OCDM and $\Lambda$CDM models do the same.  Worse,
all predict very much the wrong asymptotic velocity, as the halo mass
appropriate for the observed baryon mass gives $V_{200} = 43\kms$ when
$V_c = 84\kms$.

\placefigure{Nav}

Part of the problem here is the well known failure of CDM models to
simultaneously match the observed luminosity density and the normalization
of the Tully-Fisher relation (\eg Heyl\markcite{heyl} \etal 1995;
Frenk\markcite{carlos} \etal 1996).
Let us therefore try another approach.  Navarro\markcite{sesto} (1996a)
suggests that the Tully-Fisher relation arises because $V_c \approx V_{200}$
(though note that Navarro\markcite{NavIAU} 1996b found that halo mass
should not be well correlated with optical luminosity).  By adopting
$V_{200} = 80\kms$, we should at least come close to matching the outer
portion of the rotation curve.  This implies a much more massive halo,
$\mass_H \approx 1.2 \times 10^{11} \;\mass\solar$, and a correspondingly
lower baryon fraction, $f_b = 0.015$.

This exercise again fails (Fig~14b).  Only the lowest $c$ model comes
close to the observations, and even that predicts velocities which are too
high for the dark matter.  This is especially true in the inner parts, but
remains true even in the outer parts where the normalization was set.
Apparently, the Tully-Fisher relation does not arise form a simple equation
of $V_c$ with $V_{200}$.  But if it does not, why does a Tully-Fisher relation
with small scatter arise at all?

Note that the shapes of the observed rotation curves are not similar to the
predicted shapes.  Is it possible to fit the data with equation~(25) at
all?  The answer would appear to be no.  Fig.~14(c) gives several examples
which come reasonably close by choosing $c$ and $V_{200}$ without regard
to their cosmological origins.  The model with $c = 12$ gives a nicely flat
rotation curve for the outer points, but grossly over predicts the inner
rotation.  Lower concentration models can be made to fit the interior points,
but then get the exterior ones wrong.  Equation~(30) gives the wrong
shape, and the clear prediction of CDM is simply not realized:
\begin{equation}
\rho_{CDM}(R) \ne \rho_{obs}(R).
\end{equation}
Moreover, we do not have the freedom to fit $c$ and $V_{200}$ freely.
The virtue of the model is that these are predicted once the cosmological
parameters (especially $\Omega$, $P(k)$, and $f_b$) are stipulated.
No plausible cosmology predicts $(c,V_{200})$ which approximate the
lowest surface brightness galaxies.

It has already been noted (Moore\markcite{Moore} 1994; Flores \&
Primack\markcite{FP} 1994) that the steep interior density
distribution ($\rho_{CDM}(R) \propto R^{-1}$ at small $R$)
predicted by CDM is completely inconsistent with the few (4) analyzed
observations of dwarf galaxies.  These are all low surface brightness
systems, which is the reason they are relevant.  This problem of the
shape of the rotation curves is general and clear in all of our data.
We therefore confirm and extend the results of Moore\markcite{Moore} (1994)
and Flores \& Primack\markcite{FP} (1994).

This situation admits three possibilities:
\begin{enumerate}
\item CDM is not the solution to the mass discrepancy problem;
\item CDM is correct but $\rho_{CDM}(R)$ has not been correctly predicted; or
\item Both are correct but further physics intervenes to transform
$\rho_{CDM}(R)$ into $\rho_{obs}(R)$.
\end{enumerate}
The second possibility seems unlikely.
There is widespread agreement between independent modeling efforts
(Dubinski\markcite{Dub} 1994; Navarro\markcite{NFW} \etal 1996;
Cole \& Lacey\markcite{CL} 1996).
There is good physical reason for this agreement.
Non-baryonic CDM only interacts gravitationally, with no way to set a
preferred scale such as a core radius.  This prediction of CDM seems very
robust.

Within the framework of CDM, we are thus forced to consider the third
possibility.  This requires large scale mass redistribution
which presumably results from the behavior of the baryonic component.
The behavior of the baryons depends on such things as the hydrodynamics
of gas flows and energy input from star formation, and is much more
difficult to model than the CDM.  Various possibilities
for the effect of the baryons are generally referred to as ``feedback''
or sometimes as ``gastrophysics.''

That feedback takes place at some level will modify the prediction that
arises from CDM-only models.  However, there is no guarantee that it will
fix the problem under consideration.  If we simply add feedback parameters
to the model and tune them to match the observations, the model loses its
predictive value.

There are three basic possibilities for mass redistribution:
\begin{enumerate}
\item Contraction of the CDM following dissipation of the baryons;
\item Orbital family redistribution at small radii; and
\item Expansion of the CDM following expulsion of baryons in galactic winds.
\end{enumerate}
These might come about in a myriad of ways;
we discuss the generic aspects of each in turn below.  There is, however,
a serious objection to any of them being important to large scale redistribution
of the dark matter.  In dark matter dominated LSB galaxies,
baryons are a very small fraction of the total mass.  To alter $\rho(R)$
in the required fashion is a case of the tail wagging the dog.

The first possibility is that the dissipation of baryons during disk
formation draws the dark matter distribution further in.  This is sometimes
called the adiabatic response of the halo to the disk, and should happen
at some level (Dubinski\markcite{Dub} 1994).  It is relatively straightforward
to model, but makes the problem {\it worse}.  Halos which are initially
too concentrated become even more so.

In order to model adiabatic contraction, one generally assumes that the
baryons conserve angular momentum.  This need not be the case, as numerical
models indicate that a great deal of angular momentum can be transported
to the dark matter (Katz \& Gunn\markcite{Katz} 1991).
Such a process might result in
the redistribution of orbital families, especially at small radii where
the importance of the baryons is maximized.  Such a redistribution might
alter the cuspy nature of the initial CDM mass distribution.  This process
must transfer enough angular momentum to the dark matter to establish
a large core radius, but not so much that it fails to form a rotationally
supported baryonic disk.  Simulations suggest that while some orbital
redistribution does occur, it does not significantly impact $\rho(R)$
(Merritt\markcite{Merritt} 1997).

The third possibility is the opposite of the first.  Massive amounts
of baryons are expelled.  The dark matter follows gravitationally, thus
establishing a more diffuse mass distribution. 
A frequently invoked mechanism for expelling the baryons
is feedback due to violent star
formation (\eg Navarro\markcite{Eke} \etal 1996c).  There is
one clear prediction of this scenario: galaxies explode and gas
is lost.  Yet the dim galaxies for which the need for mass
redistribution is most severe are in fact very gas rich (McGaugh \& de
Blok\markcite{MdB} 1997).  In addition, a direct search for the
residue of baryonic blow out resulted in non-detection (Bothun \etal
1992).  It therefore seems unlikely that baryonic outflows can play
a significant role in redistributing the dark matter.

Regardless of what mechanism is invoked, two drastic events are required
to reconcile CDM with the observations.
First, the baryon fraction must change by an order of magnitude from
the universal value found in clusters.  Second, the dark matter
distribution must be radically changed from a Navarro\markcite{NFW}
\etal (1996) profile (equation~24) to something close to an isothermal
sphere (equation~20).  This is indeed a major transformation:
$\rho(R) \propto R^{-1} \rightarrow R^0$ at small $R$ and $\rho \propto
R^{-3} \rightarrow R^{-2}$ at large $R$.

In many ways, CDM is superior to other hypothesized forms of dark matter.
Most importantly, it does make some testable predictions.  Unfortunately,
these predictions persistently fail.

\subsection{Hot Dark Matter}

Another possible dark matter candidate is a massive neutrino.
Though often invoked in the context of the solar neutrino problem,
a massive neutrino does not help with the mass discrepancy problem
in LSB galaxies.  It has long been known that hot dark matter is good
at forming structure on large scales (top-down), but not at making galaxies.
In order for neutrinos to remain in dwarf galaxy halos, they would have
to be much more massive than allowed by experiment
(Lin \& Faber\markcite{LF} 1983).  It seems unlikely that
mixing hot and cold dark matter would address the problems encountered
by CDM alone, simply because the HDM component plays no role on the
small scale of galaxies.

\subsection{Baryonic Dark Matter}

Baryonic dark matter is any form of hypothesized DM composed of ordinary
matter.  There are a great variety of hypothesized forms
including brown dwarfs, very faint stars, Jupiters, and
very cold molecular gas.  The advantage of BDM is that baryons
are known to exist.  The problem is that BDM candidates are detectable, so
most have been ruled out (Carr\markcite{Carr} 1994).

The case in favor of BDM has been strengthened by the recent detection
of microlensing events (\eg Alcock\markcite{macho} \etal 1996).
However, the detection of microlensing,
a physical process which should at least occasionally occur, is not the
same as the detection of BDM MACHOs.  The clearest result of these
experiments is to rule out MACHOs as dark matter over many decades of mass,
up to and including the most reasonable possibility of brown dwarfs
(Ansari\markcite{EROS1} \etal 1996; Renault\markcite{EROS2} \etal 1997).
Only if the lensing objects are surprisingly massive ($\gtrsim 0.5 M\solar$)
can MACHOs make up a significant fraction of the halo, and only then
because this mass delimits the sensitivity range of the experiments.
Sufficient statistics have yet to accrue to demand a significant halo
mass fraction (Paczynski\markcite{pacman} 1996),
and there is no guarantee that the
events observed to date are connected with the missing mass problem.
Since particle physics analogies are often made for these experiments,
we offer one of our own.  The first thing one is likely to learn
about is the background contaminants one did not expect (\eg
Zaritsky \& Lin\markcite{ZL} 1997).

A substantial amount of BDM suffers the two serious drawbacks which
motivate CDM: the small amplitude of fluctuations in the microwave
background and $\Omega \gg \Omega_b$ from primordial nucleosynthesis.
Although $\Omega_b \gg \Omega_*$ is often invoked as an argument for
baryonic dark matter, this can not explain the dynamical mass discrepancy
or add up to the total $\Omega$  (Dalcanton\markcite{D94} \etal 1994).
To contemplate a conventional universe composed entirely of baryons,
one must somehow dismiss both of these pillars of modern cosmology.

We are not aware of viable BDM scenarios which make specific predictions.
Those which did have been ruled out.  Nevertheless, there are several
lines of argument favoring BDM.  One is that light and dark matter are
intimately coupled.  This is certainly true (paper II).
The potential for information transfer between the
distinct dynamical components is maximized if they are cut from the same cloth.
Another connections is
that the ratio of atomic gas mass to dynamical mass $\mass_{HI}/\mass_o$ seems
to be roughly constant (Bosma\markcite{Bosma} 1981).  This holds
approximately true for LSB galaxies (see Fig.~13 of de Blok\markcite{BMH}
\etal 1996) as $\mass_{HI}/L$ and \MLo\ are both related to \csb\ with
similar slopes.

A serious problem with this picture is the total failure of $f_*$ to
redress the \MLo-\lincsb\ relation (Fig.~7).  This occurs because a constant
$\mass_{HI}/\mass_o$ gives a constant shift in $\log \MLo$, not a systematic
trend as required.  The only way to salvage this picture would be
to arbitrarily make the conversion factor $\eta$ between $\mass_{HI}$ and
total gas mass an appropriate function of \csb.  That is,
since $\eta = 1.4$ strictly applies only to atomic gas, one might
suppose that molecular gas becomes systematically more important
with decreasing surface brightness.  This would have to be an enormous
effect to account for the large change in \MLo, and of course must be
very systematic with surface brightness.  This seems quite unreasonable,
and in fact the opposite is observed:  lower surface brightness galaxies
appear to have substantially less molecular gas than higher surface
brightness spirals (Schombert\markcite{CO} \etal 1990).
An additional conspiracy must now be invoked:  the $CO/H_2$ conversion
factor also varies systematically with surface brightness.
This would have to be a big effect, but the size if not the systematic
surface brightness dependence is already required if much
of the dark matter is to be hidden in molecular form
(Pfenniger\markcite{PCM} \etal 1994).

Another motivation for BDM is that the outer shape of the
rotation curve of HSB galaxies can sometimes be fit by scaling up
the contribution of the HI.  This suggests an additional undetected
gaseous component with the same distribution as the HI but substantially
more mass.  Implicit in this is that the dark matter be confined to a
disk, which has the usual problems with disk instability
(Ostriker \& Peebles\markcite{OP} 1973).  If instead the BDM is distributed
in a halo, there is no reason to draw a connection between the shape of
the HI disk and that of the total rotation curve.

We can test the notion of gas-scaling against the rotation curves of LSB
galaxies.  This sometimes works tolerably well, as long as one is free
to adjust the scaling factor from galaxy to galaxy.  However, it frequently
fails to reproduce the shape of the rotation curves at all well.
In Fig.~15, we show the {\it best\/} fit that can be
obtained by this method to the best resolved LSB
galaxy.  The shape of the actual rotation curve is completely different
from that predicted by the HI distribution.  This is often true of the
LSB data, so even though there may be some cases of HSB galaxies where the
rotation curves are reasonably well predicted by the HI distribution, this
simply is not the case in LSB galaxies.  This removes a
major argument in favor of BDM.

\placefigure{gasscale}

The various arguments in favor of BDM make no sense in the context of
the LSB galaxy observations.  This does not completely rule out the existence
of BDM, or mean that we have discovered and cataloged every baryon in the
universe.  But it does make it very difficult to construct a sensible model
with a dynamically important component of BDM.

In general, the arguments which favor BDM are based on the inferred
coupling between dark and luminous matter.  With this much we certainly
concur: the two components must be intimately related (\eg \S 6.2).
Indeed, there does exist a unique analytic formalism relating the
dynamics to the observed luminous mass distribution (papers II and III).

\section{General Constraints on Baryon Fractions}

There are a few interesting constraints that can be placed on the
presence of dark matter completely irrespective of its composition.
There is a wealth of data besides rotation curves which bear on the issue.
These place both upper and lower limits on the extent and mass of dark
matter halos, and sometimes provide measures of the actual baryon fraction.

Observations which place lower limits include
the maximum radii to which rotation curves remain flat (Sancisi \&
van Albada\markcite{SvA} 1987; Meurer\markcite{Gerhardt} \etal 1996),
the timing argument in the local group (Kahn \& Woltjer\markcite{KW} 1959;
Peebles\markcite{LGtime} \etal 1989),
the statistical motions of satellite galaxies at large separations
from $L^*$ galaxies (Zaritsky \& White\markcite{ZW} 1994), and QSO
absorption lines which imply large halos with
roughly flat rotation curves extending
to very large radii (Barcons\markcite{BLW} \etal 1995).  Evidence which
places upper limits on the extent of dark halos is more difficult to
establish, but there is the very important result that overly massive
halos suppress the formation of observed tidal tails (Dubinski\markcite{tails}
\etal 1996).  In addition, it might be possible to constrain the amount
of dark matter allowed within the optical radius by disk stability constraints.
Some dark matter is needed to stabilize disks (Ostriker \& Peebles\markcite{OP}
1973), but too much inhibits features like spiral arms which stem from
self gravity in the disk.  This criterion has been used by
Athanassoula\markcite{ABP} \etal (1987) to place minimum masses on disks
in disk-halo decompositions.  Finally, it now seems reasonably well established
that actual baryon fractions can be measured in X-ray galaxy clusters so
that the ratio of dark to luminous mass is known (White\markcite{WNEF} 1993;
White \& Fabian\markcite{WF} 1995; Evrard\markcite{EMN} \etal 1996).

The most interesting (\ie extreme) of these limits are shown in
Fig.~16 and given in Table~5.  Listed are the object, the type of data
and analysis, the linear extent required of the dark matter halo (in
both kpc and disk scale lengths where applicable),
the observed mass to light ratio \MLo, the ratio of dark to
luminous mass, the corresponding baryon fraction, and the source of
the data.  We have been as conservative as possible in assessing these
limits; most of them are quite hard.
 
\placetable{darklimit}

Listed first are those observations which place lower limits on the amount
of dark matter required.  The hardest of these limits come from the extended
\HI\ rotation curves of disk galaxies. These are observed to remain flat
as far as observed, sometimes to very large radii.  To calculate the limits,
we have used the maximum disk case, attributing as much mass as possible to
the luminous matter.  The minimum required dark mass is then just 
$V_c^2 R/G - \mass_{bar}$.  NGC 3198 is a classic
example of an HSB spiral with a rotation curve which remains flat to
11 scale lengths, nearly 3 times the extent of the bulk of the luminous mass.
The dark halo must extend at least this far, and presumably much further
since there is no hint of a turn down in $V(R)$.  However, the required amount
of dark mass is fairly modest: $\mass_{dark}$ exceeds the observed luminous
mass $\mass_{bar}$ by only a factor of a few. 
Recently, Meurer\markcite{Gerhardt}
\etal (1996) have been able to trace $V(R)$ out to 22 scale lengths in the
HSB dwarf NGC 2915.  This extraordinary radius places interesting constraints,
with the dark mass exceeding the luminous baryonic mass by at least a factor
of 19.  There are 4 LSB galaxies listed in Table~5; we have chosen the most
extreme examples with $\mass_{dark}/\mass_{bar} > 10$. There are many
more which are nearly as extreme (de Blok \& McGaugh\markcite{dBM2} 1997).
Even though $V(R)$ is not measured to large $R/h$ in LSB galaxies as in
the above 2 examples, it should be obvious from Fig.~4
that LSB galaxies provide interesting limits in this way. The
lowest surface brightness galaxies have $\MLo > 30$,
so $\mass_{dark}/\mass_{bar} > 10$
for $\mass_{bar}/L = 3$.  This is a fairly conservative number in the $B$-band;
$\ML* \approx 2$ is more realistic.
We do not need to make any assumption about \ML*\ though:  the
dynamical mass of the baryonic disk is constrained by the rotation curve.

\placefigure{MdMbRh}

We stress that the limits obtained from the maximum disk
decompositions of the rotation curves provide very hard lower limits
on $\mass_{dark}/\mass_{bar}$ and $R_H$.  Maximum disk does not
generally return realistic values for \ML*, especially in LSB galaxies
(de Blok \& McGaugh\markcite{dBM2} 1997).  A more reasonable estimator
based on disk velocity dispersions (Bottema 1993\markcite{Bott1},
1997\markcite{Bott2}) gives
nearly a factor of 2 less in HSB galaxies and even less in LSB
galaxies.  Also, we have of course not observed to the edge of the
halo.  It is difficult to construct a model in which the effects of an
outer edge to the halo does not lead to observable consequences unless
$R_H > 2 R_o$.  The point is that the limits in
Table~5 are both hard and conservative.  A more realistic estimate
gives a result 3 or 4 times more extreme.

Aside from \HI\ rotation curves, there are a number of other observations
which place lower limits on the extent of dark matter halos.  
Barcons\markcite{BLW} \etal (1995) present observations of two systems with
QSO absorption lines in ionized gas many tens of kpc from the
centers of disk galaxies, very nearly along the major axis.  The gas is
apparently associated with the galaxies, though the geometry and orbital
orientation are unknown.  Nevertheless, nearly flat rotation seems to
persist out to very large radii.  If interpreted in the most obvious way,
this implies limits even more extreme than those
derived from bona-fide rotation curves.

On larger scales, the timing argument in the local group
(\eg Peebles\markcite{LGtime} \etal 1989) implies yet more dark matter.
The same sort of result follows from the motions of satellite galaxies
around $L^*$ galaxies, which show no evidence of an edge to the halos
to the largest scales probed, $\gtrsim 200$ kpc (Zaritsky \& White\markcite{ZW}
1994; Zaritsky\markcite{Z97} \etal 1997).  Though statistical
in nature, both of these approaches robustly require $\MLo > 100$.
For $\mass_{bar}/L = 3$, this means
$\mass_{dark}/\mass_{bar} > 33$.
This is generally consistent with the impression given by the rotation
curves:  dark matter halos are large and massive.

The lower limits just discussed are challenging to obtain.  Perhaps even
more difficult to acquire are upper limits on the extent of dark matter
halos.  Nevertheless, a few constraints can be placed.  Recently,
Dubinski\markcite{tails} \etal (1996) have pointed out that massive
halos suppress the formation of tidal tails observed around merger remnants.
This constrains the dark to luminous mass ration rather severely;
tidal features do not form unless $\mass_{dark}/\mass_{bar} \lesssim 10$. 
This upper limit is smaller than the hard lower limits imposed above,
$\mass_{dark}/\mass_{bar} > 20$.

The results of Dubinski\markcite{tails} \etal (1996) are based on $N$-body
simulations which one might simply choose to dismiss as model dependent.
However, further modeling (Mihos\markcite{tails2} \etal 1997) shows
that there is no plausible variation of parameters which changes the essential
result.  There is a very good, simple reason for this.  If dark matter
halos are large and massive, the baryonic matter is buried at the bottom
of a deep potential well.  Even in a violent merger, the largest velocities
imparted to some of the stars are insufficient to climb out of this potential
well.  This very effectively suffocates the formation of the large, linear
features observed in merger remnants to extend over tens and sometimes even
hundreds of kpc.  Systems with $\mass_{dark}/\mass_{bar} \gtrsim 10$ suppress
such features before they get started.

Strictly speaking, it is the gradient of the potential which this modeling
exercise constrains, not just the dark to luminous mass ratio
(Mihos\markcite{tails2} \etal 1997).  It is possible to contrive very
tenuous halo distributions which might weaken this limit, essentially by
placing lots of mass at large radii where it does not participate in the
merger dynamics.  The tidal tails argument therefore provides the weakest
of the limits discussed so far, and the only upper limit, so
one might be tempted to equivocate.  However, it does not appear that
the argument can simply be dismissed.  The limit it imposes is
separated from the others by a factor of 2 and points in the
{\it opposite\/} direction.  Models contrived specifically to evade
the tidal tails limit seem unlikely to satisfy other constraints.

Another way to constrain the ratio of halo to disk mass, at least within
the radius of the disk, comes from disk stability.  Purely Newtonian disks
suffer from unchecked instabilities; perturbations like bars grow
exponentially and destroy the disk in a few dynamical times.
The survival of spiral disks over a Hubble time requires some stabilizing
influence.  One possibility is to embed the disks in dynamically hot,
spherical dark matter halos (Ostriker \& Peebles\markcite{OP} 1973),
though it should be noted that altering the effective force law can also have
a stabilizing effect (Christodoulou\markcite{stab} 1991).  The problem
with dark matter halos is that they can provide too much stability.
Observed features like bars and spiral arms occur because of self-gravity
in the disk.  If the halo is too dominant, this is negligible and one must
invoke non-dynamical origins for the observed features. 
Athanassoula\markcite{ABP} \etal (1987) used this fact to constrain
\ML*\ in their
rotation curve decompositions.  In HSB galaxies, this gives reasonable
results, the relevant effects being relatively small ($\lesssim 50\%$
difference between the maximum disk and the minimum disk required to
sustain spiral structure).  Translating this into a limit on the
dark halo mass within the radius of the optical disk gives
$\mass_{dark}(R<4h)/\mass_{bar} \lesssim 3$ or 4.

LSB galaxies are so dark matter dominated, even within the optical extent of
the disk, that features due to disk self gravity should simply not be evident
(Mihos\markcite{MMB} \etal 1997).   The minimum disk
required to form spiral features by the criteria of
Athanassoula\markcite{ABP} \etal (1987) exceeds the maximum disk allowed
by the rotation curve (\eg Quillen \& Pickering\markcite{QP} 1997).
Though rarely pretty grand design spirals,
LSB galaxies do have spiral features (de Blok\markcite{BHB} \etal 1995;
McGaugh\markcite{LSBmorph} \etal 1995).  This is not an isolated problem
of a few strange morphologies; $> 80\%$ of the 198 LSB galaxies in the
primary list of Schombert\markcite{LSBcat} \etal (1992) are spirals.

Again, one might be inclined to equivocate.  Spiral structure is not well
understood, and might not have an origin internal to disk dynamics at all.
It is not likely that the spiral features can be attributed to interactions
with companions, as LSB galaxies are usually very isolated
(Bothun\markcite{small} \etal 1993; Mo\markcite{Mo} \etal 1994).
The more extreme LSB galaxies have
$\mass_{dark}(R<4h)/\mass_{bar} > 10$, whereas dynamical
spiral features seem to require $\mass_{dark}(R<4h)/\mass_{bar} < 4$.
Considerable work remains to be done to better quantify these effects,
but the discrepancy is already large.
There is a serious problem with invoking dark matter halos to stabilize disks
if spiral structure is driven by disk self-gravity as often supposed.

So far we have only discussed limits which can be placed on the amount
of dark matter required by various observations.  Recently, it has become
possible to estimate actual baryon fractions in X-ray clusters and
groups of galaxies (White\markcite{WNEF} \etal 1993; White \&
Fabian\markcite{WF} 1995; Pildis\markcite{PBE} \etal 1995).
This procedure has a number of uncertainties, but the essential result
seems to be quite robust (Evrard\markcite{EMN} \etal 1996).  This
gives a surprisingly large amount of baryonic mass, $f_b \approx 0.09$
for the value of the Hubble constant adopted here.  This resides
between the limits discussed above.  The clusters give
$\mass_{dark}/\mass_{bar} \approx 10$, while rotation curves, satellites
and the local group all require
$\mass_{dark}/\mass_{bar} > 20$ and tidal tails imply
$\mass_{dark}/\mass_{bar} \lesssim 10$.

At this juncture, it is unclear how to proceed.  One option is to
selectively disbelieve some combination of the results imposing these
contradictory limits.  Note that it is not possible to arrive at a single
universal baryon fraction by dismissing a single limit; at least several
observations must be altered.  Another option
is to conclude that $f_b$ varies arbitrarily
from halo to halo.  We could perhaps salvage a universal $f_b$ by making
some or all of the dark matter baryonic, and simply varying the fraction
which becomes luminous.  The distinction between these two latter possibilities
is small.  In both cases, we must randomly vary the amount of luminous matter
in a way stipulated only by observation.  Things that form tidal tails
happen to have small halos.  Galaxies with satellites happen to have large
halos.  The Tully-Fisher relation somehow manages to
ignore these random, large fluctuations in the luminous to dark mass ratio
and remain universal with little scatter.
Structure formation theories lose all potential predictive power
since $f_b$ becomes a free parameter for each and every halo.

It is not clear whether the contradictory limits imposed by the various
observations can be reconciled.  Further work needs to concentrate on
this point and on the fine-tuning problems that arise from the systematics
of rotation curves.  The current situation also poses a philosophical dilemma:
what would be required to falsify the dark matter hypothesis?

\acknowledgments We thank all the many, many people who have discussed
and debated these issues with us with varying degrees of patience and
credulity.  We are particularly grateful to Renzo Sancisi and Vera Rubin
for many enlightening conversations.  Most of all, we would like to
thank Thijs van der Hulst for his contributions to our work.
We also thank Peter van Dokkum for the lively debate which clarified
the virtues and vices of DD and SH, and brought out the implicit assumptions
we all were making.  We are also grateful to Chris Mihos, Roelof Bottema,
Eric Schulman, Moti Milgrom, and the referee for their comments.
We would both like to thank the Kapteyn Institute and the Department of
Terrestrial Magnetism of the Carnegie Institution of Washington for their
strong support and warm hospitality.

\appendix

\section{Color Terms and the Slope of the Tully-Fisher Relation}

In the discussion of the physical interpretation of the Tully Fisher relation
(\S 5), we adopted the near infrared luminosity as an indicator of stellar
mass.  This results in a slope sometimes attributed to the viral theorem,
\begin{equation}
L \propto V_c^4.
\end{equation}
This particular slope is not specific to dark matter in general, as it
depends on the assumptions which are necessarily made (\S 5).

Here we show that band pass dependent deviations of the slope from the
adopted value have no significant impact on our discussion of
the physical interpretation of the Tully Fisher relation.
The full requirement on the relation between mass to light ratio,
surface brightness, and luminosity imposed by the Tully-Fisher relation
in any band pass $j$ is
\begin{equation}
\Upsilon_j \propto \Sigma_j^{-1/2} L_j^{y_j/2}
\end{equation}
where $y_j = (4 - x_j)/x_j$ measures the band pass dependent
deviation of the slope of the Tully-Fisher relation from the adopted
value of 4.  This is a small effect compared to the fine balancing act which
must be performed by $\Upsilon$ and $\Sigma$.

Consider the residual color term between two bands $j$ and $k$.
The residual effect on the mass to light ratio is
\begin{equation}
\Delta \log \left(\frac{\Upsilon_k}{\Upsilon_j}\right) = \frac{4}{5 x_j}
\left[ (m_j-m_k) +\left(1-\frac{x_j}{x_k}\right) M_k \right]
\end{equation}
where $(m_j-m_k)$ is the color.  For example, if $x_V = 3$ and $x_H = 4$,
this becomes
\begin{equation}
\Delta \log \left(\frac{\Upsilon_H}{\Upsilon_V}\right) = 0.27[(V-H)+0.25M_H]
\end{equation}
so a 1 mag.\ color change corresponds to a 0.27 dex shift in the
ratio of $\Upsilon$, and 1 mag.\ in luminosity to a shift of only
0.07 dex.

Systematic changes in the slope of the Tully-Fisher relation
with band pass can plausibly be attributed to modest systematic
variation of the stellar population or extinction with luminosity.
The observed trends operate in the sense expected for stellar populations
(brighter galaxies tend to be redder) and the luminosity-metallicity
relation (brighter galaxies should contain relatively more dust).  These are
small effects which have no real impact on the $\Upsilon$-$\Sigma$ relation.
The increment for increment shift expected in the Tully-Fisher relation
with surface brightness does not happen.  This
requires fine-tuning over nearly 5 magnitudes in \csb\ irrespective
of the precise slope of the Tully-Fisher relation.

\clearpage

\begin{deluxetable}{lrrrrrr}
\tablewidth{0pt}
\tablecaption{Data \label{data}}
\tablehead{
\colhead{Galaxy}  & \colhead{$M_B$} & \colhead{\csb} & \colhead{$h$} &
\colhead{$V_c$} & \colhead{$i$} & \colhead{$\MLo$} }
\startdata
F563--1         & $-$17.3  & 23.5  &  4.3 & 111  & 25  & 36.6 \nl
F563--V2        & $-$18.2  & 22.1  &  2.1 & 111  & 29  &  8.2 \nl
F568--1         & $-$18.1  & 23.8  &  5.3 & 116  & 26  & 24.3 \nl
F568--3         & $-$18.3  & 23.1  &  4.0 & 119  & 40  & 16.2 \nl
F568--V1        & $-$17.9  & 23.3  &  3.2 & 124  & 40  & 20.8 \nl
F571--V1        & $-$17.0  & 24.0  &  3.2 &  73  & 35  & 15.6 \nl
F574--1         & $-$18.4  & 23.3  &  4.7 & 100  & 65  & 11.8 \nl
F583--1         & $-$16.5  & 24.0  &  1.6 &  88  & 63  & 18.6 \nl
F583--4         & $-$16.9  & 23.8  &  2.7 &  67  & 55  & 12.0 \nl
UGC~~128        & $-$18.8  & 24.2  &  9.2 & 130  & 55  & 28.8 \nl
UGC 6614        & $-$20.3  & 24.5  & 15.8 & 204  & 36  & 30.2 \nl
\tableline
DDO~~154        & $-$13.8  & 23.2  &  0.5 &  48  & 70  & 20.8 \nl
DDO~~168        & $-$15.2  & 23.4  &  0.9 &  55  & 58  & 13.5 \nl
NGC~~~55        & $-$18.6  & 21.5  &  1.6 &  87  & 65  &  2.6 \nl
NGC~~247        & $-$18.0  & 23.4  &  2.9 & 108  & 72  & 12.8 \nl
NGC~~300        & $-$17.8  & 22.2  &  2.1 &  97  & 79  &  8.9 \nl
NGC~~801        & $-$21.7  & 21.9  & 12.0 & 222  & 81  &  7.4 \nl
NGC 1560        & $-$15.9  & 23.2  &  1.3 &  79  & 75  & 21.2 \nl
NGC 2403        & $-$19.3  & 21.4  &  2.1 & 136  & 67  &  4.4 \nl
NGC 2841        & $-$21.7  & 21.1  &  4.6 & 323  & 69  &  6.0 \nl
NGC 2903        & $-$20.0  & 20.5  &  2.0 & 201  & 78  &  4.8 \nl
NGC 2998        & $-$21.9  & 20.3  &  5.4 & 214  & 63  &  2.7 \nl
NGC 3109        & $-$16.8  & 23.1  &  1.6 &  67  & 62  &  8.2 \nl
NGC 3198        & $-$19.4  & 21.6  &  2.6 & 157  & 64  &  6.7 \nl
NGC 5033        & $-$20.2  & 23.0  &  5.8 & 222  & 30  & 14.3 \nl
NGC 5533        & $-$21.4  & 23.0  & 11.4 & 273  & 75  & 14.1 \nl
NGC 5585        & $-$17.5  & 21.9  &  1.4 &  92  & 76  &  7.1 \nl
NGC 6503        & $-$18.7  & 21.9  &  1.7 & 121  & 66  &  4.9 \nl
NGC 6674        & $-$21.6  & 22.5  &  8.3 & 266  & 78  &  8.1 \nl
NGC 7331        & $-$21.4  & 21.5  &  4.5 & 241  & 65  &  4.3 \nl
UGC 2259        & $-$17.0  & 22.3  &  1.3 &  90  & 63  & 10.0 \nl
UGC 2885        & $-$22.8  & 21.9  & 13.0 & 298  & 68  &  5.3 \nl
\enddata
\end{deluxetable}

\begin{deluxetable}{ll}
\tablewidth{0pt}
\tablecaption{Other LSB Galaxies \label{notused}}
\tablehead{
\colhead{Galaxy}  & \colhead{Limitation} }
\startdata
F561--1   & $i < 25\arcdeg$ \nl
F563--V1  & gross asymmetry \nl
F564--V3  & gross asymmetry \nl
F565--V2  & $V(R)$ still rising \nl     
F567--2   & $i < 25\arcdeg$ \nl
F571--8   & $i = 90\arcdeg$ \nl         
F571--V2  & no $B$-band data \nl        
F574--2   & $V(R)$ still rising \nl
F577--V1  & gross asymmetry \nl
F579--V1  & gross asymmetry \nl         
UGC 1230  & $i < 25\arcdeg$ \nl
UGC 5005  & no $B$-band data \nl        
UGC 5209  & unresolved \nl
UGC 5750  & no $B$-band data \nl        
UGC 5999  & no $B$-band data \nl        
\enddata
\end{deluxetable}

\begin{deluxetable}{lcr}
\tablewidth{0pt}
\tablecaption{Model \label{prosandcons}}
\tablehead{
\colhead{Test}  & \colhead{DD} & \colhead{SH} }
\startdata
Age & $\sqrt{}$ & ? \nl
$\phi(\csb)$ & ? & ? \nl
$\xi(r)$ & $\sqrt{}$ & X \nl
TF & X & $\sqrt{}$ \nl
$V(R)$ & $\sqrt{}$ & X \nl
\enddata
\end{deluxetable}

\clearpage
\begin{deluxetable}{lcc}
\tablewidth{0pt}
\tablecaption{Concentration \label{Navconc}}
\tablehead{
\colhead{Model}  & \colhead{$f_b = 0.09$} & \colhead{$V_{200} = 80$} }
\startdata
SCDM & 59 & 52 \nl
OCDM & \phn 9 & \phn 8 \nl
$\Lambda$CDM & 12 & 11 \nl
\enddata
\end{deluxetable}

\begin{deluxetable}{lcrrrccl}
\tablewidth{0pt}
\tablecaption{Limits on the Extent of Dark Halos \label{darklimit}}
\tablehead{
\colhead{Object} & \colhead{Type} & \colhead{$R$ (kpc)} & \colhead{$R$ ($h$)}
& \colhead{\MLo} & \colhead{$M_{dark}/M_{bar}$} & $f_b$ & \colhead{Refs.} }
\startdata
& Lower Limits & & & & & & \nl
NGC 3198 & \HI\ $V(R)$ & $> \phn 30$ & $> 11$ & $> \phn 18$
& $\phn \phn > \phn 2.7 \phn $ & $< 0.27$ & 1 \nl
NGC 2915 & \HI\ $V(R)$ & $> \phn 15$ & $> 22$ & $> \phn 76$ & $> 19$ 
& $< 0.05$ & 2 \nl
F568--V1 & \HI\ $V(R)$ & $> \phn 19$ & $> \phn 6$ & $> \phn 41$ & $> 11$
& $< 0.08$ & 3,4 \nl
F571--8 & \HI\ $V(R)$ & $> \phn 15$ & $> \phn 4$ & $> \phn 49$ & $> 20$
& $< 0.05$ & 3,4 \nl
UGC 5750 & \HI\ $V(R)$ & $> \phn 21$ & $> \phn 6$
& $> \phn 10$\tablenotemark{a} & $> 11$ & $< 0.08$ & 3,5 \nl
UGC 5999 & \HI\ $V(R)$ & $> \phn 15$ & $> \phn 3$
& $> \phn 35$\tablenotemark{a} & $> 11$ & $< 0.08$ & 3,5 \nl
1704+6068 & QSO abs. & $> \phn 82$ & $> 35$ & $> \phn 75$
& $> 25$\tablenotemark{b} & $< 0.04$ & 6 \nl
2135$-$1446 & QSO abs. & $> \phn 64$ & $> 16$ & $> 123$
& $> 41$\tablenotemark{b} & $< 0.03$ & 6 \nl
Local Group & Timing & $\sim 700$ & \nodata & $> 100$
& $> 33$\tablenotemark{b} & $< 0.03$\tablenotemark{b} & 7 \nl
Satellites & Statistical & $> 200$ & \nodata & $> 100$
& $> 33$\tablenotemark{b} & $< 0.03$\tablenotemark{b} & 8,9 \nl
\tableline
& Upper Limits & & & & & & \nl
Tidal Tails & Model & $\sim 100$ & \nodata & $< 30$\tablenotemark{b}
& $< \phn 10$\tablenotemark{c} & $> 0.09$ & 10,11 \nl
Disk Stability & Theory & \nodata & $< 4$ & $< 12$\tablenotemark{b}
& $< \phn 4$\tablenotemark{d} & \nodata & 12,13 \nl
\tableline
& Measurement & & & & & & \nl
X-ray Clusters & $\beta$-model & $> 700$ & \nodata
& $> 100$\tablenotemark{e} & $11 \pm 4$\tablenotemark{f}
& $0.09 \pm 0.03$\tablenotemark{f} & 14,15 \nl
\enddata
\tablenotetext{a}{$R$-band: $\MLo = \mass/L_R$}
\tablenotetext{b}{Assumes $\mass_{bar}/L = 3$}
\tablenotetext{c}{Constrains total mass}
\tablenotetext{d}{Constrains mass within disk}
\tablenotetext{e}{Includes optical luminosity only}
\tablenotetext{f}{Scaled to $H_0 = 75\;{\rm km}\,{\rm s}^{-1}\,{\rm Mpc}^{-1}$;
errors represent full range of data}
\tablerefs{1.~Sancisi \& van Albada 1987 2.~Meurer \etal 1996
3.~de Blok \& McGaugh 1997 4.~de Blok \etal 1996 5.~van der Hulst \etal 1993
6.~Barcons \etal 1995 7.~Peebles \etal 1989 8.~Zaritsky \& White 1994
9.~Zaritsky \etal 1997
10.~Dubinski \etal 1996 11.~Mihos \etal 1997a 12.~Athanassoula \etal 1987
13.~Mihos \etal 1997b 14.~White \& Fabian 1995 15.~Pildis \etal 1995}
\end{deluxetable}

\clearpage

\figcaption[TFSB.ps]{The Tully-Fisher relation for
spiral galaxies over a large range in surface brightness.
The B-band relation is shown; the same result is obtained in all bands.
In (a), the ``authentic'' Tully-Fisher relation is shown with the abscissa
being the velocity width of the 21 cm line observed by single dish radio
telescopes.  Corrections have been applied as per Tully \& Foqu\'e (1985)
as discussed by Zwaan etal (1995).  In (b), the ``intrinsic'' Tully-Fisher
relation is plotted with $V_c$ measured from full rotation curves plotted
along the abscissa.  The velocity measurements in (a) and (b) are completely
independent, and there is little overlap between the samples.
The lines are fits to the data; though there is a perceptibly different
slope, to a good approximation $W_{20}^c$ measures $\sim 2 V_c$.
Open symbols in (a) are an independent sample (Broeils 1992) taken to define
the Tully-Fisher relation (solid line).  Solid symbols are galaxies binned by
surface brightness:  stars: $\csb < 22$;
squares: $22 < \csb < 23$;  triangles: $23 < \csb < 24$;
circles: $\csb > 24$.  Clearly, galaxies
fall on the same Tully-Fisher relation irrespective of surface brightness.
\label{figTF}}

\figcaption[u128n2403.ps]{Log-log plots showing the shapes of the rotation
curves of two galaxies, one of high surface
brightness (NGC~2403; open circles) and one of low surface brightness
(UGC~128; filled circles).  The two galaxies have very nearly the
same asymptotic velocity $V_c$ and luminosity,
as required by the Tully-Fisher relation.  However, they have
central surface brightnesses which differ by a factor of 13.
Even though the asymptotic velocities are similar,
$V_c$ occurs at a very small radius in the high surface brightness
galaxy, but not until a very large radius in the LSB galaxy.
Indeed, the rotation curve of UGC 128 is still rising at the last measured
point of NGC 2403, which occurs at $R \approx 9 h$.
\label{figcomp}}

\figcaption[R2MB.ps,R2SB.ps]{The radius of dark matter domination $R_{2:1}$
in disk galaxies of different (a) absolute magnitude and (b) central surface
brightness.  Dark matter domination is defined to occur when the mass
discrepancy reaches a factor of two, even attributing as much mass to the
disk as possible (maximum disk).  That is, $R_{2:1}$ is the radius where
$\mass_{T}(R_{2:1})/\mass_{disk}(R_{2:1}) \ge 2$.  Bright galaxies do not
require much dark matter until quite large radii, whereas dim galaxies are dark
matter dominated down to nearly $R = 0$.  The mass discrepancy does not set
in at any particular length scale.
\label{R2}}

\figcaption[MLM.ps]{The observed dynamical mass to light ratio \MLo
(in $\mass\solar/L\solar$) plotted against a) luminosity, b)
central surface brightness, and c) scale length.
Error bars are plotted for a nominal inclination uncertainty of
$\sigma_i = 3\arcdeg$.  The very strong correlation in (b)
is related to the Tully-Fisher relation (Zwaan \etal 1995) and
can not be a selection effect.
\label{figMLM}}

\figcaption[TFSBresid.ps]{The residuals about the Tully-Fisher relation
as a function of surface brightness.  The residual in luminosity is shown
in (a), while that in velocity is shown in (b).  The triangles
are data where only the linewidth $W_{20}^c$ has been measured while
the circles are galaxies with $V_c$ measured from a full rotation curve.
Also shown as solid lines are the prior expectations for the relation between
rotation luminosity or velocity and surface brightness for a fixed mass to
light ratio.  More diffuse, lower surface brightness galaxies require
a higher mass and luminosity to achieve the same rotation velocity, or should
peak at much lower velocities at the same mass.  Though these statement
follow directly from $V^2 = G\mass/R$ and the assumption of constant \MLo,
neither are true --- galaxies rigorously adhere to the Tully-Fisher relation
(dashed lines) rather than following the expected trend (solid lines).
\label{TFresid}}

\figcaption[fSB.ps]{The fraction of baryonic mass in the form of stars $f_* =
\mass_*/\mass_b$ plotted against \lincsb\ (see McGaugh \& de Blok 1996).
There is a relation:  $f_* \propto \lincsb^{0.17}$, but this is not
sufficient to explain the \MLo-\lincsb relation.
\label{figFstar}}

\figcaption[MLMwogas.ps]{The \MLo-\csb\ relation, with and without gas.
The solid triangles include all mass within $R_o = 4h$ (identical to Fig.~4b),
while the open squares are the same quantity with the gas mass subtracted
($\mass_o^c = R_o V_c^2 G^{-1} - 1.4 \mass_{HI}$; the factor 1.4 is the standard
correction for helium and metals.)  The downward apex of each triangle
points at the corresponding square for the same galaxy.  Correcting for
$f_*$ has no impact on the \MLo-\csb\ relation other than a very small
downwards shift.
\label{figMLMcorr}}

\figcaption[spin.ps]{The surface brightness distribution (data points from
various sources) together with the distribution expected from the variation
of spin parameters.  Solid line: Efstathiou \& Jones (1979).  Dashed line:
Eisenstein \& Loeb (1995).  Theory predicts a very broad distribution
with curvature inconsistent with observations.  Worse, a cut-off must
be inserted by hand to reconcile the high surface brightness end of
$\phi(\csb)$.
\label{spin}}

\figcaption[ToyGal.ps]{Toy galaxy model rotation curves with $V$ and $R$
in normalized physical units.
Each panel represents the expectations of the simple models described
in the text.  The dashed line is the contribution of the halo and the
dotted line the contribution of the disk to the total rotation curve
(solid line).  Inset in each panel is a schematic representation of the
model.  Top panels are a fiducial HSB galaxy of high surface density
residing in a high density halo.  Lower panels are the expectations
for LSB galaxies in two scenarios.  On the left is the ``density begets
density'' (DD) hypothesis with a diffuse LSB galaxy residing in a diffuse
halo.  On the right is the ``Same Halo'' (SH) case with a diffuse LSB
galaxy extending out to near the edge of a dense halo.  The expected shapes
of the rotation curves are strikingly different:  V(R) rises slowly and
does not reach the same $V_c$ as an HSB of the same mass in the DD case,
while in the SH case this happens after a rapid rise.
\label{Toygal}}

\figcaption[reality.ps]{The model LSB galaxy rotation curves from Fig.~9
confronted with data for galaxies of the appropriate surface brightness.
Both models fail to predict the shapes of the actual rotation curves.
These rise gradually as expected in the DD model, but reach a $V_c$
dictated by the Tully-Fisher relation as expected in the SH model.
\label{cruelreality}}

\figcaption[DalcJ.ps]{The relative luminosity density of disk galaxies
$J(\csb)$ as a function of surface brightness, as
predicted by Dalcanton \etal (1997; line) and as
observed (data as per Fig.~8; see McGaugh 1996).
\label{DalcJ}}

\figcaption[RORA.ps,RORB.ps]{The rate of rise of disk galaxy rotation curves
as measured by the radius $R_{34}$ where $V(R_{34}) = \threequarters V_c$.
(a) $R_{34}/h$
\vs absolute magnitude and (b) $R_{34}/h$ \vs disk central surface brightness.
Only well resolved galaxies (at least 8 beams/diameter) are plotted.
Bright galaxies have rapidly rising rotation curves, with $V(R)$ frequently
reaching $\threequarters V_c$ before one scale length.  Dimmer galaxies
have rotation curves which rise more gradually, sometimes not reaching
$\threequarters V_c$ for
three scale lengths.  Also plotted are the predictions of the model of
Dalcanton \etal (1997) for no luminosity-surface brightness correlation
(solid line) and for $L \propto \lincsb^{1/3}$ as they predict
(dashed line).  These predictions are obtained from their Fig.~1
and depend on two parameters, the halo mass $\mass_H$ and the spin
parameter $\lambda_s$.  Points along the lines are labeled by
$\log(\mass_H)$ (in solar masses) in (a) and by $\lambda_s$ in (b).
\label{ROR}}

\figcaption[CenOsA.ps,CenOsB.ps]{The predictions of Cen \& Ostriker (1993)
confronted with the data.  The halo mass (a) and the cold gas mass (b)
are shown as a function of stellar mass.  The solid line represents the mean
prediction while the dashed line shows the expected 1$\sigma$ dispersion.
The squares in (a) are the data discussed here.  The circles in (b) are
data for spiral galaxies from McGaugh \& de Blok (1997) while the triangles
are data for elliptical galaxies from Wiklind \etal (1995).
Cen \& Ostriker (1993) predict far too strong a deviation from constant
$\mass_T/\mass_*$ in (a) (the dotted line is drawn for $f_b = 0.05$).
For the cold gas mass in (b), the prediction fares even worse.  Too little
cold gas is retained in galaxies at the present epoch, the slope of the
predicted trend is orthogonal to the data, and the gas content of bright
galaxies is off by five orders of magnitude.
\label{CenOsfig}}

\figcaption[Navfb.ps,NavV80.ps,Navfit.ps]{The rotation curve of F583--1
compared with the form predicted for appropriate CDM halos
(Navarro \etal 1996).  The solid points are the observed rotation curve,
while the open points are the rotation curve of the dark matter with the
baryonic component subtracted from the total.  The open points have been
offset slightly in $R$ for clarity, as the baryons contribute very little
at any radii.  Lines show the predictions of various cosmologies (Table~4)
assuming (a) $f_b = 0.09$, the value indicated by clusters, and (b)
$V_{200} = 80\kms$ (\ie a Tully-Fisher normalization which ignores the
consequences for the baryon fraction.)  None of the predictions are
satisfactory, so in (c) we test whether any Navarro \etal (1996) profile
can fit the observations, regardless of cosmology.  Even treating both
$c$ and $V_{200}$ as completely free parameters, no fit can be obtained.
CDM predicts the wrong shape for galaxy halo density profiles.
\label{Nav}}

\figcaption[F5831.gasscale.ps]{The best fit we could attain for
the rotation curve of a well resolved LSB galaxy, F583--1, by
scaling the contribution of the \HI\ component.  The stars are
assumed to contribute maximally, but this matters little to the
fit.  Clearly, there is no scaling which reproduces the shape
of the observed rotation curve.  Though the \HI\ distribution
may be a good indicator of the shape of the rotation curves in some
HSB galaxies, it generally is not in LSB galaxies.  This nullifies one
of the arguments in favor of baryonic dark matter.
\label{gasscale}}

\figcaption[MdMbRh.ps]{A graphical representation of the limits from
Table~5.  One would expect a universal baryon fraction, but no such
value emerges unless only a small subset of the data are considered
(\eg clusters of galaxies).  When all data are considered,
one finds many contradictory measurements and nonoverlapping limits.
Note that even within the same type of object, contradictory limits
occur.  In disk galaxies, flat rotation curves require a lot of dark
matter, while disk stability prefers a more moderate amount.  For
groups of galaxies, the Local Group requires much more dark matter
than is measured in external groups.  A broad distribution of baryon fractions
seems to be required, but this contradicts the small scatter in the
Tully-Fisher relation.
\label{MdMbRh}}

\clearpage
\begin{figure}
\plotone{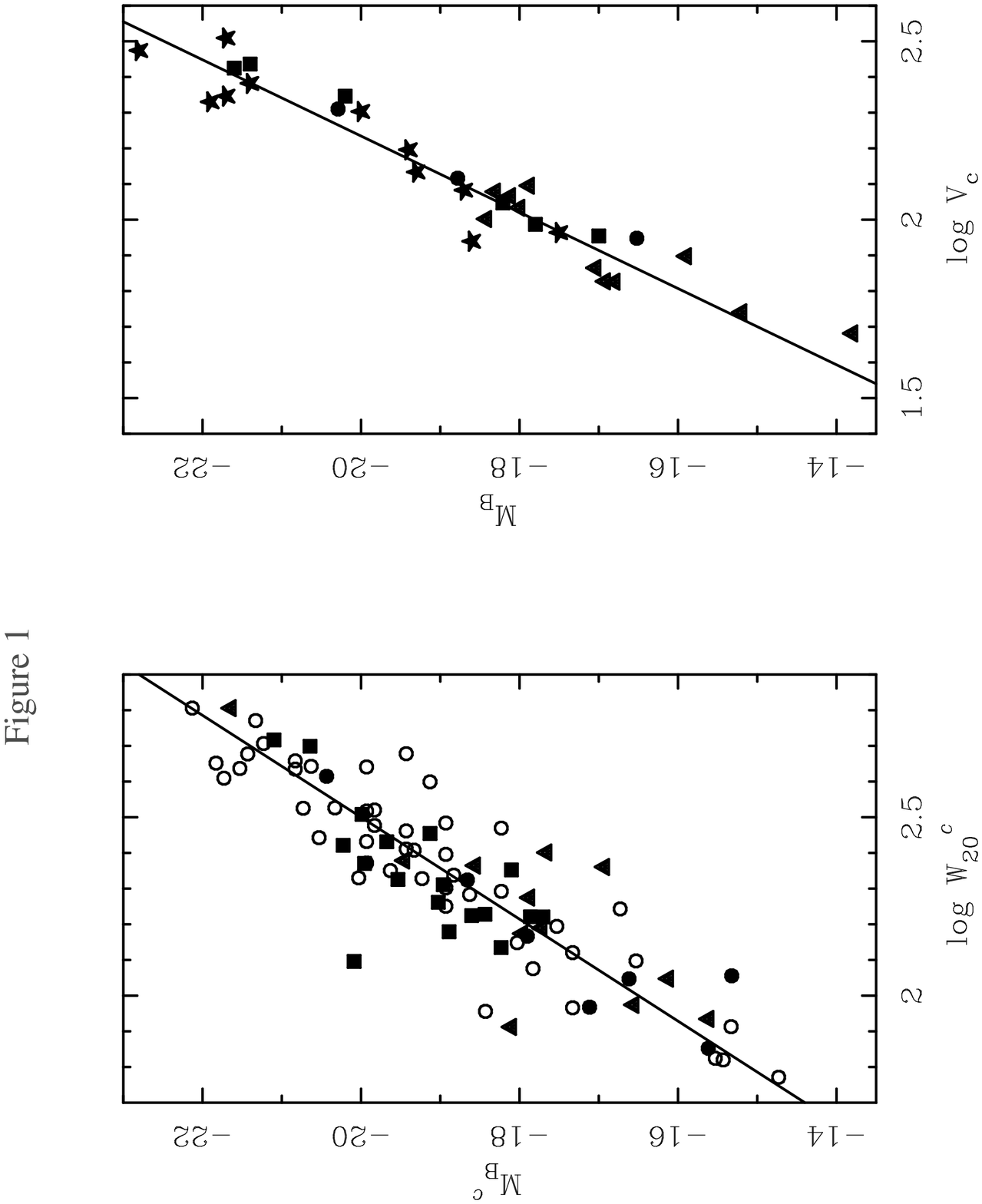}
\end{figure}

\clearpage
\begin{figure}
\plotone{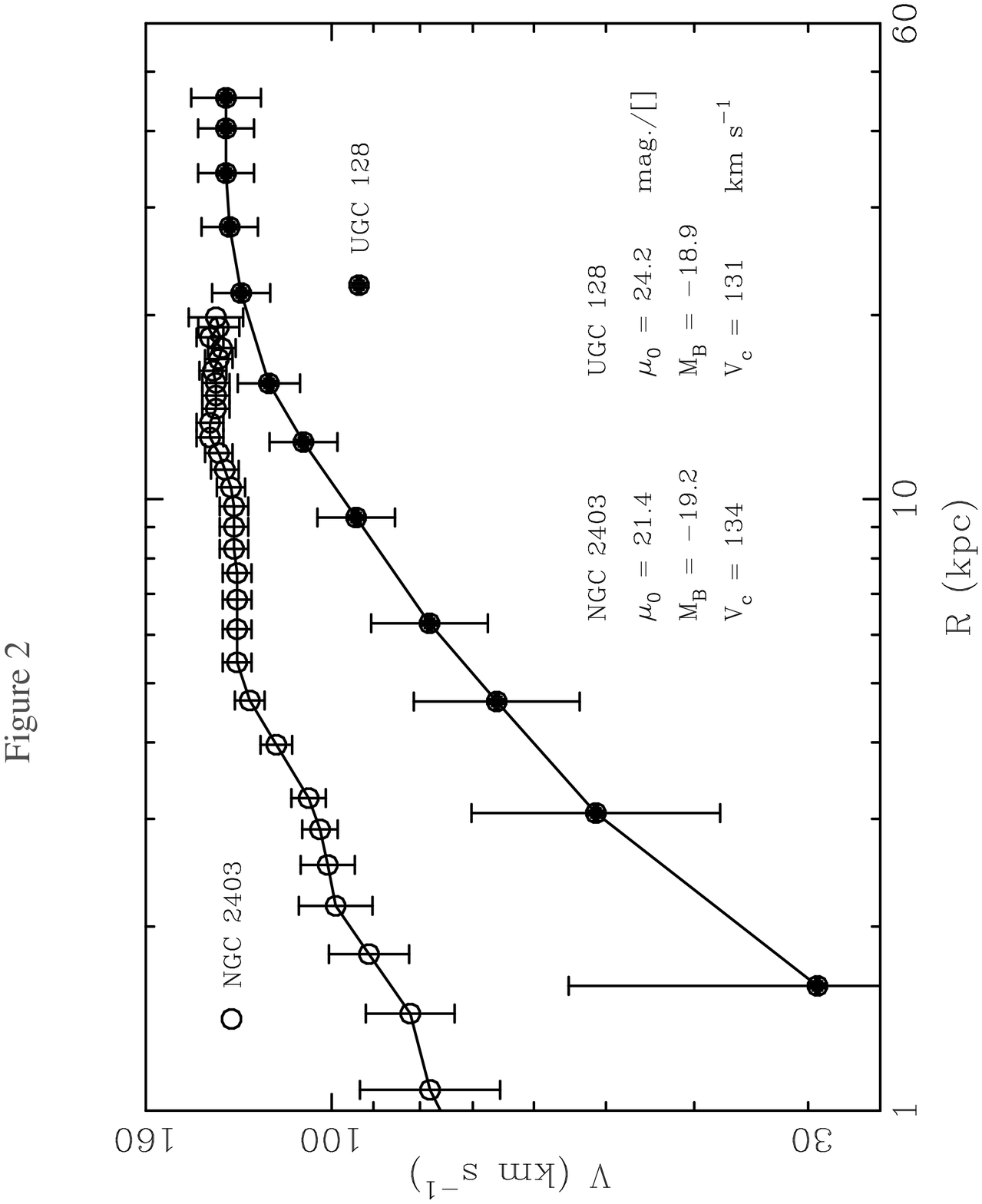}
\end{figure}

\clearpage
\begin{figure}
\plotone{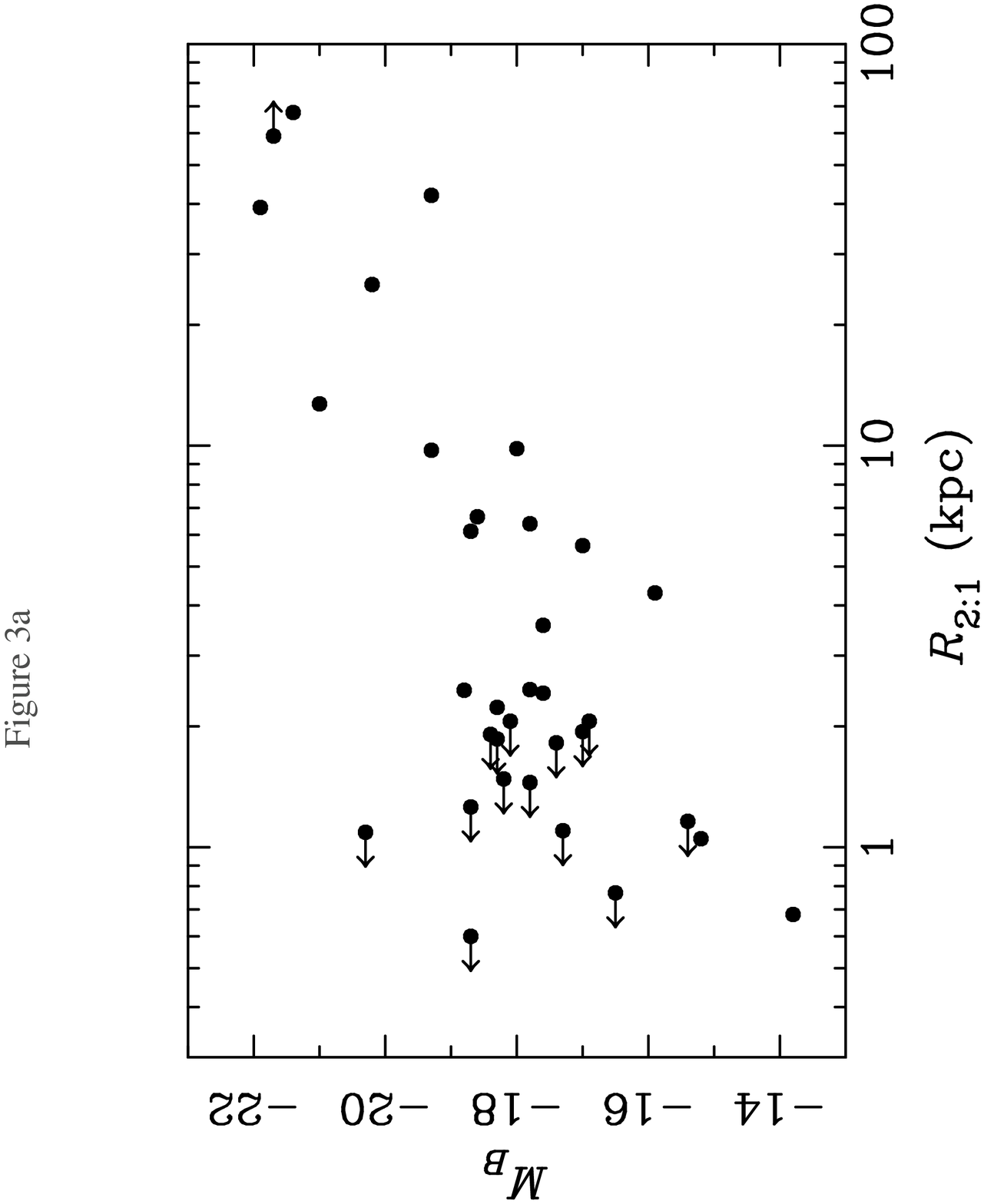}
\end{figure}

\clearpage
\begin{figure}
\plotone{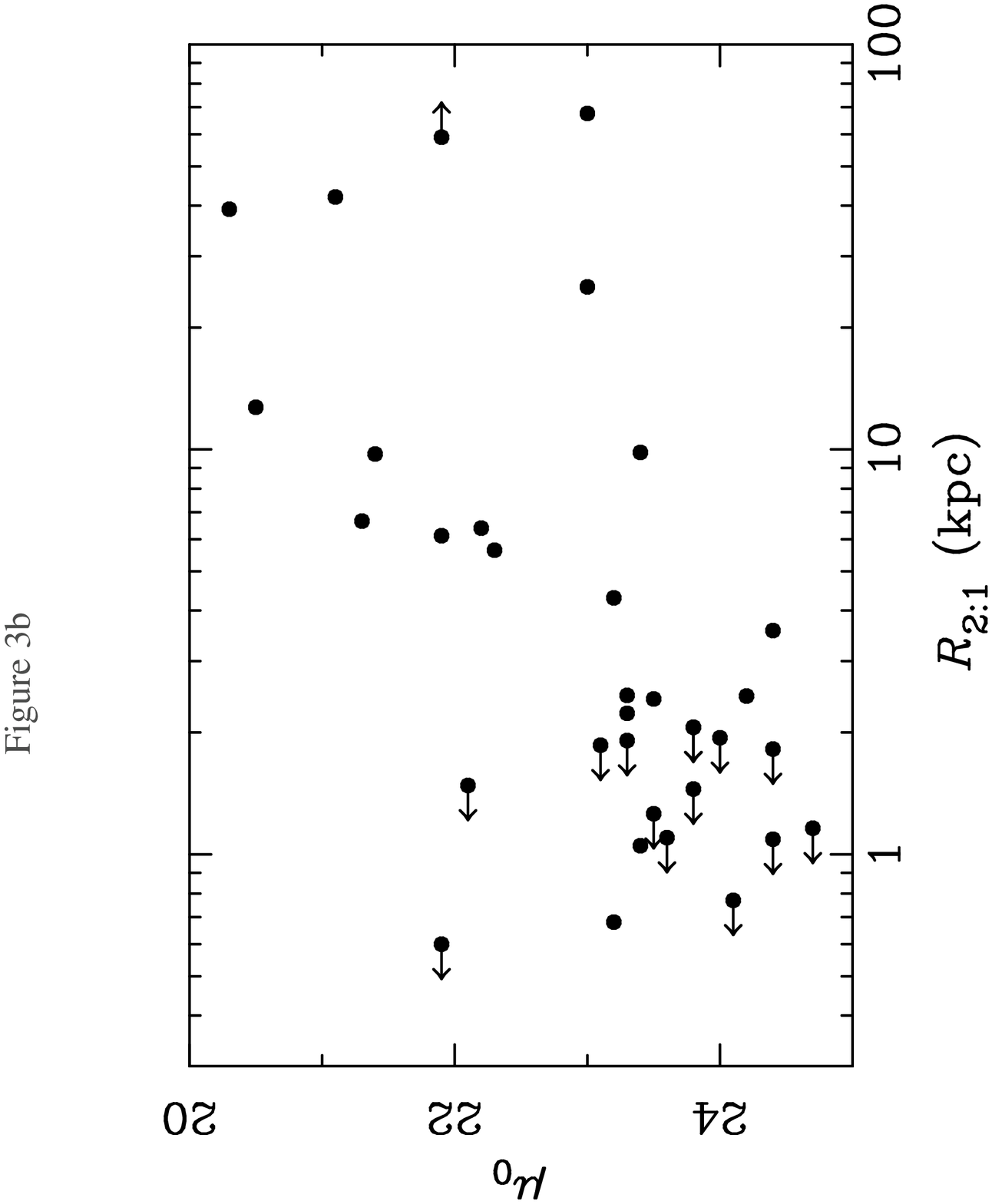}
\end{figure}

\clearpage
\begin{figure}
\plotone{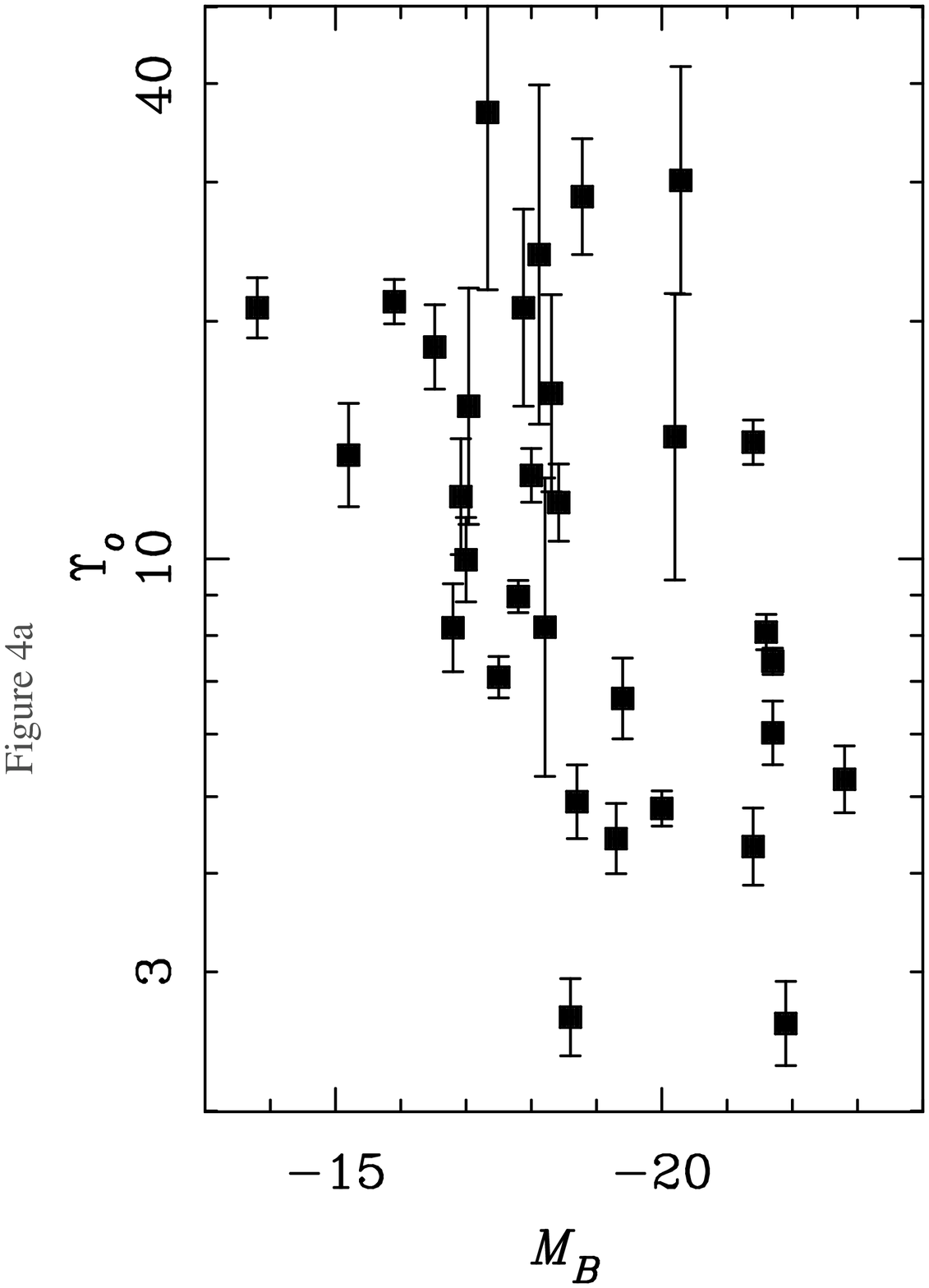}
\end{figure}

\clearpage
\begin{figure}
\plotone{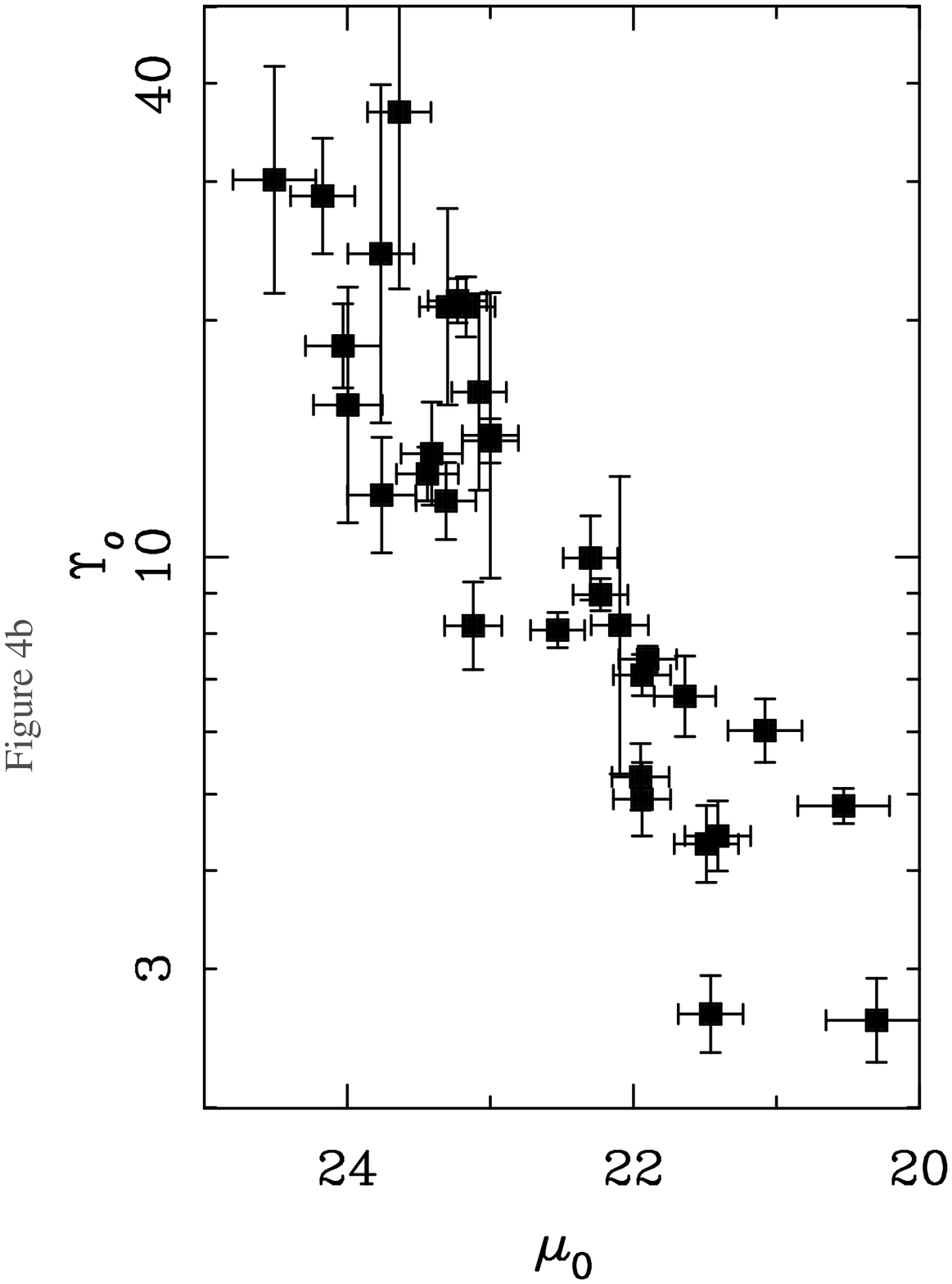}
\end{figure}

\clearpage
\begin{figure}
\plotone{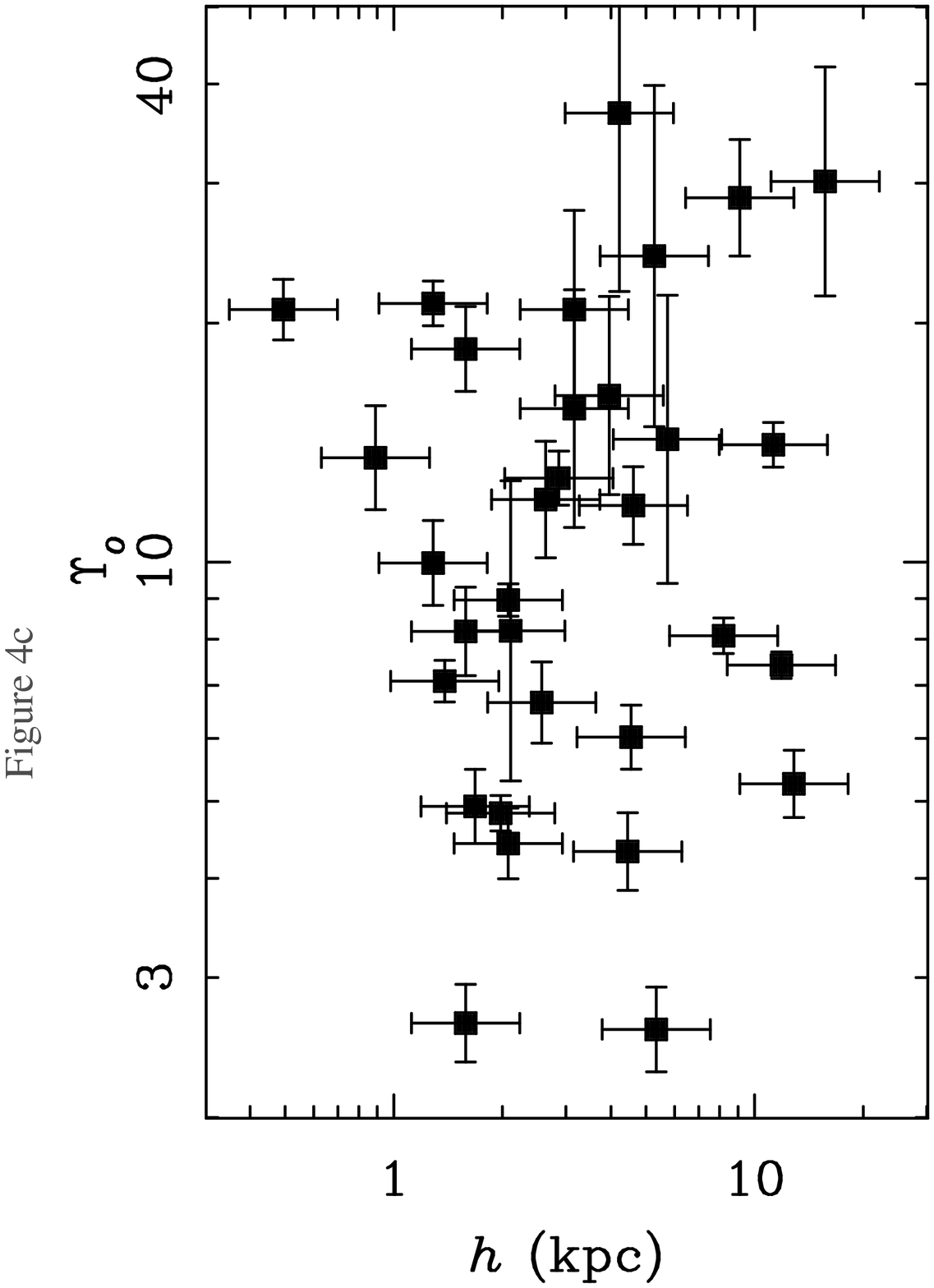}
\end{figure}

\clearpage
\begin{figure}
\plotone{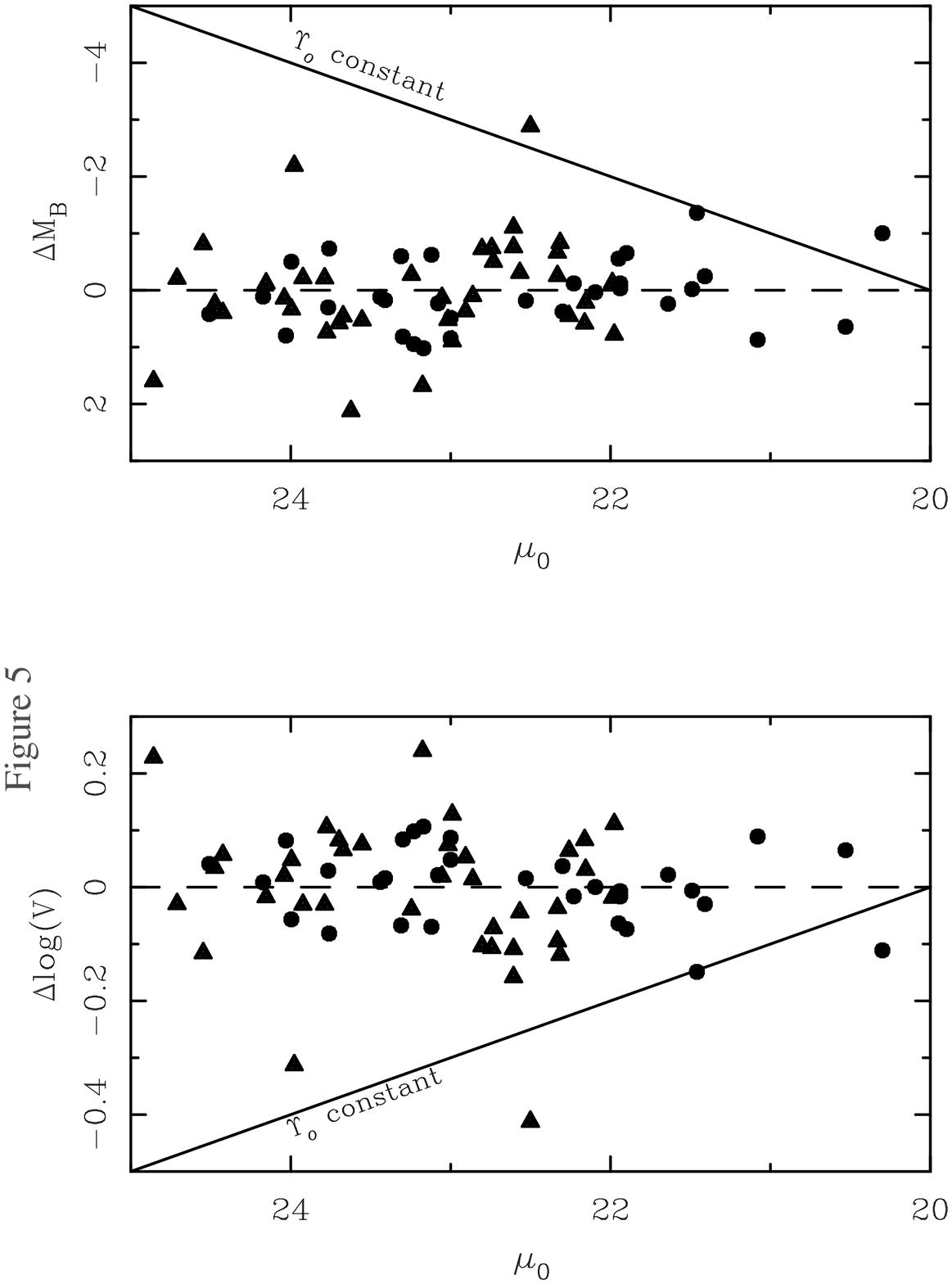}
\end{figure}

\clearpage
\begin{figure}
\plotone{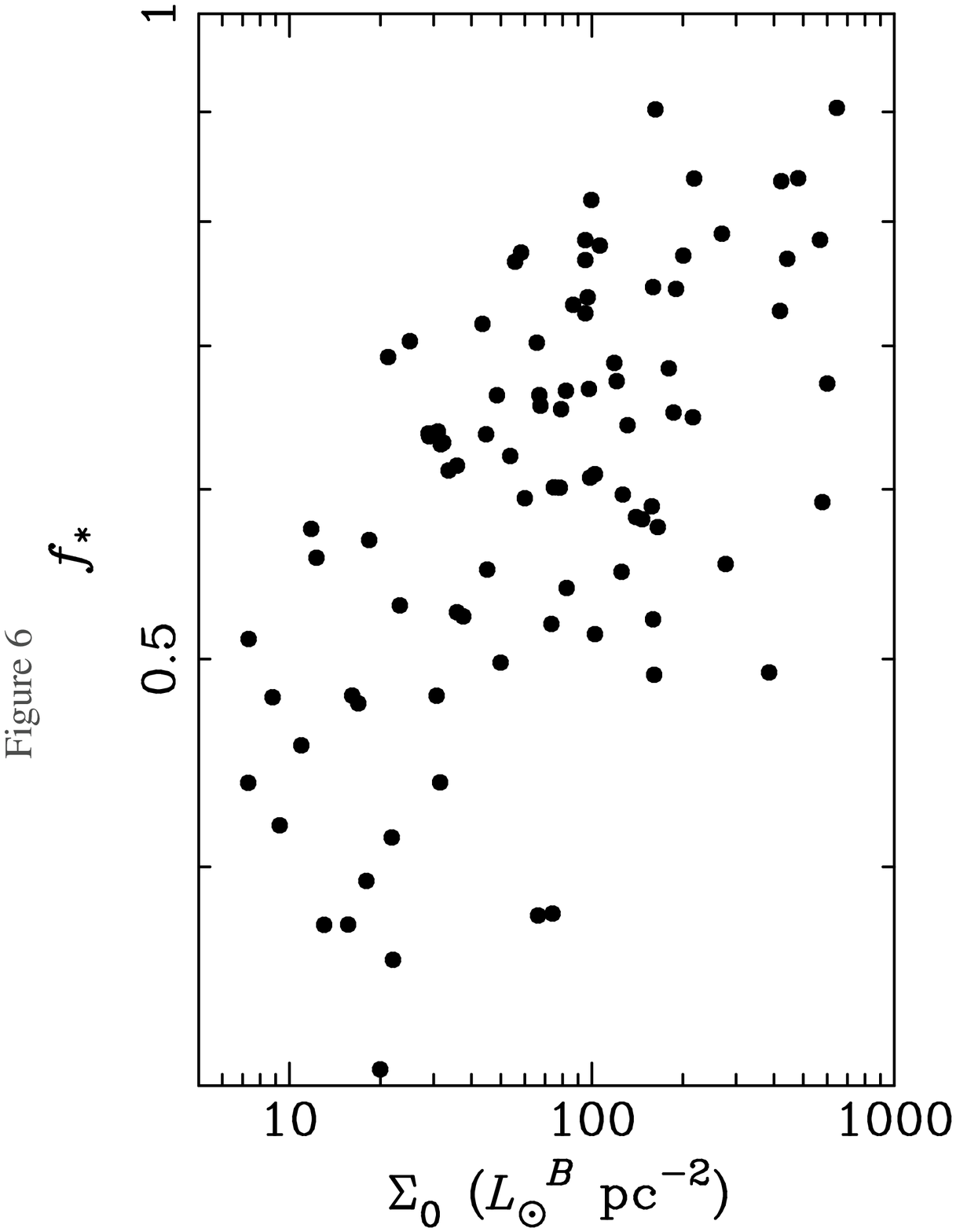}
\end{figure}

\clearpage
\begin{figure}
\plotone{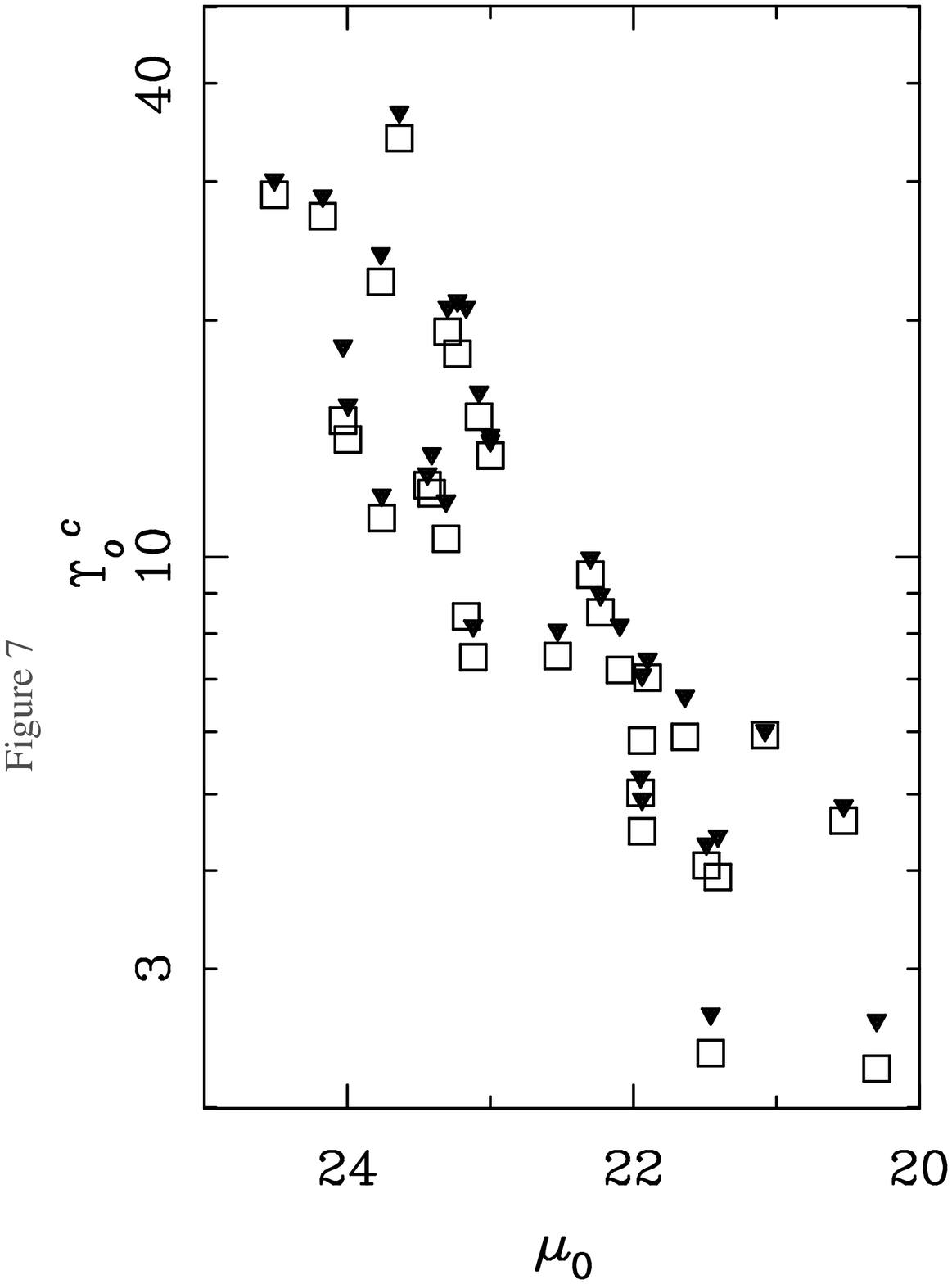}
\end{figure}

\clearpage
\begin{figure}
\plotone{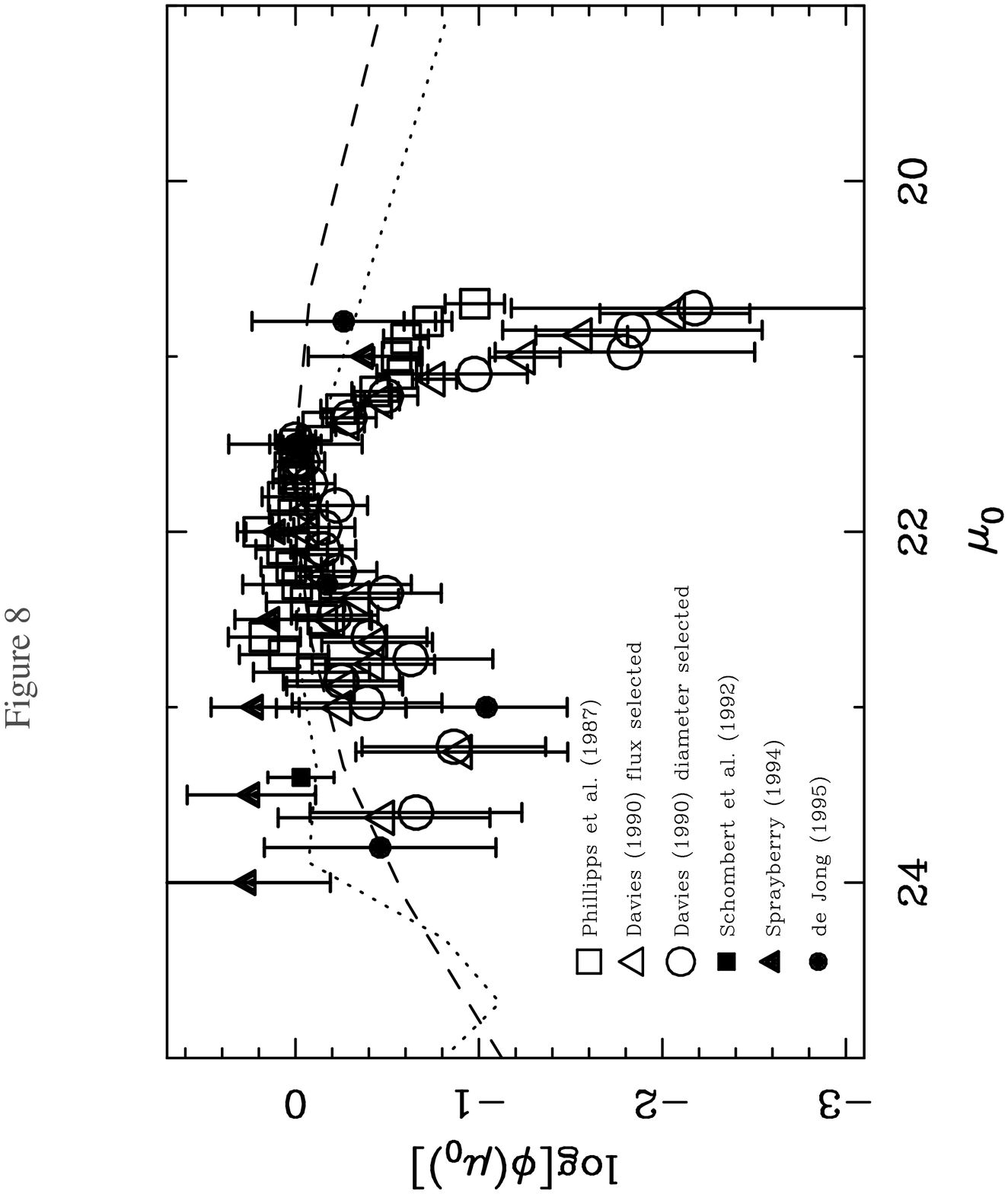}
\end{figure}

\clearpage
\begin{figure}
\plotone{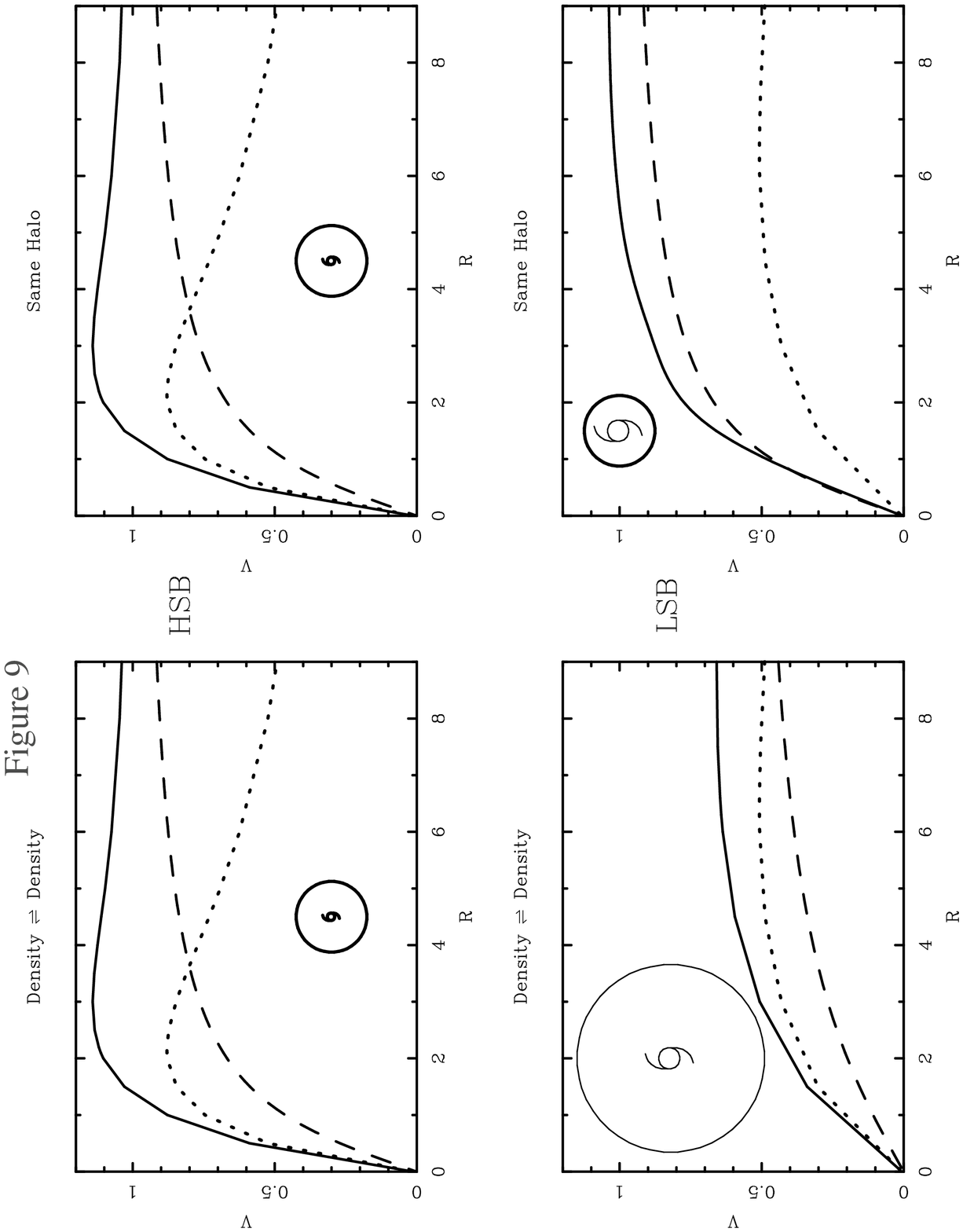}
\end{figure}

\clearpage
\begin{figure}
\plotone{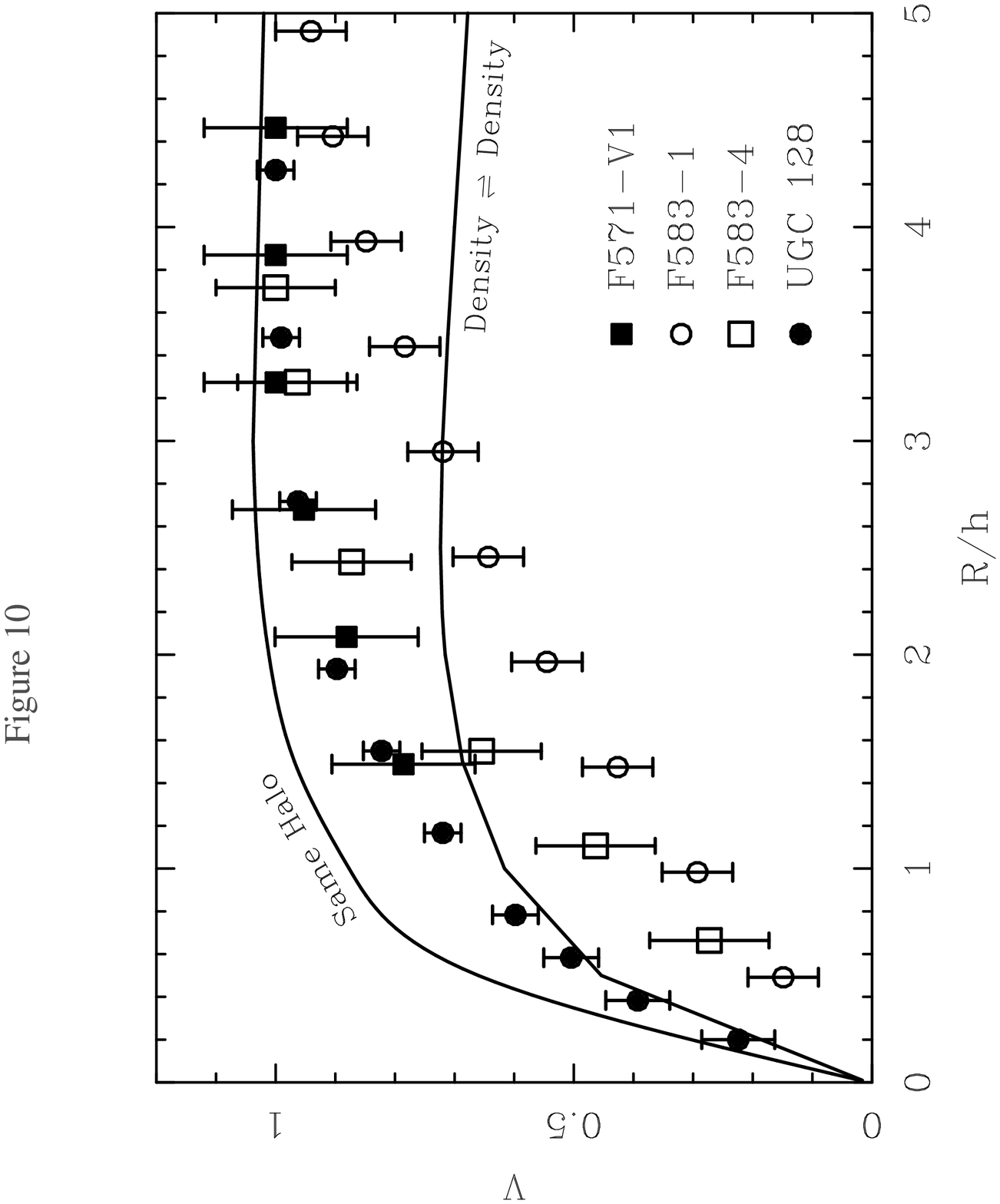}
\end{figure}

\clearpage
\begin{figure}
\plotone{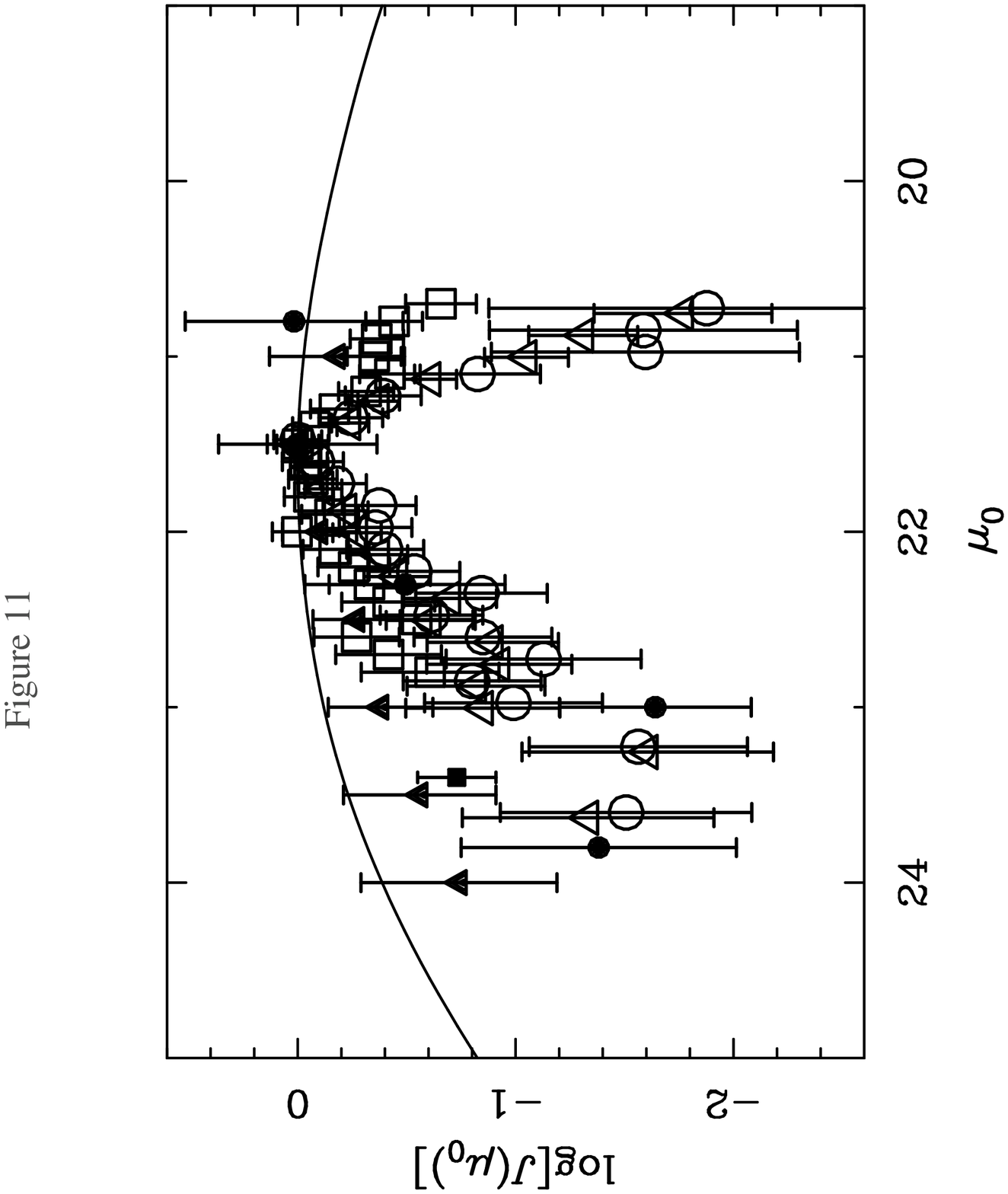}
\end{figure}

\clearpage
\begin{figure}
\plotone{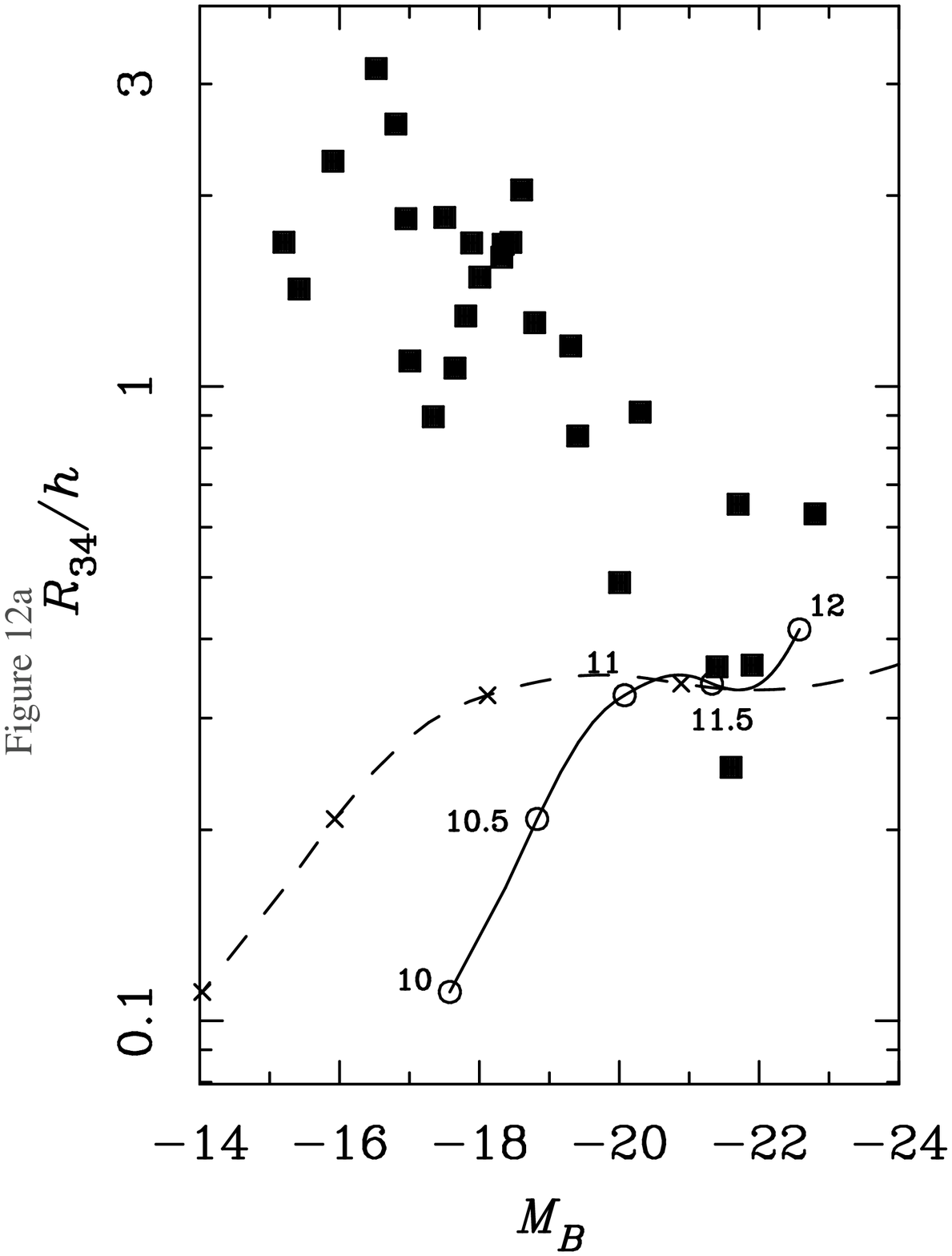}
\end{figure}

\clearpage
\begin{figure}
\plotone{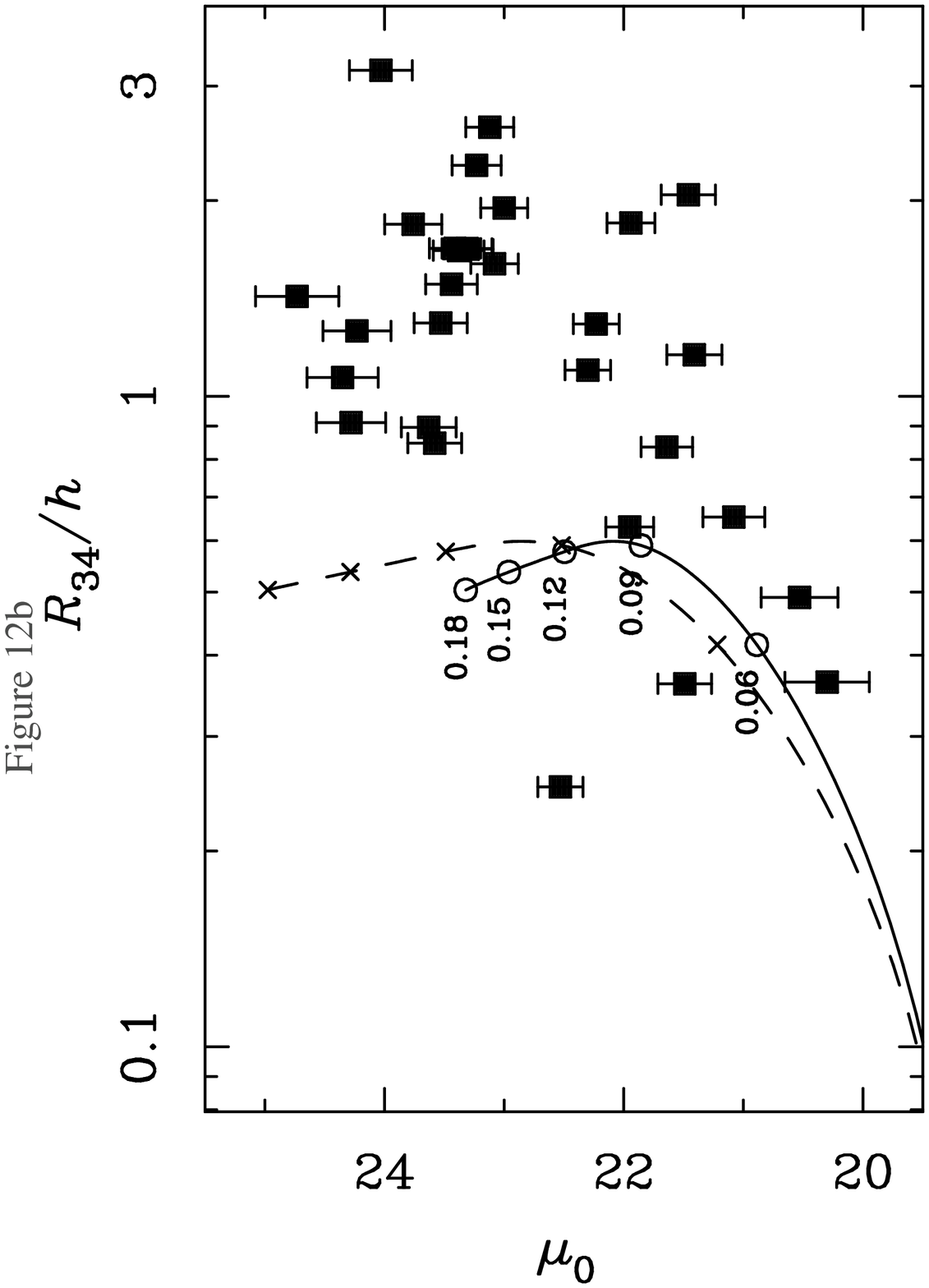}
\end{figure}

\clearpage
\begin{figure}
\plotone{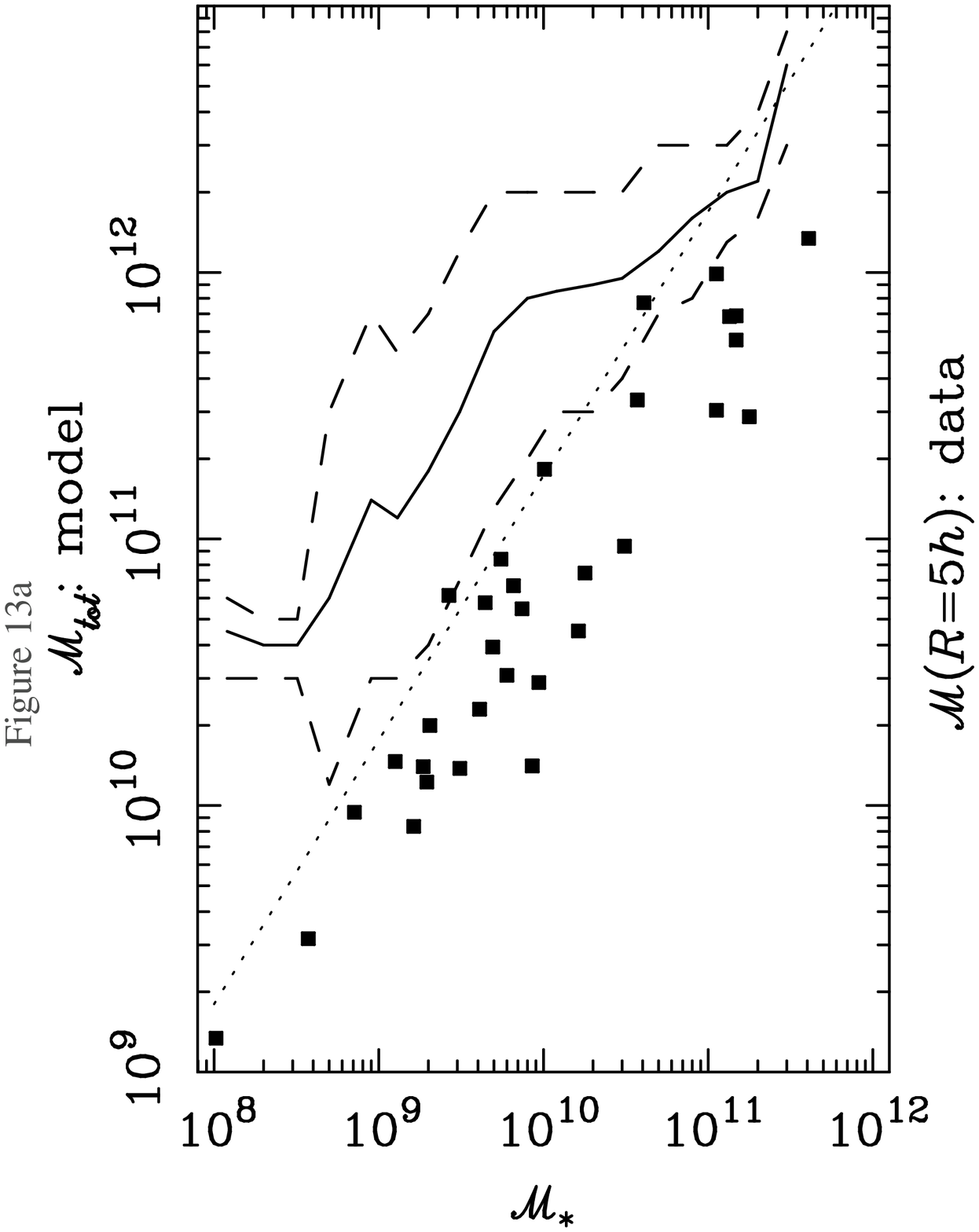}
\end{figure}

\clearpage
\begin{figure}
\plotone{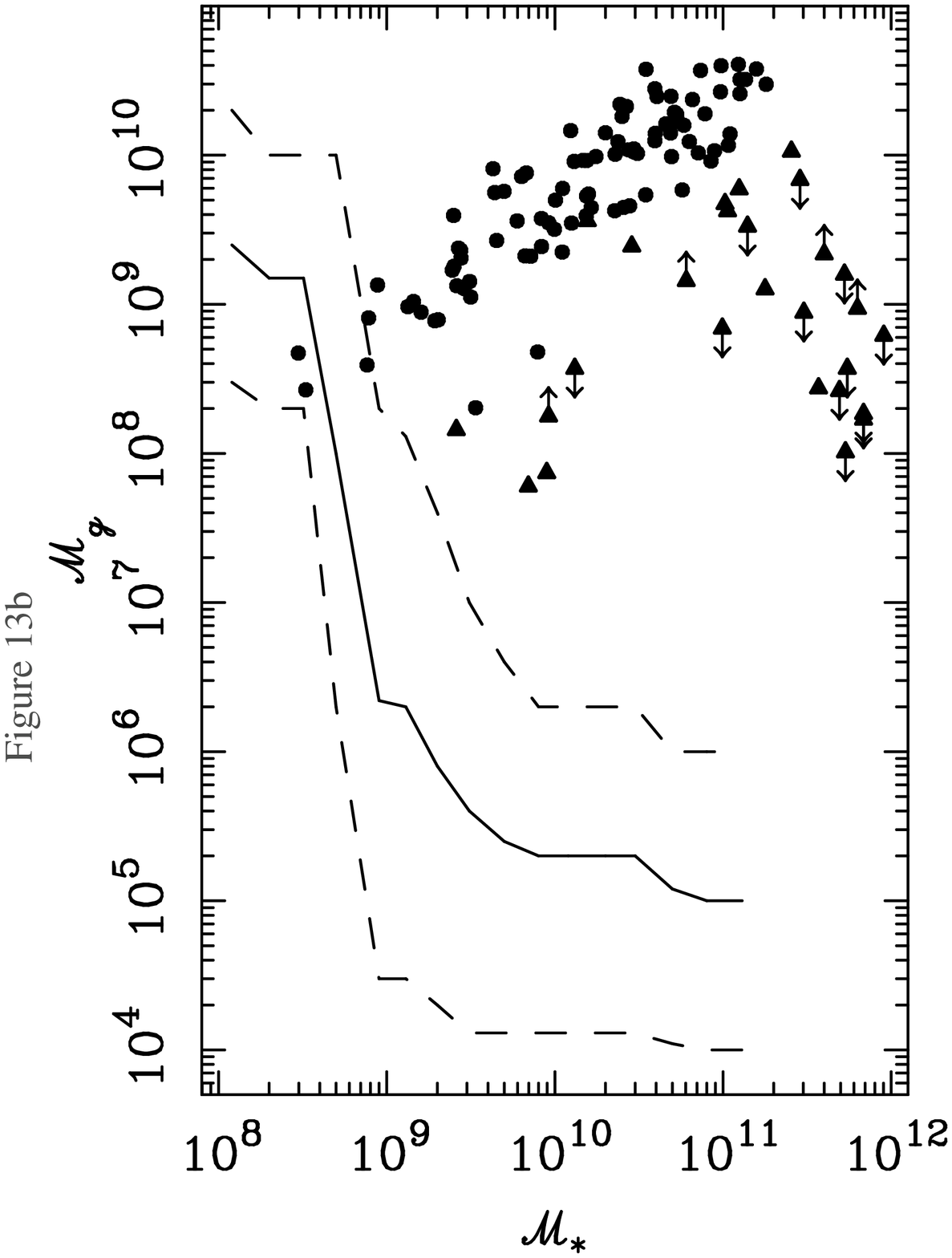}
\end{figure}

\clearpage
\begin{figure}
\plotone{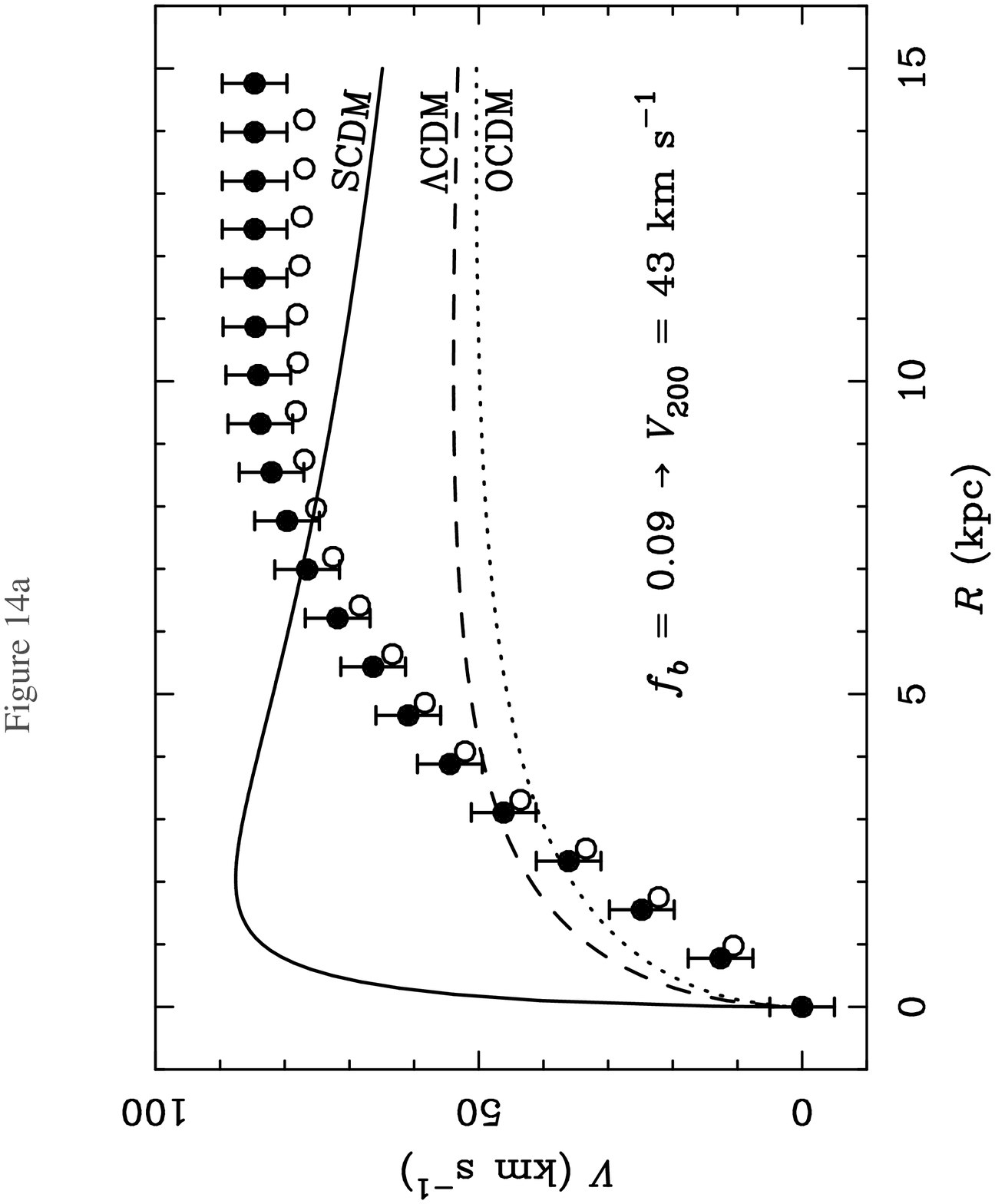}
\end{figure}

\clearpage
\begin{figure}
\plotone{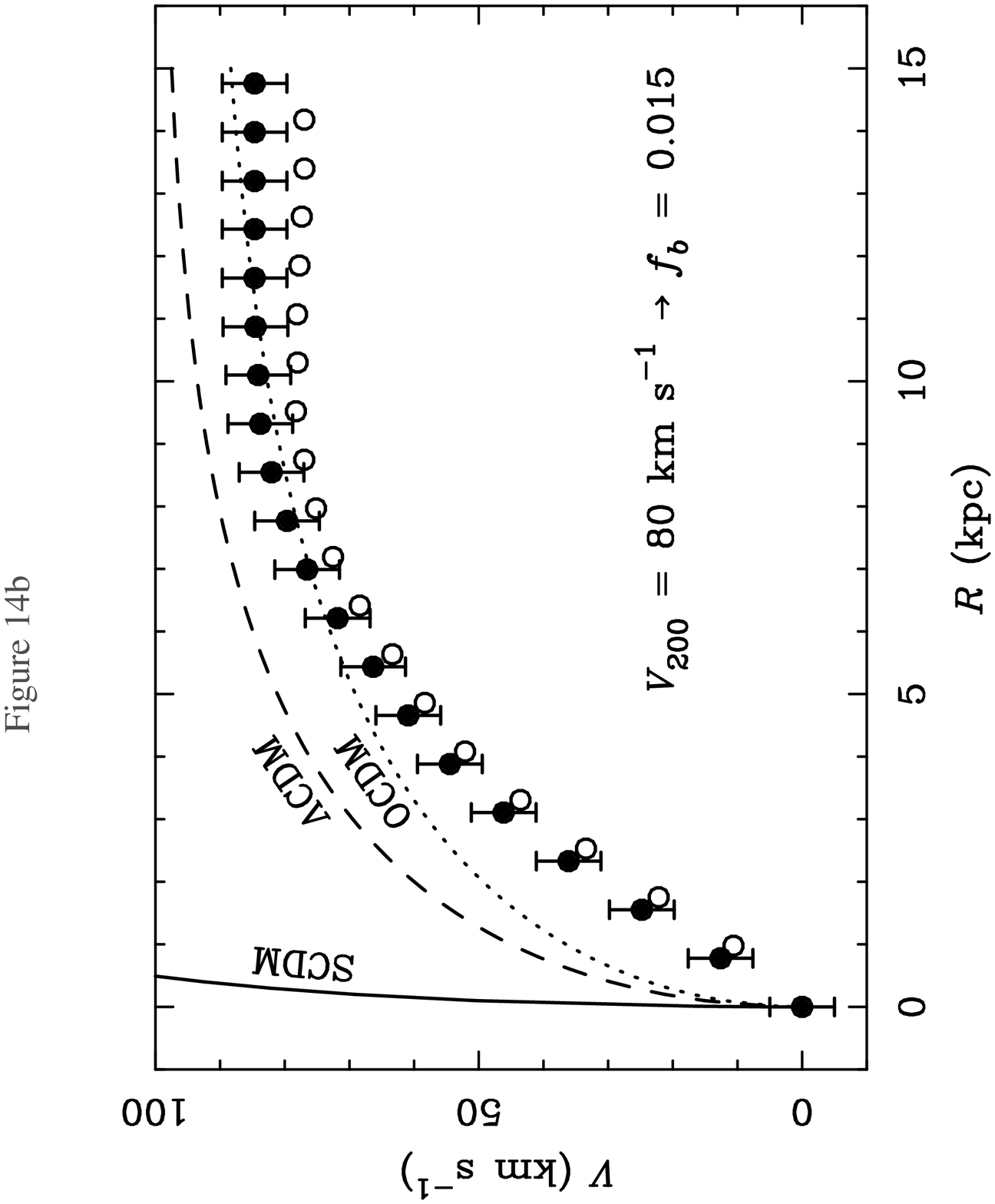}
\end{figure}

\clearpage
\begin{figure}
\plotone{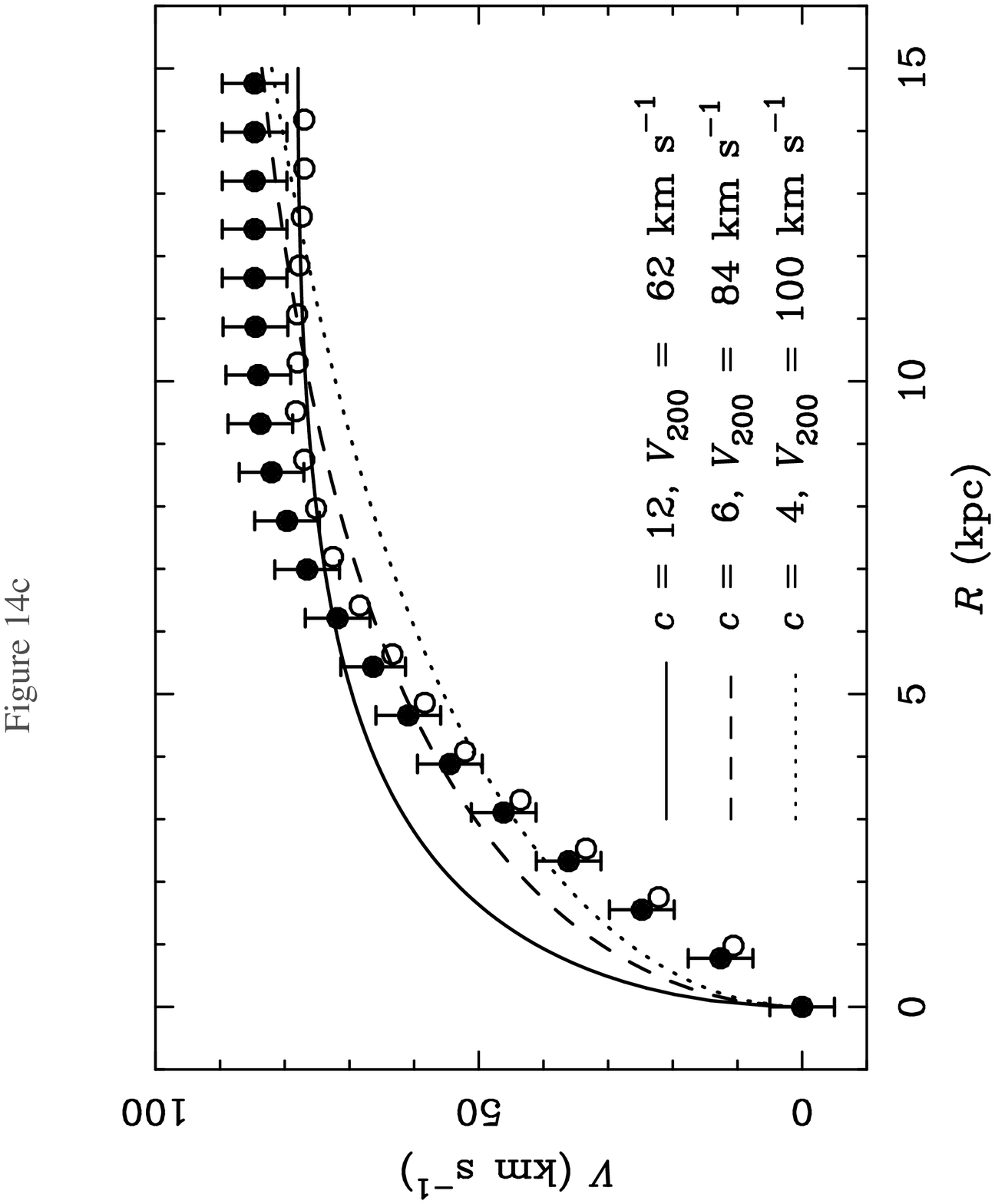}
\end{figure}

\clearpage
\begin{figure}
\plotone{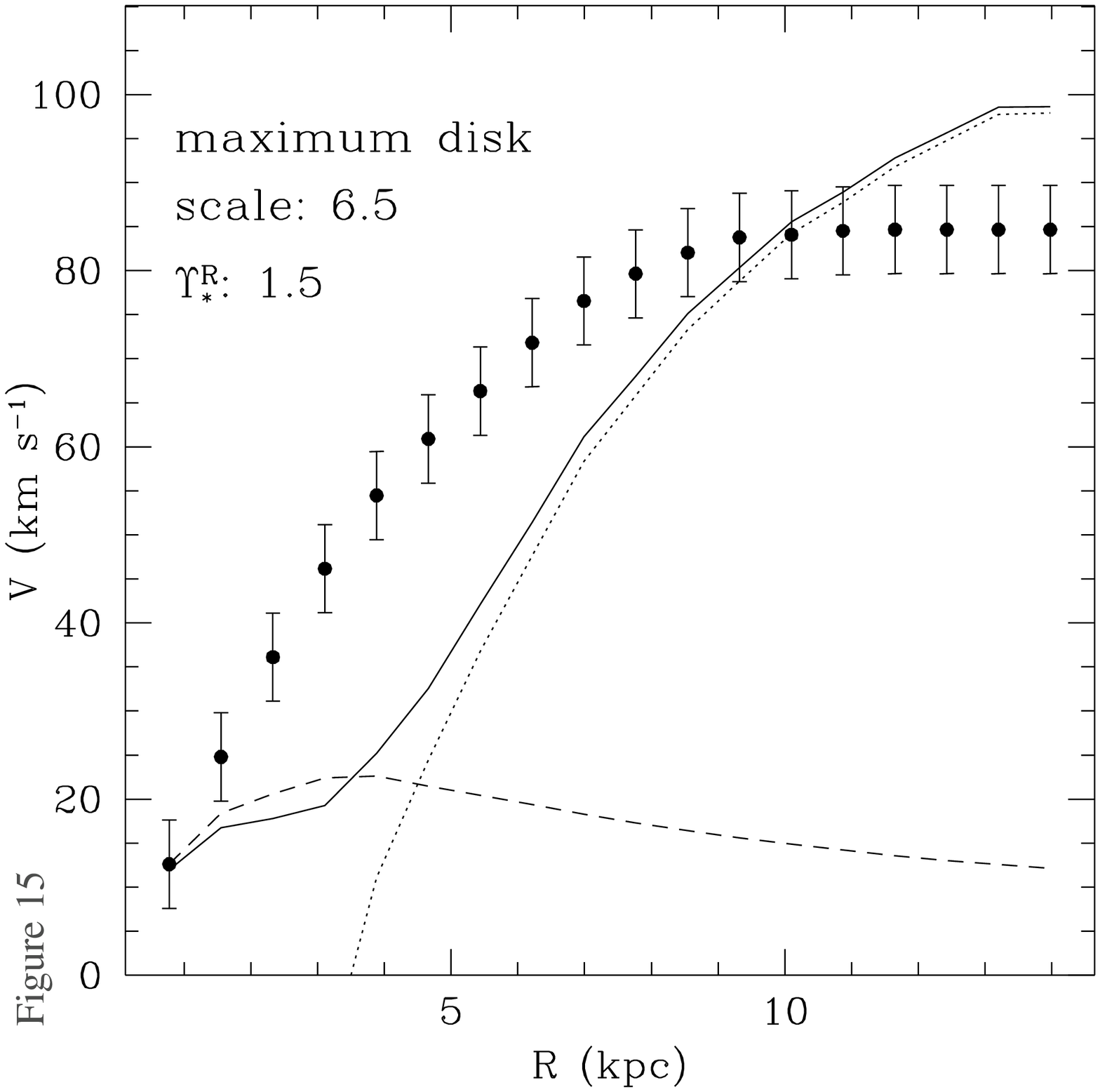}
\end{figure}

\clearpage
\begin{figure}
\plotone{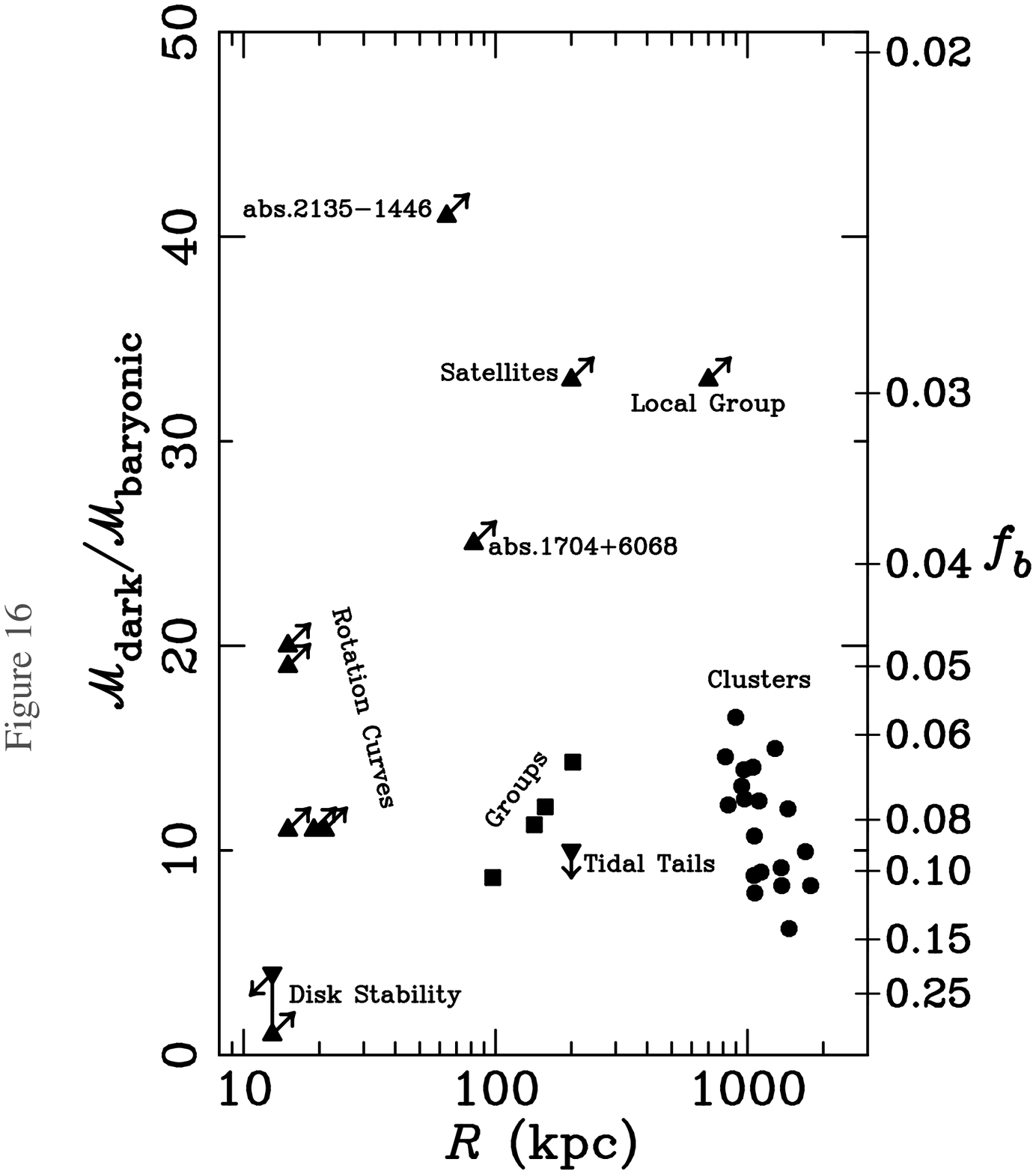}
\end{figure}


\clearpage

\begin{references}

\reference{oldtf} Aaronson, M., Huchra, J., \& Mould, J. 1979, \apj, 229, 1

\reference{macho} Alcock, C. \etal 1996, \apj, 463, 67

\reference{AvdK} Andredakis, Y. C., \& van der Kruit, P. C. 1992, \aap,
	265, 396

\reference{EROS1} Ansari, R. \etal 1996, \aap, 314, 94

\reference{ABP} Athanassoula, E., Bosma, A., \& Papaioannou, S. 1987,
        \aap, 179, 23

\reference{BFG} Bahcall, J. N., Flynn, C., \& Gould, A. 1992, \apj, 389, 234

\reference{BLW} Barcons, X., Lanzetta, K. M., \& Webb, J. K. 1995, \nat,
        376, 321

\reference{BE} Barnes, J. \& Efstathiou, G. 1987, \apj, 319, 575

\reference{BBS} Begeman, K. G., Broeils, A. H. \& Sanders, R. H. 1991,
        \mnras, 249, 523

\reference{COBE} Bennett, C. L. \etal 1994, \apj, 436, 423

\reference{Bosma} Bosma, A. 1981, \aj, 86, 1825

\reference{malin1} Bothun, G.~D., Impey, C.~D., Malin, D.~F., \& Mould, J.~R.
      1987, \aj, 94, 23

\reference{small} Bothun, G.~D., Schombert, J.~M., Impey, C.~D.,
        Sprayberry, D., \& McGaugh, S.~S. 1993, \aj, 106, 530

\reference{Bott1} Bottema, R. 1993, \aap, 275, 16

\reference{Bott2} Bottema, R. 1997, \aap, 328, 517

\reference{broeils} Broeils, A. H. 1992, Ph.D. thesis, University of Groningen

\reference{sharc} Burke, D. J., Collins, C. A., Sharples, R. M.,
	Romer, A. K., Holden, B. P., \& Nichol, R. C. 1997, \apj, 488, 83

\reference{BRTF} Burstein, D., Rubin, V. C., Thonnard, N., \& Ford, W. K.
        1982, \apj, 253, 70

\reference{Carr} Carr, B. 1994, \araa, 32, 531

\reference{CenOs} Cen, R., \& Ostriker, J. P. 1993, \apj, 417, 415

\reference{stab} Christodoulou, D. M. 1991, 372, 471

\reference{CL} Cole, S. \& Lacey, C. 1996, \mnras, 281, 716

\reference{Cour} Courteau, S. 1996, \apjs, 103, 363

\reference{nucleo} Copi, C., Schramm, D. N., \& Turner, M. S. 1995,
        Science, 267, 192

\reference{D94} Dalcanton, J. J., Canizares, C. R., Granados, A.,
        Steidel, C. C., \& Stocke, J. T. 1994, \apj, 424, 550

\reference{D95} Dalcanton, J. J., Spergel, D. N., \& Summers, F. J. 1995,
        astro-ph/9503093

\reference{D97} Dalcanton, J. J., Spergel, D. N., \& Summers, F. J. 1997,
        \apj, 482, 659

\reference{omega} Davis, M., Nusser, A., \& Willick, J. A. 1996, \apj, 473, 22

\reference{dBM1} de~Blok, W.~J.~G., \& McGaugh, S.~S. 1996, \apj, 469, L89

\reference{dBM2} de~Blok, W.~J.~G., \& McGaugh, S.~S. 1997, \mnras, 290, 533

\reference{pIII} de~Blok, W.~J.~G., \& McGaugh, S.~S. 1998, in preparation
	(paper III)

\reference{BMH} de~Blok, W.~J.~G., McGaugh, S.~S., \& van~der~Hulst, J.~M.
        1996, \mnras, 283, 18

\reference{BHB} de~Blok, W.~J.~G., van~der~Hulst, J.~M.,
        \& Bothun, G.~D. 1995, \mnras, 274, 235

\reference{dJ2} de Jong, R. S. 1996a, \aaps, 118, 557

\reference{dJ3} de Jong, R. S. 1996b, \aap, 313, 45

\reference{DGLHS} Donahue, M., Gioia, I., Luppino, G., Hughes, J. P.,
	\& Stocke, J. T. 1997, astro-ph/9707010

\reference{Dub} Dubinski, J. 1994, \apj, 431, 617

\reference{DC} Dubinski, J., \& Carlberg, R. G. 1991, \apj, 378, 496

\reference{tails} Dubinski, J., Mihos, J. C., \& Hernquist, L. 1996,
        \apj, 462, 576

\reference{EJ} Efstathiou, G., \& Jones, B. J. T. 1979, \mnras, 186, 133

\reference{EL} Eisenstein, D. J., \& Loeb, A. 1995, \apj, 439, 520

\reference{EMN} Evrard, A. E., Metzler, C. A., \& Navarro, J. F. 1996,
        \apj, 469, 494

\reference{omegab} Fields, B. D., Kainulainen, K., Olive, K. A., \&
        Thomas, D. 1996, New Astronomy, 1, 57

\reference{powerspec} Fisher, K. B., Davis, M., Strauss, M. A., Yahil, A.,
        \& Huchra, J. P. 1993, \apj, 402, 42

\reference{FP} Flores, R.~A., \& Primack, J.~R. 1994, \apj, 427, L1

\reference{FPBF} Flores, R.~A., Primack, J.~R., Blumenthal, G. R., \&
        Faber, S. M. 1993, \apj, 412, 443

\reference{F70} Freeman, K.~C. 1970, ApJ, 160, 811

\reference{carlos} Frenk, C. S., Baugh, C. M., Cole, S. 1996,
        in IAU Symposium No. 171: New Light on Galaxy Evolution, eds.
        Bender, R. \& Davies, R.~L. (Dordrecht:  Kluwer), 247

\reference{GMCSLQ} Governato, F., Moore, B., Cen, R., Stadel, J.,
        Lake, George, Quinn, T. 1997, New Astr., 2, 91

\reference{H93}Hernquist, L. 1993, \apjs, 86, 389

\reference{heyl} Heyl, J. S., Cole, Shaun, Frenk, Carlos S., \&
	Navarro, J. F. 1995, \mnras, 274, 755

\reference{HSFRWH} Hoffman, G. L., Salpeter, E. E., Farhat, B., Roos, T.,
        Williams, H. \& Helou, G. 1996, \apjs, 105, 269

\reference{HvA} Huizinga, J. E., \& van Albada, T. S. 1992, \mnras, 254, 677

\reference{KW} Kahn, F. D., \& Woljter, L. 1959, \apj, 130, 705

\reference{nick} Kaiser, N. 1986, \mnras, 222, 323

\reference{Katz} Katz, N., \& Gunn, J. E. 1991, \apj, 377, 365

\reference{K87} Kent, S. M. 1987, \aj, 93, 816

\reference{KG} Kuijken, K. \& Gilmore, G. 1989, \mnras, 239, 605

\reference{7sam} Lynden-Bell, D., Faber, S. M., Burstein, D. Davies, R. L.,
        Dressler, A., Terlevich, R. J., \& Wegner, G. 1988, \apj, 326, 19

\reference{lake} Lake, G., \& Feinswog, L. 1989, \aj, 98, 166

\reference{mythesis} McGaugh, S.~S. 1992, Ph.D. thesis, University of Michigan

\reference{myOH} McGaugh, S.~S. 1994, \apj, 426, 135

\reference{me} McGaugh, S.~S. 1996a, \mnras, 280, 337

\reference{IAU} McGaugh, S.~S. 1996b, in IAU Symposium No. 171:
        New Light on Galaxy Evolution, eds. Bender, R. \& Davies, R.~L. 
        (Dordrecht:  Kluwer), 97

\reference{MB} McGaugh, S.~S., \& Bothun, G.~D. 1994, \aj, 107, 530

\reference{MdB} McGaugh, S.~S., \& de Blok, W. J. G. 1997, \apj, 481, 689

\reference{pII} McGaugh, S.~S., \& de Blok, W. J. G. 1998, companion paper
	(paper II)

\reference{LSBmorph} McGaugh, S.~S., Schombert, J.~M. \& Bothun, G.~D. 1995,
        \aj, 109, 2019

\reference{Merritt} Merritt, D. 1997, \apj, 486, 102

\reference{Gerhardt} Meurer, G. R., Carignan, C., Beaulieu, S. F., \&
        Freeman, K. C. 1996, \aj, 111, 1551

\reference{tails2} Mihos, J. C.,  Dubinski, J., \& Hernquist, L. 1997,
        astro-ph/9708009

\reference{MMB} Mihos, J. C., McGaugh, S. S., \& de Blok, W. J. G. 1997,
        \apj, 477, L79

\reference{Mo} Mo, H.~J., McGaugh, S.~S., \& Bothun, G.~D. 1994, \mnras, 267,
        129

\reference{Moore} Moore, B. 1994, \nat, 370, 629

\reference{MABHHRS} Mould, J. R., Akeson, R. L., Bothun, G. D., Han, M.,
        Huchra, J. P., Roth, J., \& Schommer, R. A. 1993, \apj, 409, 14

\reference{sesto} Navarro, J. F. 1996a, astro-ph/9610188

\reference{NavIAU} Navarro, J. F. 1996b, in IAU Symposium No. 171:
        New Light on Galaxy Evolution, eds. Bender, R. \& Davies, R.~L.
        (Dordrecht:  Kluwer), 255

\reference{Eke} Navarro, J. F., Eke, V. R., \& Frenk, C. S. 1996,
	\mnras, 283, L72

\reference{NFW} Navarro, J. F., Frenk, C. S., \& White, S. D. M. 1996,
        \apj, 462, 563

\reference{NSz} Navarro, J. F., \& Steinmetz, M. 1997, \apj, 478, 13

\reference{OS} Ostriker, J. P. \& Steinhardt, P. J. 1995, \nat, 377, 600

\reference{pacman} Paczynski, B. 1996, \araa, 34, 419

\reference{firstlambda} Peebles, P.~J.~E. 1971, \aap, 11, 377

\reference{LGtime} Peebles, P.~J.~E., Melott, A. L., Holmes, M. R., \&
        Jiang, L. R. 1989, \apj, 345, 108

\reference{PBE} Pildis, R. A., Bregman, J. N., \& Evrard, A. E. 1995, \apj,
        443, 514

\reference{PSE} Pildis, R. A., Schombert, J. M., \& Eder, J. 1997,
        \apj, 481, 157

\reference{URC} Persic, M., Salucci, P., \& Stel, F. 1996, \mnras, 281, 27

\reference{PCM} Pfenniger, D., Combes, F., \& Martinet, L. 1994,
        \aap, 285, 79

\reference{PW} Prochaska, J. X., \& Wolfe, A. M. 1997, \apj, 487, 73

\reference{QP} Quillen, A. C., \& Pickering, T. E. 1997, \aj, 113, 2075

\reference{EROS2} Renault, C. \etal 1997, \aap, 342, L69

\reference{Rhee} Rhee, M.-H. 1996, Ph.D. thesis, University of Groningen

\reference{RS} Richter, O.-G., \& Sancisi, R. \aap, 290, 9

\reference{roman} Romanishin, W., Krumm, N., Salpeter, E. E., Knapp, G. R.,
        Strom, K. M., \& Strom, S. E. 1982, \apj, 263, 94

\reference{RB} R\"onnback, J., \& Bergvall, N. 1994, \aap, 108, 193

\reference{rosat} Rosati, P., Della Ceca, R., Norman, C., \& Giacconi, R.
	1998, \apj, 492, L21

\reference{RFT} Rubin, V. C., Ford, W. K., \& Thonnard, N. 1980, \apj, 238, 471

\reference{SH} Salpeter, E. E., \& Hoffman, G. L. 1996, \apj, 465, 595

\reference{SvA} Sancisi, R. \& van Albada, T. S. 1987, in IAU Symp. No. 117:
        Dark Matter in the Universe, eds. Knapp, G. \& Kormendy, J.
        (Dordrecht: Reidel), 67

\reference{S90} Sanders, R.~H. 1990, \aapr, 2, 1

\reference{warps} Scharf, C., Jones, L. R., Ebeling, H., Perlman, E.,
	Malkan, M., Wegner, G. \apj, \apj, 477, 79

\reference{CO} Schombert, J.~M., Bothun, G.~D., Impey, C.~D., \& Mundy, L.~G.
        1990, \aj, 100, 1523

\reference{LSBcat} Schombert, J.~M., Bothun, G.~D., Schneider, S.~E., \&
        McGaugh, S.~S. 1992, \aj, 103, 1107

\reference{lsbtf} Sprayberry, D., Bernstein, G.~M., Impey, C.~D.,
        \& Bothun, G.~D. 1995b, ApJ, 438, 72

\reference{SB} Steinmetz, M., \& Bartelmann, M. 1995, \mnras, 272, 570

\reference{VT} Trimble, V. T. 1987, \araa, 25, 425

\reference{TF} Tully, R. B., \& Fisher, J. R. 1977, \aap, 54, 661

\reference{TV} Tully, R. B., \& Verheijen, M. A. W. 1997, \apj, 484, 145

\reference{TWV} Tyson, J. A., Wenk, R. A., \& Valdes, F. 1990, \apj, 349, L1

\reference{vAS} van Albada, T. S., \& Sancisi, R. 1986, Phil. Trans.
	R. Soc. A, 320, 447

\reference{vdH} van der Hulst, J.~M., Skillman, E.~D., Smith, T.~R.,
        Bothun, G.~D., McGaugh, S.~S. \& de Blok, W.~J.~G. 1993, \aj, 106, 548

\reference{vdK} van~der~Kruit, P.~C. 1987, \aap, 173, 59

\reference{vdM} van der Marel, R. P. 1991, \mnras, 253, 710

\reference{VMOK} Vogt, S. S., Mateo, M., Olszewski, E. W., \& Keane, M. J.
        1995, AJ, 109, 151

\reference{WF} White, D. A., \& Fabian, A. C. 1995, \mnras, 273, 72

\reference{WK} White, R. E., \& Keel, W. C. 1992, \nat, 359, 129

\reference{WNEF} White, S. D. M., Navarro, J. F., Evrard, A. E. \& Frenk, C. S.
        1993, \nat, 366, 429

\reference{WCH} Wiklind, T., Combes, F., \& Henkel, C. 1995, \aap, 297, 643

\reference{ZL} Zaritsky, D. \& Lin, D. N. C. 1997, \aj, 114, 2545

\reference{Z97} Zaritsky, D., Smith, R., Frenk, C., \& White, S. D. M.
        1997, \apj, 478, 39

\reference{ZW} Zaritsky, D. \& White, S. D. M. 1994, \apj, 435, 599

\reference{ZHBM} Zwaan, M. A., van der Hulst, J. M., de Blok, W. J. G., \&
        McGaugh, S. S. 1995, \mnras, 273, L35

\reference{ZH} Zwicky, F., \& Humason, M. L. 1964, \apj, 139, 269

\end{references}
\end{document}